\documentclass[english, 11pt, openright, twoside]{report}
\usepackage{babel, cite, amsmath, amsfonts, amssymb, mathrsfs, bbm, bm, graphicx, pdfsync, textcomp}
\usepackage[T1]{fontenc}
\usepackage[latin9]{inputenc}
\usepackage[a4paper,total={12cm,20cm},centering,includehead]{geometry}
\usepackage{fancyhdr}
\usepackage[Sonny]{fncychap}

\usepackage{hyperref}
\hypersetup{
	bookmarks=true,
	colorlinks=true}

\newcommand{\HRule}{\rule{\linewidth}{0.5mm}}
\newcommand{\mc}[1]{\mathcal{#1}}

\newcommand{\ul}[1]{\underline{#1}}
\newcommand{\Oslash}{\mc{O} \hspace{-1.4ex}/\hspace{.5ex}}
\newcommand{\Lslash}{\mbox{\it\L}}
\renewcommand{\exp}[1]{\mbox{exp} \left [ #1 \right ]}

\newcommand{\Y}{\text{\it\textyen}}
\newcommand{\T}{\text{\TH}}
\newcommand{\HB}{{\hbar_*}}

\begin{document}
\pagenumbering{roman}
\pagestyle{empty}
\begin{titlepage}
\begin{center}

% Upper part of the page
\begin{tabular}{c c}
\raisebox{-.5\height}{\includegraphics[width=1.5cm]{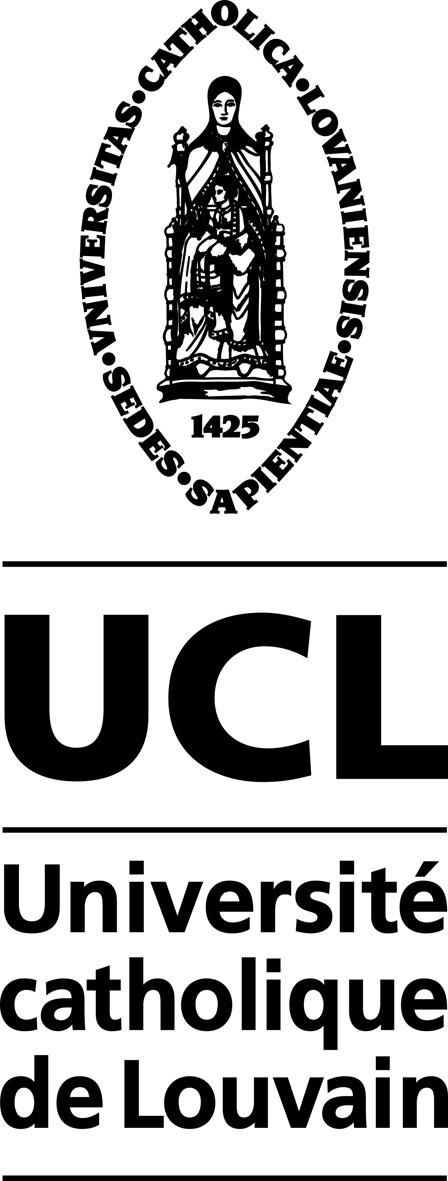}}
 &
\begin{minipage}[!h]{9.8cm} \textsf{ \textbf{
\flushright \small Universit\'e catholique de Louvain\\
Secteur des Sciences et Technologies\\
Institut de Recherche en Math\'ematique et Physique\\
Centre for Cosmology, Particle Physics and Phenomenology\\}}
\end{minipage}
\end{tabular}\\[3cm]

%\includegraphics[width=3cm]{files/logo}
%\begin{center}
%\textsf{ \textbf{
%Universit\'e catholique de Louvain\\
%Secteur des Sciences et Technologies\\
%Institut de Recherche en Math\'ematique et Physique\\
%Centre for Cosmology, Particle Physics and Phenomenology\\}}
%\end{center}

% Title
\HRule \\[0.4cm]
{ \huge \bfseries Two-dimensional quantum dilaton gravity and the quantized cosmological constant}\\[0.4cm]

\HRule \\[1cm]
Doctoral dissertation presented by\\
Simone \textsc{Zonetti}\\
in fulfilment of the requirements for the degree of Doctor in Sciences
\\[1cm]
% Author and supervisor
{\large
\begin{tabular}{ p{5.5cm} p{5.5cm} }
\emph{Supervisor:} & \hfill \emph{Members of the Jury:} \\
Prof. Jan \textsc{Govaerts} & \hfill Prof. Andr\'e \textsc{F\"uzfa}\\
& \hfill Prof. Jean-Marc \textsc{G\'erard}\\
& \hfill Prof. Jan \textsc{Govaerts}\\
& \hfill Prof. Daniel \textsc{Grumiller}\\
& \hfill Prof. Fabio \textsc{Maltoni}\\
& \hfill Prof. Christophe \textsc{Ringeval}
\end{tabular}}

%\begin{tabular}{p{5.5cm} p{5.5}}
%\begin{minipage}{5.5cm}
%\begin{flushleft} \large
%\emph{Supervisor:}\\
%Prof. Jan \textsc{Govaerts}
%\end{flushleft}
%\end{minipage} &
%\begin{minipage}{5.5cm}
%\begin{flushright} \large
%\emph{Members of the Jury:} \\
%Prof. Andr\'e \textsc{F\"uzfa}\\
%Dr. Jean-Marc \textsc{G\'erard}\\
%Prof. Jan \textsc{Govaerts}\\
%Dr. Daniel \textsc{Grumiller}\\
%Prof. Fabio \textsc{Maltoni}\\
%Prof. Christophe \textsc{Ringeval}
%\end{flushright}
%\end{minipage}
%\end{tabular}

% Bottom of the page
\end{center}
\end{titlepage}
\cleardoublepage
\pagestyle{fancy}
%\chapter*{Acknowledgements}
%\input{files/thx}
\tableofcontents
\chapter{Introduction}
	\pagenumbering{arabic}
	Beauty, elegance, simplicity. The one ultimate goal of science is nothing but finding a unified theory that would embrace the whole Universe under a few simple principles\footnote{And even fewer numbers.}. Depending on the personal attitude, we have come a long way, or we are not even close.\\
One thing can not be doubted: we are on the right path. Modern science has managed not only to unveil many of the mysteries of Nature, but it has proved that different phenomena are often described within a single theoretical framework. And mathematics is the language to use to this purpose.\\
In the same way as advances in philosophy determine the need of new vocabulary in order to describe the depths of human thought, exact sciences, and physics in particular, require and generate new mathematics. With a difference: physicists have to deal with the most severe and obscure of mentors. Numbers.\\
Numbers and data\footnote{When measured properly!} do not have tolerance for theories that cannot reproduce them. To fit the data we often have to sacrifice the beauty of a newly found theory, bending our aesthetic taste to the greater good of a theory that lives in the real world. It is frustrating to have only a glimpse of this elegance, before it crumbles under blows of sigma's.\\
On the other hand numbers and data are always suggesting us something, even if they do it in a way often hard to understand. When they fit our theories they are gratifying us, when our theories do not fit them it means that something is amiss.\\
The frustration is then overcome by the feeling that beyond lies perfection. Notably, beyond the frustration of seeing Newton's law of universal gravitation unable to explain the precession of the perihelion of Mercury lies General Relativity.\\
This work humbly attempts to contribute to what is, perhaps, the most ambitious endeavour in theoretical physics today: Quantum Gravity. 
\smallbreak
Once again we have a number. A very small one\footnote{for example in Planck units.}. So small that it is ironic how much trouble we run into because of it: the cosmological constant is, roughly, related to the rate at which the Universe is expanding.\\
We also have not one, but two beautiful theories: General Relativity and Quantum Field Theory. If we look at these two theories at the apex of their perfection, for instance considering a semi-classical model comprising General Relativity and a Supersymmetric Standard Model of particle physics, and we wonder about the cosmological constant we have a result worth of their perfection: the cosmological constant vanishes.\\
But our number is not zero, it is just very small. We also know - from other numbers! - that Supersymmetry, if it is there, is a broken symmetry. We can try to accommodate things, but we soon realize that we do not have much luck. By trying to predict the value of the cosmological constant using General Relativity and Quantum Field Theory we get what has been called ``the worst theoretical prediction in the history of physics''. A number that differs from the experimental one up to a factor $10^{120}$.\\
Beyond our frustration in seeing General Relativity and the Standard Model struggling with one single number, hopefully lies Quantum Gravity.
\smallbreak
This thesis is structured as follows:
\begin{itemize}
\item[*] Chapter \ref{ch:cc} is a general review on the cosmological constant problem. After a brief summary of the historical background and observations, we will discuss the relation between the cosmological constant and the vacuum, in General Relativity as well as in the Standard Model of particle physics.
We will then review how a quantized cosmological constant appears in the case of one-dimensional gravity, and discuss the potential generalization of this scenario.
\item[*] Chapter \ref{ch:dmax} discusses the classical treatment of dilaton-Maxwell gravity. We introduce the general model and find a dual formulation in terms of decoupled Liouville fields. We obtain its classical solutions, with particular attention to the role of the cosmological constant, and we discuss the behaviour of the space-time curvature and the presence of singularities. In the last part of this chapter we formulate the theory in the Hamiltonian and BRST formalisms, leading the way to quantization.
\item[*] Chapter \ref{ch:qt} deals with the quantum theory. The algebra of constraints is quantized and central extensions in the commutation relations are identified and eliminated. We discuss the action of the quantum constraints and determine the equations for the cosmological constant. Finally we describe the spectrum of the cosmological constant for the lower excitations of the model.
\item[*] Chapter \ref{ch:end} contains a summary and a final discussion of the results and perspectives of this work.
\end{itemize}
\cleardoublepage

\pagestyle{empty}
\vspace{-4cm}
\begin{flushright}
\textit{est igitur ni mirum id quod ratione sagaci\\
quaerimus, admixtum rebus, quod inane vocamus.}\\[.4cm]
\small{Che sia dunque fr\`a corpi il v\^oto sparso,\\
Bench\`e mal noto a' nostri sensi infermi,\\
Per l'addotte ragioni \`e chiaro e certo.\\[.4cm]
That which we're seeking with sagacious quest \\
Exists, infallibly, commixed with things- \\
The void, the invisible inane.}\footnote{Titus Lucretius Carus \textit{de rerum natura} (Liber I, vv. 369-370)\\Translations: Alessandro Marchetti, William Ellery Leonard.}
\end{flushright}
\chapter{The cosmological constant problem}
	\label{ch:cc}
	\pagestyle{fancy}
	%The cosmological constant has always been a troubled character in theoretical physics and a great number of studies have been devoted to find a solution to the issues it  poses \cite{Carroll:2000fy,Bousso:2007gp,Rugh:2002,Carroll:1992,Weinberg:1989,Weinberg:2000}. \\
%This chapter is dedicated to a general review of the subject, with a focus on the elements more relevant to our purpose, guiding the reader towards the core of this thesis.\\
%After a historical introduction we will discuss on the value of the cosmological constant. Then we will look at its relation with the vacuum, in General Relativity alone and in the semi-classical theory in which the Standard Model of particle physics interacts with Einstein's theory.\\
%In the second last Section of this Chapter, we will review the case of one-dimensional gravity coupled to a generic matter system. In this setting one is able to obtain a mechanism that determines the value of the cosmological constant through the realization of the quantum symmetry of the model.\\
%Finally, we will give an outline of the original work contained in this manuscript.
This chapter contains a non exhaustive review of the cosmological constant and the coincidence problems. It is based on few excellent reviews \cite{Weinberg:1989,Weinberg:2000,Carroll:1992,Peebles:2002gy,Carroll:2000fy,Bousso:2007gp,Rugh:2002,Bousso:2012dk}, to which we refer for further details and a broader overview of the subject.
\section{Historical background}
\subsubsection{$\Lambda$ in Newtonian cosmology}
It was already in the 1890's that a sort of cosmological constant term made its first appearance, as a modification of the Newtonian theory of gravity. It was in fact realized by von Seeliger \cite{Seeliger:1895fk} and Neumann \cite{Neumann:1896lr} that in the standard Newtonian cosmology an infinite cosmos filled with a static and uniform distribution of matter was not possible, as the integrals giving Newton's force and potential are formally divergent (see \cite{Norton:1999lr} for a modern review).\\
A ``conservative'' solution proposed to save Newton's theory and to conclude that no static and homogeneous Universe is permitted. The matter distribution, for instance, could have been peaked at a given point in space (geocentrism seems hard to kill!) and falling off more rapidly than $r^{-3}$ (the so called ``island Universe''), so that the integrals converge.\\
Alternatively Neumann proposed to include an additional factor $e^{-\sqrt{\Lambda}r}$ in Newton's potential, while von Seeliger's idea was to consider a modification of Newton's law in the form:
\begin{equation}\label{vonseeligereq}
\nabla^2\varphi-\Lambda\varphi = 4 \pi G \rho
\end{equation}
In either cases by making $\Lambda$ sufficiently small its effects on the gravitational dynamics would have been visible only at large distance scales.\\
Einstein himself embraced this idea, which he applied later in General Relativity to accommodate the picture of a static Universe.
\subsubsection{Introducing $\Lambda$ in General Relativity}
To better understand the important dynamical role played by the cosmological constant in General Relativity and the original reason to introduce it, we can focus our attention on the large scale dynamics of space-time.
Let us consider Einstein's equations (in 3+1 dimensions) in their original form, without any cosmological constant term:
\begin{equation}\label{einsteineq}
  R_{\mu\nu} - \frac{1}{2}Rg_{\mu\nu} = 8\pi GT_{\mu\nu}\ .
\end{equation}
At cosmological scales the Universe can be considered very well approximated by a spatially homogeneous and isotropic manifold, so that one can take as a solution the famous Friedman-Lema\^{i}tre-Robertson-Walker line element:
\begin{equation}
  {\rm d}s^2 = -{\rm d}t^2 + \bar{a}(t)^2\left[
  {\frac{{\rm d}r^2}{1-kr^2}} + r^2 {\rm d}\Omega^2\right]\ ,
\end{equation}
where ${\rm d}\Omega^2 = {\rm d}\theta^2 + \sin^2\theta {\rm d} \phi^2$ is the metric on a round two-sphere.  The parameter $k$ is a curvature parameter, which takes the values $+1$, $0$, or $-1$ for positive, flat and negative spatial curvature, respectively. The scale factor $\bar{a}(t)$ in front of the spatial portion of the line element determines the size of the spatial submanifold as a function of time.\\
It is useful to consider a normalized scale factor $a(t)=\bar{a}(t)/a_0$, where $a_0$ is the scale factor as measured at a given time.\\
With an additional, but still rather general approximation we can consider the gravitational sources, \emph{i.e.} matter and radiation fields, to be modelled by a perfect fluid, so that the energy-momentum tensor on the r.h.s. of Einstein's equations can be taken in the form
\begin{equation}\label{perfectfluid}
  T_{\mu\nu} = (\rho +p)U_\mu U_\nu + p g_{\mu\nu}\ ,
\end{equation}
where $U^\mu$ is the fluid four-velocity, $\rho$ is the density and $p$ the pressure.\\
In order to reproduce the FLRW solution, the gauge fixing (\emph{i.e.} the choice of reference frame) to be considered is the one in which the normal to the spatial hypersurfaces is the normalized fluid four-velocity (\emph{i.e.} the so-called co-moving frame). In this way \eqref{einsteineq} reduce to the two Friedman equations:
\begin{subequations}\label{friedmaneq}
\begin{align}
\left(\frac{\dot a}{a}\right)^2 &= \frac{8\pi G}{3}\rho - \frac{k}{a^2a_0^2}\ ,\\
\frac{\ddot a}{a} &= -\frac{4\pi G}{3}(\rho + 3p)\ .
\end{align}
\end{subequations}
Einstein was guided by the idea of finding a static solution, \emph{i.e.} $\dot{a}=0$, to these equations, since according to the observations of the time ``the most important fact that we draw from experience is that the relative velocities of the stars are very small as compared with the velocity of light''\cite{Einstein:1917fk}.\\
He was also looking for a connection in his model between mass distribution and geometry and, following Mach's work, he expected matter to set inertial frames.
With these ideas in mind he found out that no static Universe can be obtained from \eqref{friedmaneq} for a non-negative pressure $p$. This was of course quite unsettling, as ordinary astronomical matter always give positive contributions to the pressure.\\
A modification of \eqref{einsteineq} seemed necessary. The simplest addition to be made, using only the metric and its derivatives, is a term linear in $g_{\mu \nu}$. So the cosmological constant made its first appearance in Einstein's equations in 1917:
\begin{equation}\label{einsteineqcc}
  R_{\mu\nu} - \frac{1}{2}Rg_{\mu\nu} +\Lambda g_{\mu \nu}= 8\pi GT_{\mu\nu}\ .
\end{equation}
It is worth noticing, however, that the Newtonian limit of these equations is not \eqref{vonseeligereq}, but rather $\nabla^2\varphi+\Lambda = 4 \pi G \rho$. In this way also the Friedman equations \eqref{friedmaneq} acquire an extra term, and read:
\begin{subequations}\label{friedmaneqcc}
\begin{align}
\left(\frac{\dot a}{a}\right)^2 &= \frac{8\pi G}{3}\rho - \frac{k}{a^2a_0^2}+ \frac{\Lambda}{3}\ ,\\
\frac{\ddot a}{a} &= -\frac{4\pi G}{3}(\rho + 3p)+ \frac{\Lambda}{3}\ .
\end{align}
\end{subequations}
A static solution can be easily found, with all of $\rho, p, \Lambda$ non negative and $k=1$, and takes the name of ``Einstein static universe''. This seemed very appealing also for another reason: the fact that Einstein's static universe is spatially closed obviated the need to specify any boundary conditions at infinity, a feature which Einstein found to be attractive: ``Boundary conditions presuppose a definite choice of the system of reference, which is contrary to the spirit of relativity'' \cite{Einstein:1952fj}.\\
But both staticity and the relation of matter and inertial frames were about to tremble with new results from both the theoretical and the observational side.
Already in 1917 de Sitter proposed a solution of Einstein's equations \eqref{einsteineqcc} which would ensure staticity without the need of any matter distribution. With de Sitter's choice of coordinates the solution appears to be static:
\begin{equation}
ds^2=dr^2+\lambda^2 sin^2\left (\frac{r}{\lambda}\right )\left [d\psi^2+sin^2\left (\psi \right )d\theta^2\right ]-cos^2\left( \frac{r}{\lambda}\right )dt^2
\end{equation}
where $\lambda$ is a constant. While this would have satisfied Einstein's desire of a static cosmological model, at the same time it was clearly stating that the link between matter and inertial frames was a misconception.\\
At the same time Keeler, Slipher and Campbell were observing that distant objects exhibit for the greater part redshifted spectra, as if they are all receding from Earth. At first this discovery found an explanation in the so-called de Sitter static universe, in which the coordinate system is time independent but test objects would not be at rest.\\
Soon enough the expansion of the Universe was discovered by Lema\^itre and Hubble and finally the assumption of staticity dropped. But it was too late: the cosmological constant term had already taken its place as a legitimate addition to Einstein's equations and $\Lambda$ itself needed to be considered as a free parameter to be tuned by observations and explained theoretically.
\section{Observations and experiments}\label{sec:obs_exp}
Even before getting to precise and sophisticated astronomical measurement, it is possible to consider some simple arguments that can give an idea of the value we should expect for $\Lambda$. In the following we will adopt the conventional ``natural units'' by defining $c=\hbar=G=1$.
Dimensionally the cosmological constant is an inverse length squared. Empirically, one can consider it to introduce a length scale $r_\Lambda \sim |3/\Lambda|^{1/2}$. Above such scale the effects of $\Lambda$ would dominate the gravitational dynamics.\\
Another natural scale to be accounted for is the Planck scale:
\begin{equation}
l_P \sim 10^{-33} cm\ .
\end{equation}
As we see that General Relativity provides a very good description of space-time well above $l_P$, with no sign of a cosmological constant, we can already infer that $r_\Lambda \gg l_P$.\\
Let us consider, for instance, a matter free universe and a positive cosmological constant. We can find that the only possible isotropic solution of Einstein's equation is de Sitter (dS) space, which exhibits a cosmological horizon $\sim r_\Lambda$ \cite{hawking:1973}. A cosmological horizon is the largest observable distance scale, \emph{i.e.} the boundary of the causally connected region for a given observer. The presence of matter will only decrease the horizon radius \cite{Gibbons:1977ceh}. As we can roughly consider the current cosmological horizon to be $r_c \sim 10^{27} cm  \gg l_P$, by the inverse relation between $r_\Lambda$ and $\Lambda$ we can see that the cosmological constant has to be small.\\
On the other hand for a negative cosmological constant there is a timescale $t_\Lambda \sim |3/\Lambda|^{1/2}$ by which the Universe would collapse on itself, disregarding the presence of spatial curvature \cite{Edwards:1972lr}. As we see that the Universe is much older than the Planck time $t_P \sim l_P/c$ we can again conclude that $\Lambda$ is small in Planck units.\\
Putting together these empirical considerations, we can claim that:
\begin{equation}
3t_c^{-2} < \Lambda <3r_c^{-2} \ ,
\end{equation}
where $t_c,r_c$ are the observed scales. \\
In Planck units these scales are $t_c = r_c \sim 10^{60}$, so that these simple arguments put quite a stringent bound on the value of the cosmological constant
\begin{equation}\label{empiricalCC}
|\Lambda| < 10^{-120}\ .
\end{equation}
As stated in \cite{Bousso:2007gp}:
\begin{quote}
\textit{These conclusions did not require cutting-edge experiments: knowing only that the world is older than 5000 years and larger than Belgium would suffice to tell us that $|\Lambda|\ll 1$} 
\end{quote}
\subsubsection{Measuring the cosmological constant}\label{sec:measuring_cc}
It is now known that the cosmological constant is non vanishing. This was first discovered in 1998 with the measurement of the apparent luminosity of distant supernovae \cite{Perlmutter:1998np,Riess:1998cb} which indicated an accelerated expansion of the Universe in a way consistent with a positive cosmological constant \cite{tegmark:2004}:
\begin{equation}
\Lambda = (1.48 \pm 0.11) \times 10^{-123}
\end{equation}
in Planck units, and not consistent with $\Lambda=0$.

A general and preliminary formulation of the cosmological constant problem\footnote{Often called the \emph{old} cosmological constant problem, in contrast with the \emph{new} problem, known also as the coincidence problem. With no ambiguity we will always refer to the former as cosmological constant problem and to the latter as coincidence problem.} is then readily stated: (how) is it possible to predict the value of the cosmological constant in a solid theoretical framework?\\
Our intuition is that this has to do with the real nature of space-time. Quoting Guth \cite{Guth:1980zm}:
\begin{quotation}
\textit{The reason $\Lambda$ is so small is of course one of the deep mysteries of physics. The value of $\Lambda$ is not determined by the particle theory alone, but must be fixed by whatever theory couples particles to quantum gravity}.
\end{quotation}

\section{The cosmological constant and the vacuum}\label{sec:cc_vac}
It is easy to see from \eqref{friedmaneqcc} that if one considers an expanding Universe ($\dot{a} > 0$), the rate of expansion is slowed by the presence of matter and (positive) pressure, while the rate of expansion is speeded up by the presence of a positive cosmological constant. In this way $\Lambda>0$ acts as a cosmic repulsive force, as it is clear also by considering \eqref{einsteineqcc} in the absence of matter.\\
In particular one can take the example of the Schwarzschild solution, in the presence of an unspecified cosmological constant $\Lambda$:
\begin{equation}
ds^2 = \frac{dr^2}{1-\frac{2M}{r}-\frac{\Lambda}{3}r^2} - \left ( 1 - \frac{2M}{r} -\frac{\Lambda}{3}r^2  \right )dt^2 +d\Omega^2
\end{equation}
and take its Newtonian limit, so that the Newtonian potential is:
\begin{equation}
\varphi = -\frac{M}{r}-\frac{\Lambda}{6} r^2 .
\end{equation}
So even for $M=0$ a particle moving in such a field feels a repulsive (resp., attractive) radial force for a positive (resp., negative) $\Lambda$.\\
Effects are in principle present also in Solar system measurements. For instance $\Lambda \neq 0$ provides an additional perihelion shift for Mercury of $\Delta \simeq \Lambda \times 10^{42} cm^2$ seconds of arc per century \cite{Rindler:1969kx}, so that for a sufficiently small value of the cosmological constant the effect might be undetectable.\\
In a more field theoretical example let us consider a single scalar field $\phi$, with a potential $V(\phi)$ and an action
\begin{equation}
S = \int d^4x\ \sqrt{-g}\left[-\frac{1}{2} g^{\mu\nu}
\partial_\mu\phi \partial_\nu\phi - V(\phi)\right],
\end{equation}
where $g$ is the determinant of the metric tensor. Then the energy-momentum tensor is
\begin{equation}
T_{\mu\nu} = -\partial_\mu\phi \partial_\nu\phi - \frac{1}{2} (g^{\rho\sigma}\partial_\rho\phi \partial_\sigma\phi) g_{\mu\nu} - V(\phi)g_{\mu\nu}\ .
\end{equation}
The ground state of the theory is then the one in which no kinetic contributions are present, \emph{i.e.} $\partial_\mu\phi =0$, so that $T_{\mu\nu} =- V(\phi_0)g_{\mu\nu}$ (as also suggested by Lorentz invariance), where $\phi_0$ minimizes $V$. It is easy then to interpret such a energy-momentum as the one of a perfect fluid \eqref{perfectfluid} with $\rho_{vac}=V(\phi_0),\ p_{vac}=-V(\phi_0)$.\\
On the other hand it is clear that it contributes to Einstein's equations in the same way a cosmological constant, so that a link between vacuum energy and $\Lambda$ is established and a total cosmological constant could be defined in terms of a ``bare'' contribution $\Lambda_0$ and a ``matter'' one $-8 \pi G \rho_{vac}$.\\

While this is quite trivial for classical physics, when quantum mechanics comes into play things get quite more complicated, as it was realized already in the early '70s \cite{Zeldovich:1971fk,Linde:1974at,Veltman:1974au}.\\
In particular the temptation of setting $\Lambda=0$ (tuning for example the bare cosmological constant) looses its appeal when one realizes that the quantum fluctuations of the vacuum contribute to the energy-momentum tensor in a way that gives essentially a cosmological constant term. And this turns an interesting issue about the tuning of a parameter into a very challenging problem in quantum field theory and, perhaps, quantum gravity.\\
In the usual harmonic oscillator interpretation of Quantum Field Theory (QFT) every mode of every field carries a zero point energy which contributes to the energy-momentum tensor of the vacuum. Equivalently, from the perspective of Feynman diagrams, the vacuum is in fact filled with virtual particle-antiparticle loops, so that from the quantum mechanical point of view the ``vacuum'' is not ``empty''. This becomes problematic when one considers the many contributions to $\rho_{vac}$ coming from the different fields of the chosen QFT, which sum up and are usually proportional to the fourth power of the cutoff scale of the model.
\section[The cosm. const. and the energy density of the quantum vacuum]{The cosmological constant and the energy density of the (quantum) vacuum}\label{sec:cc_q_vac}
To better understand the role of the vacuum it is important to carefully consider its properties in the quantum field theories describing known particles and in the current cosmological models.\smallbreak
The Standard Model of Particle Physics is one of the most successful physical theories ever conceived. In the last three decades plenty of theoretical predictions have been confirmed by experiments and many of the outcomes of experiments have been incorporated in the model.\\
It describes matter on a fixed background space-time as bound states of leptons and quarks which are interacting through three fundamental quantum interactions: the so-called electromagnetic, weak and strong interactions. The first and second of these find a beautiful description in the unified electroweak theory (Glashow-Salam-Weinberg theory), while the theory of strong interactions, quantum chromodynamics (QCD), stands on its own. An additional coupling of the fields to the Higgs field(s) plays a fundamental role in generating the masses of all particles.\\
Each sector of the Standard Model has its own vacuum energy density, which will contribute to the cosmological constant in the sense described previously. In addition, any additional quantum field still to be discovered gives also its contribution.
\subsection{Quantum Electrodynamics}
In classical field theories, as for instance classical electromagnetism, physical configurations are described by infinitely many degrees of freedom, namely fields that take values at every point of space-time.\\
In the quantization procedure these fields (or more precisely, their components in the case of tensor fields) are replaced by quantum operators, which have to obey specific commutation relations. This means that in general the product of two operators might not be commutative any longer and non vanishing commutators will be proportional to the reduced Planck constant $\hbar$ or its powers. This ensures that in the limit $\hbar\rightarrow0$ the classical commutative product is recovered.\\
For the free electromagnetic field the classical Hamiltonian density has a quantum counterpart in which classical fields are replaced by quantum operators: $\hat{\mc{H}}=\frac{1}{2}\left (\hat{E}^2 + \hat{B}^2 \right )$.\\
The vacuum state $\vert\Omega\rangle$ of the theory is taken to be the one that minimizes the energy, so that one has $\langle \Omega \vert \hat{E} \vert \Omega \rangle = 0$ and $\langle \Omega \vert \hat{B} \vert \Omega \rangle = 0$.
However $\langle \Omega \vert \hat{E}^2 \vert \Omega \rangle \neq 0$ and $\langle \Omega \vert \hat{B}^2 \vert \Omega \rangle \neq 0$, and the total energy of the vacuum state can be expressed in terms of wave numbers $\omega_k$ of the plain wave expansion of $E$ and $B$ as:
\begin{equation}
\langle \Omega \vert \hat{\mc{H}} \vert \Omega \rangle = \delta^3_k(0)\int d^3k\frac{1}{2}\hbar \omega_k
\end{equation}
The divergent Dirac's delta in momentum space $\delta^3_k(0)$ can be regularized by quantizing the system in a finite volume $V$, to be taken later on to infinity, with suitable boundary conditions. This allows to exploit the equivalence between field modes and quantum harmonic oscillators. In this way the 3-dimensional Dirac $\delta$ can be written in integral form $(2\pi)^3\delta^3(k) = \int d^3x e^{ikx}$ and produces a factor $V$ when $k=0$ is imposed.\\
The divergent integral can be kept finite by introducing an ultra-violet cutoff, so that a finite result for the energy density is obtained in the limit $V\rightarrow\infty$:
\begin{equation}\label{QEDvac}
\rho_{vac} = \frac{\hbar}{8\pi^2}\omega^4_{max}
\end{equation}
Furthermore, when EM interactions are taken into account, one expects additional contributions to the vacuum energy density proportional to powers of the fine structure constant $\alpha_{QED} = 1/137$. These contributions are generated, as mentioned before, by all those (virtual) processes which involve no asymptotic states, \emph{i.e.} no incoming and outgoing particles which can be detected by experiments, and can be called, roughly speaking, electron-positron loops (or zero-point functions). These processes are completely overlooked in the calculation of scattering amplitudes, where only energy differences are relevant,  \emph{because these loops produce a shift of the zero-point energy}.
On the other hand when dealing with General Relativity, and the cosmological constant, for the reasons explained in the previous paragraph, they cannot be ignored, \emph{because these loops produce a shift of the zero-point energy}.\smallbreak
To estimate the contribution of QED to the vacuum energy then one can roughly consider \eqref{QEDvac} and the electroweak scale $\lambda_{EW} \sim 100 \ GeV$. This is the energy scale at which QED is unified with the weak interactions in the Electroweak model. We get a rough estimate, in Planck units:
\begin{equation}
\rho_{vac} \sim 10^{-69} 
\end{equation}
which is already some 50 orders of magnitude of discrepancy with the expected value \eqref{empiricalCC}. 
If one wants to \emph{roughly} extrapolate such a \emph{rough} result up to the Planck scale, assuming that the Quantum Field Theory  framework remains valid up to such scale (with no supersymmetry), and so considering a cutoff energy of the order of the Planck scale itself, the quite discouraging and expected result is $\rho_{vac} \sim 1$, so that there are $120$ orders of magnitude between (an extremely rough) prediction and (empirical, but reasonable) expectations.
\subsection{Electroweak theory and the Higgs sector}
As mentioned above, QED is unified with the Weak interactions in the Electroweak (EW) theory.\\
The masses of particles are generated through a ``spontaneous symmetry breaking'' mechanism, a situation in which the symmetries of the dynamics (and hence of the Lagrangian) are not present in the vacuum state.\\
In particular one can consider a coupling of the massless EW theory to an Higgs sector (whose exact form has still to be determined in fact, allowing for quite a variety of models). The vacuum expectation value of the Higgs field(s) is non-vanishing when the symmetry is broken, and the masses will be proportional to such value, times coupling constants.\\
For instance in the simplest Higgs sector, comprising a single complex scalar field, one considers, following symmetry considerations and the requirement of renormalizability, a potential in the form:
\begin{equation}\label{higgspot}
V(\phi) = V_0 -\mu^2\phi^2+g_H\phi^4
\end{equation}
where $g_H$ is a self coupling constant and $\mu$ is related to the vacuum expectation value of the Higgs field itself $\mu^4=4g_H^2\langle \phi \rangle^4$.
In turn $\langle \phi \rangle$ can be inferred from the Fermi coupling constant and estimated to be $\sim 250\ GeV$\cite{weinberg:qtf2}.\\
The potential above is minimized for $\phi^2=\mu^2/2g_H$, where $V_{min}=V_0 - \mu^4/4g_H = \rho_{vac}^H$. In the assumption of $V_0=0$, the Higgs coupling can be estimated as $g_H\sim\alpha_{QED}^2$\cite{weinberg:qtf2}.\\
Finally one gets an estimate for the vacuum energy density of:
\begin{equation}
\rho_{vac}^H = - \mu^4/4g_H \sim -10^5\ GeV^4 \sim -10^{-71}
\end{equation}
which again is $\sim 50$ orders of magnitude off.\\ It is clear that all these estimates are strongly model dependent, and for instance one could assume $V_0= \mu^4/4g_H$, and obtain a vanishing $\rho_{vac}^H$. This, however, would require an extreme fine tuning.
\subsection{Quantum Chromodynamics}
Due to its particular behaviour, by which the theory is highly non perturbative at low energies, and asymptotically free at higher scales, the study of the vacuum in QCD is a challenging issue. It is expected that quarks and gluons, the fermions and gauge bosons of the theory, form condensates at low energies, so that the vacuum expectation value of the fields is non vanishing.\\ While the estimates of the vacuum energy density are strongly model dependent, they can be generally considered to be in the form of a factor times $\lambda_{QCD}^4$, where $\lambda_{QCD}$ is the scale at which perturbation theory is no longer applicable in QCD. This characteristic scale can be taken to be approximately of the order of $\sim 10^{-1} GeV$.\\
Again, a rough estimate of $\rho_{vac}^{QCD}$ gives:
\begin{equation}
\rho_{vac}^{QCD} \sim 10^{-80}
\end{equation}
which one more time is very different from the expected value.\smallbreak
It is clear then that there are many different known contributions to the vacuum energy. They are not correlated with one another nor with a ``bare'' cosmological constant which can appear in Einstein's equations \eqref{einsteineqcc}.\\
In particular the quantum fluctuations of the vacuum in the Standard Model are dozens of orders of magnitude larger than the empirical bound \eqref{empiricalCC}. They contribute with different signs but they would have to cancel to better than a part in $10^{120}$ at present times.
\subsection{Phase transitions in the early Universe}
During its evolution, in the standard theory of the Big Bang, the Universe has rapidly expanded and cooled down. In this process it has passed through some critical temperatures, corresponding to the characteristic scales of phase transitions. Such transitions are connected with symmetry breakings, by which the vacuum looses part of its symmetric properties.\\
In this way we can think that a highly symmetric vacuum was present in the very early stages of the evolution of the Universe, while now we are left with a less symmetric one, because of the chain of phase transitions that has occurred.\\ This generally involves a symmetry breaking of the Grand Unified Theory ($\sim 10^{14}\ GeV$ \cite{Kolb:1990vq}, depending on the model assumptions),  Electroweak theory ($\sim 10^2\ GeV$) and QCD ($\sim10^{-1} GeV$).\\
If the vacuum energy density is considered as a cosmological constant this implies a series of different cosmological constant values throughout the history of the Universe. In this sense then tuning a bare cosmological constant to cancel out all contributions and reproduce the observed value of $\Lambda$ is increasingly difficult, also considering that one should account for higher order corrections to the lowest order estimates of the vacuum energy density in every sector of the Standard Model.
An additional issue is that little is known from the observational point of view on the value of the cosmological constant at earlier stages of the evolution of the Universe.\\
In the original formulation of cosmic inflation \cite{Sato:1980yn,Guth:1980zm} a large value of the vacuum energy during the GUT phase transition is needed to drive the inflationary process. Thus the use of spontaneous symmetry breaking to account for inflation requires a positive cosmological constant, which can be obtained for instance by tuning the constant value of the Higgs potential \eqref{higgspot} to a positive value to cancel the present negative vacuum energy density. It has to be said however that while inflation is very successful in solving many issues in cosmology, e.g. the dilution of monopoles and the horizon problem, it does not provide any further understanding of the cosmological constant problem.
\subsection{On measurability of the vacuum energy}
In Quantum Field Theory, and in the particular example of QED, both the zero-point energy and the higher order contributions to the vacuum energy density are a direct consequence of the quantization procedure.\\
By choosing a suitable ordering prescription for the quantum operators (\emph{i.e.} normal ordering) it is possible to remove all zero-energy contributions, order by order.\\
In spite of this, there are effects which seem to rely on the notion of the absolute value of the vacuum energy, as the Casimir effect (by which two uncharged conducting plate are subject to a force due to the ``polarization'' of the vacuum), the Lamb shift (a difference in the energies of electron orbitals due to the interaction with the vacuum) and the anomalous magnetic moment of the electron (higher order contributions to the magnetic moment of a charged particle). It is in principle possible to explain such effects on a different basis: the Casimir effect for instance could be described in terms of fluctuations of the constituents of the two conducting plates rather than fluctuations of a pre-existent vacuum. An example is Schwinger's \emph{Source theory}, where the Casimir effect is derived without notions of quantum fields and zero-point energy and higher order effects might be similarly explained \cite{Rugh:1999rt}.\\
It is therefore not clear what the vacuum energy in Quantum Field Theory is, and if it can be measured.\smallbreak
The question of the measurability of field components in QED was addressed already by Bohr and Rosenfeld in the 1930s \cite{Bohr:1983ys}. They argued that defining fields at specific space-time points is an unphysical idealization, and one should rather consider average values of field components over finite space-time region. In this way they obtain that in the limit of point-localized test particles the measured field strengths diverge.\\ As a result it is unclear how fluctuations of the test bodies and fluctuations of the fields interplay and it is not possible to determine whether ``the field fluctuations are already present in empty space or only created by the test bodies''. This translates in an ambiguity in the definition of vacuum energy density.\smallbreak
It could therefore be that the most direct measure of the properties of the quantum vacuum is indeed measuring the cosmological constant. Being a classical measurement, it would not be affected by the sort of ambiguities described above, which are confined to the quantum ``domain''. With this idea in mind one could claim that the large difference between observations on $\Lambda$ and Quantum Field Theory predictions could be a hint that there is no information on the vacuum energy to be extracted from QFTs in fixed flat space-time, and that in fact those effects (e.g. the Casimir effect metioned above) which seem to rely on this concept are to be explained from a different perspective \cite{Rugh:2002}.\\
%A more drastic approach would be to replace Quantum Field Theory with Schwinger's source theory, in which the vacuum is ``empty''.
The possibility that our understanding of the vacuum in Quantum Field Theory might not be as good as we think is intriguing indeed. However, in the development of this work, we will not consider this possibility, and rather focus on investigating the effects of quantum gravity in the framework of QFTs.

\section{The cosmological constant and the coincidence problems}

\subsection{The cosmological constant problem}

By comparing the observations of Section \ref{sec:obs_exp} with the theoretical results of Sections \ref{sec:cc_vac} and \ref{sec:cc_q_vac} the cosmological constant problem stated in Section \ref{sec:measuring_cc} can be reformulated in a more precise manner: ``Why is the measured cosmological constant so much smaller than the expected contributions to it from quantum fluctuations?'' Equivalently, assuming the estimate of $\Lambda$ from quantum fluctuations described above is accurate\footnote{In a recent review \cite{Martin:2012p5779} an alternative procedure for the calculation of the contributions of the quantum vacuum to the value of the cosmological constant is proposed, reducing the order of magnitude of the predicted cosmological constant. However such value is still far from the measured value of $\Lambda$, therefore the cosmological constant problem still stands.}: ``Why does the approximate equality $\Lambda_0 \simeq -8\pi G \rho_{vac}$ hold good to an accuracy of somewhere between 60 to 120 decimal places?''

\subsection{The coincidence problem}
We mentioned earlier that the presence of a cosmological constant determines a particular time scale $t_{\Lambda }\sim |3/\Lambda| ^{1/2}$. With the measured value of $\Lambda$ this scale is of order $t_\Lambda \sim 10^{60}$, which is of the same order of magnitude than the age of Universe today $t_{U} \sim 13.7\,\text{Gyrs}\sim 10^{60}$.\\
This is what has been called the \emph{coincidence problem}: ``Why is $%
t_{\Lambda }\sim t_{U}$ today?''\\
The epoch at which we, as obervers, have access to measuring physical quantities is constrained by the requirement that the Universe is old enough for typical stars to have produced the heavy elements required for our own very existence \cite{Dicke:1957zz}. This time scale is estimated with a combination of the constants of nature: $t_{*}\sim \alpha _{em}^{2}/Gm_{p}m_{e}$ \cite{Barrow:1986lr}, where $m_p$ and $m_e$ are the masses of the proton and the electron, respectively. Naturally, one expects that $t_U\sim O(1)t_{*}$, which is indeed the case.\\
This unexplained coincidence of two fundamental time scales, $t_{\Lambda }$ and $%
t_{\ast }$, both determined by fundamental constants, is puzzling. The coincidence problem is then simply: ``Why is $t_{\Lambda }\sim t_{*}$?''\\
This is an interesting question to answer since either we live at a special epoch $t_{U}$ when, by chance, $t_{\Lambda }\sim O(t_{U}\sim t_{*})$, or there is some deep reason, related to the solution of the cosmological constant problem, why $\Lambda $ takes this specific value.\\
In addition vacuum energy does not redshift like matter. In past epochs vacuum energy was negligible with respect to matter, while in the future matter will be diluted and vacuum energy will dominate the gravitational dynamics. The two are comparable only at a specific time, which incidentally is the epoch in which we are making this observations.

\subsection{Solving problems}
In recent years, in the field of cosmology, a great effort has been devoted to the solution of the coincidence problem rather than the cosmological constant problem itself. A common line of thinking is that indeed there could be a dynamical mechanism that ensures that $\rho _{vac}+\Lambda /8\pi G = 0$ exactly and the observed effective cosmological constant is the result of some other mechanism. In dark energy models, for instance, the effective cosmological constant is not actually constant. Instead there is a additional field (dark energy) whose energy density, at the present day and with the measurement accuracy available, mimics perfectly a cosmogical constant, driving the acceleration of the expansion of the Universe. While this can alleviate the coincidence problem, it still requires a good amount of fine tuning to ensure that at a time scale $t_*$ the Universe is dark energy dominated.\\
Another approach would modify General Relativity at scales comparable with the size of the Universe in a way that we would perceive as a cosmological constant. In both cases future experimental results and observations could confirm or falsify these approaches\cite{Bousso:2012dk}.\\
In this line of thinking we can also include examples of modified gravitational dynamics in which the metric tensor is in fact a composite object, and the modified dynamics provides a more natural way of justifying the fine tuning that shields the vacuum energy of QFT from contributing to the cosmological constant\cite{Kimpton:2012rv}.

\subsubsection{The landscape of string theory}
At present, the approach that seems to provide the best explanation for the value of the cosmological constant is the so-called \emph{landscape of string theory}. We will only summarize these arguments and we refer to the literature for further details\cite{Polchinski:2006gy,Bousso:2012dk}.\\ 
In string theory consistency conditions on the quantum dynamics (\emph{i.e.} anomaly cancellations) require space-time to be ten-dimensional. To avoid contradiction with current observations six of the spatial dimensions have to be effectively small, so that they do not affect high-energy experimental data.\\
The general class of compact six-dimensional manifolds is given by Calabi-Yau manifolds, which have been extensively studied in the past decades. Trying to estimate the number of different possible choices of compactification of the six extra dimensions one roughly gets a number of order $\sim 10^{500}$. There is then a number $\sim 10^{500}$ of different vacua in string theory, each of which has a different physical content and a different low energy behaviour.\\
In particular the value of the vacuum energy for a given choice of compactification behaves as a random variable, receiving contributions from all fields. The spectrum of $\Lambda$ then is very dense, with a spacing of order $10^{-500}$, and we can expect it to range, in absolute value, from $0$ to $1$ (in Planck units). This means that while there is a small fraction $\sim 10^{-123}$ of vacua with $|\Lambda| \le 10^{-123}$, the absolute number is still large, $\sim 10^{377}$. There is then a considerable quantity of vacua with a cosmological constant compatible with observations. So, how do we end up in this particular vacuum, with this specific value for $\Lambda$? \\
Let us assume that the Universe starts in a vacuum with $\Lambda>0$. Since the spectrum is symmetric around $0$ this is not an extremely strong restriction. This is a de Sitter Universe and it will expand exponentially in a homogeneous and isotropic way. In a classical theory this expansion will continue indefinitely, unperturbed. Quantum mechanically, however, all fields, and in particular the ones living in the six extra dimensions, are able to fluctuate, with the possibility of tunnelling effects that would modify the vacuum energy\footnote{These processes are the analogues of the Schwinger effect. Given two charged metallic plates there are chances that quantum fluctuation of the EM field will generate an electron-positron pair. The two particles, under specific conditions, will be attracted by the plates and reduce the charge on them upon collision, hence reducing the strength of the electric field. This process can be seen as the tunnelling of the system between states of different vacuum energy. In string theory the tunnelling involves the fluxes of the fields in the extra dimensions.} in a finite region, nucleating a ``bubble'' of space-time with a different value for the cosmological constant.\\
In particular we can imagine that the decay of some field in the extra dimensions generates a ``bubble'' with a vacuum energy which is smaller than the one of the original vacuum. Energy conservation will be guaranteed by the expansion of the domain wall, which will interpolate between the two different vacua, together with the creation of matter and radiation inside the ``bubble'' itself.\\ This can also be seen as a first-order phase transition and its occurrence is generally suppressed by the exponential of the action, making nucleation a relatively rare process.
In particular, if the original vacuum undergoes eternal inflation, the volume lost to vacuum decay is small on average, and the original vacuum keeps expanding. In addition in a de Sitter metric all observers are limited in their observations by a cosmological horizon. The new vacua cannot expand faster than light, so that observers in different vacua will be causally disconnected, and each vacuum is a distinct Universe at their eyes.\\
It is therefore possible that infinitely many bubble Universes, with different vacua, nucleate from an initial vacuum, generating the so-called Multiverse. In this picture all possible vacua are generated by vacuum decay. Universes with $\Lambda<0$ will collapse in a Big Crunch in a time scale $t_\Lambda$, while Universes with $\Lambda>0$ will undergo eternal inflation an nucleate themselves additional Universes.\\
We can now turn our attention to determining which of these Universes allow for observers. Given a Universe with a non vanishing cosmological constant, the maximum area of the past light-cone of any point $p$ is roughly given by $A\sim |\Lambda|^{-1}$ ($A\sim\Lambda^{-2}$ if $\Lambda<0$ and the Universe is open). This, in turn, is a rough upper bound on the entropy of the causal past of $p$. Therefore for Universes with $\Lambda \sim 1$ there is no more than few bits of information in any causally connected region and observers, in the most general sense, require to be constituted by more than few bits. We can then claim than observers can only be located in regions with $|\Lambda | \ll 1$.\\
This, however, does not tell us why we see $\Lambda\sim 10^{-123}$. In order to make quantitative predictions in this context it is first necessary to deal with the ``measure'' problem. We refer to the literature \cite{Bousso:2006ev,Bousso:2007kq,Bousso:2010vi,Bousso:2010zi} for further details. Let us restrict ourselves to the case of a positive cosmological constant. Any observer living at a time $t_*$ after the nucleation of his Universe, is bound by the causal patch measure \cite{Bousso:2006ev} to observe a cosmological constant of the order\begin{equation}
\Lambda \sim t_*^{-2} \ .
\end{equation}
So that in the case of our Universe, with $t_*\sim 13,7\,\text{Gyrs}$, the predicted value for the cosmological constant is in good agreement with the measured value. The issue of the measure of regions with non-positive cosmological constant is still open.\smallbreak
Summarizing, the landscape of string theory provides some arguments that suggest that the value of the cosmological constant that we observe, and the fact that we observe it precisely at this epoch of the evolution of our Universe, are no coincidence, but arise naturally if we allow for eternal inflation and the nucleation of bubble Universes. It is then natural that observers would find themselves in those Universes with the value of $\Lambda$ which admits their existence.

\section{What is missing in the current approaches?}

The main concern with the standard QFT analysis of the vacuum energy, in our opinion, is represented by the semi-classical framework. The vacuum energy is calculated on a flat background geometry, discarding completely any contribution which could be given by the gravitational sector. While it is true that the vacuum fluctuations of a quantum gravitational field are suppressed by the Planck scale with respect to the vacuum fluctuations of the matter fields, the dynamics of quantum gravity is completely ignored, discarding the possibility of some mechanism induced by quantum gravity that might explain the value of $\Lambda$ that we measure today.\smallbreak
In the landscape of string theory there are different issues, in our view, that are source of concern.\\
On a very general level solving the cosmological constant and coincidence problems with the introduction of six extra dimension, extended objects as strings and branes, Calabi-Yau manifolds and all the complex dynamics of string theory is hard to justify if one considers Occam's Razor as a generally applicable principle. Additional predictions have to be produced in the context of string theory, strengthening the experimental support of the theory.\\
Moreover there is the possibility that independent signatures of the Multiverse might never be observable, questioning whether the theory is in fact falsifiable. The risk is that postulating the existence of some ``father'' Universe whose signature is impossible to detect might in fact provide a way of sweeping under the rug some challenging questions. It is worrisome that accepting this explanation of the value of the cosmological constant, superseding the doubts that it raises, might in fact refrain us from discovering a deep principle of physics that indeed determines the cosmological constant without the need of the Multiverse.\\
Furthermore, the prediction of the value of the cosmological constant is fundamentally of a statistical nature. We assume that infinitely many bubble Universes nucleate and, for anthropic reasons, we are living in a Universe with the ``right'' value for the cosmological constant. A similar argument is used to discuss the value of the distance between Earth and the Sun. There is apparently no specific reason that fixes the Astronomical Unit to its value, but on the other hand if Earth would not be inside the habitable zone of our star there would not be observers. The fundamental difference with the Multiverse approach, however, is that we are able to measure the distance of other planets in other solar systems, justifying a posteriori our statistical description. This is not possible in the Multiverse and we are entitled to measure the cosmological constant in one Universe only.\\
Finally, even if we pick one specific vacuum in the landscape of string theory, the cosmological constant should still be determined by the contributions to the vacuum energy of the different fields living in the ten-dimensional bubble Universe. Therefore we should still be able to obtain, in principle, the value of the $\Lambda$ by accounting for all contributions. The Multiverse provides us with an answer on why we measure what we measure, but how this value is built is still a mystery.

\section{Quantum gravity and the quantized cosmological constant}
While the coincidence problem remains a very interesting and fascinating issue in modern physics, we will not attempt its resolution in this thesis, rather focusing on the cosmological constant problem.\\
As briefly discussed in the previous Section, it is important, in our opinion, to investigate how the cosmological constant value is determined, independently from knowing the reason why it has its specific value. Our intuition is that there should be some principle that would allow us to calculate the value of $\Lambda$ in a fully quantized model.\\
It is interesting to see that in the case of quantum gravity coupled with matter it is natural to determine the cosmological constant through the realization of the symmetries at the quantum level.

\subsection[The cosm. const. in 1d gravity coupled with matter is quantized]{The cosmological constant in one-dimensional gravity coupled with matter is quantized}
A first attempt to take into account the (quantum) gravitational contribution to the determination of the cosmological constant was put forward in \cite{Govaerts:2004ba} for a generic system possessing one-dimensional (time) reparametrization invariance.
Using phase space variables $(q^n,p_n)$, with the symplectic structure defined by the Poisson brackets $\{q^n,p_m\}=\delta^n_m$, one can consider an action principle in the form: 
\begin{equation}\label{1dsystem}
S=\int dt \left [\dot{q}^n p_n - \lambda \left ( H(q,p) - \Lambda \right )\right ] \ ,
\end{equation}
where $\lambda=\lambda(t)$ is an arbitrary function of time, $\Lambda$ is an arbitrary real parameter and $H$ is a given matter system that we wish to couple to gravity. It is straightforward to see that $\lambda$ is a Lagrange multiplier enforcing the single first-class constraint of the model:\begin{equation}
\phi = H(q,p) - \Lambda=0 \ ,
\end{equation}
and enters the equations of motion as:
\begin{equation}
\dot{q}^n=\lambda \frac{\partial H}{p_n}\qquad\dot{p}_n=-\lambda \frac{\partial H}{q^n} \ .
\end{equation}
The system is then constrained, and the total Hamiltonian \begin{equation}
H_T = \lambda \left (H-\Lambda\right )=\lambda \phi
\end{equation} is vanishing on the constraint surface, as one would expect from a reparametrization invariant system. The only constraint $\phi$ is also the generator of such symmetry.\\
On the other hand $\lambda^2$ itself can be seen as a one-dimensional metric on the world-line, hence the interpretation of \eqref{1dsystem} as a matter system coupled to gravity in one (time) dimension. Using the equations of motion to eliminate the $p$'s in favour of $\dot{q}$'s, it is easy to determine the Lagrangian form for \eqref{1dsystem}:
\begin{equation}
S = \int dt \lambda(t)\left [L\left (q^n, \lambda^{-1}\dot{q}^n\right )+\Lambda \right ] \ ,
\end{equation} so that indeed one has $ds^2 = dt^2 \lambda(t)^2$ on the world-line, and $\Lambda$ is a cosmological constant.\\
By absorbing $\lambda$ into the definition of a proper time coordinate $\tau(t)$, it is possible to see that disregarding the specific choice of parametrization for the world-line the solutions for the matter system coupled with gravity are the same of a uncoupled one, with the addition of the constraint $\phi=0$. This condition forces the free parameter $\Lambda$ to take values in the spectrum of $H$ itself, in particular by fixing its value to be exactly the energy of the matter system as calculated with the initial conditions for all $q$'s and $p$'s, $E := H(q_i^n, p_{n,i})$.\\
The quantization of the system proceeds trivially, as no quantum operators have to be introduced for the gravitational sector, which is pure gauge. The only ``remnant'' of it is given by the quantum constraint $\hat{\phi}=\hat{H}-\Lambda$. Following Dirac's approach to constrained systems, only states annihilated by the quantum operators corresponding to the constraints of the classical system are physical
\begin{equation}
\hat{\phi} | \psi_{phys} \rangle = 0\ .
\end{equation}
This in turn gives that for any physical states the matrix elements $ \langle \psi_{phys} | \hat{H} | \psi_{phys} \rangle$ have to vanish. Then, turning things around, for a given matter state $|\psi\rangle$ to be physical, the cosmological constant  $\Lambda$ is constrained to take a specific value, namely:
\begin{equation}
\Lambda = \langle \psi | \hat{H} | \psi \rangle \ .
\end{equation}
On the other hand, once a value for $\Lambda$ is chosen, the subset of degenerate quantum states $|\psi\rangle$ with $\langle \psi| \hat{H} |\psi\rangle = \Lambda$ is identified as the set of the physical states.\\
Thus, even without any dynamics in the gravitational sector, the requirement of the realization at the quantum level of the classical symmetry of the model naturally determines the cosmological constant. It is with this idea in mind that we will try to address the cosmological constant problem in the case of two-dimensional models.

\subsection{The cosmological constant in a quantum theory of gravity}\label{sec:ccinqg}
Considering the framework set in this chapter, and in particular the one-dimensional case just described, our general aim is to see how the cosmological constant problem might be dealt with in the presence of a quantum theory of gravity and in this Section we will try to infer what general features we can expect in this setting.\\
In a completely general way the diffeomorphism invariance, in the Hamiltonian formulation, provides classical constraint equations on phase space, some of which will also include the cosmological constant:
\begin{equation}
\mc{H}^\mu \left (\Lambda, \dots\right )=0 \ ,
\end{equation}
where the dots indicate dependence on any additional field included in the theory. Additional gauge symmetries provide additional constraints, so that $\mu=1,\dots,d+N$, with $d$ is the number of space-time dimensions and $N$ the number of additional gauge symmetries. When turned into quantum constraint operators, following Dirac's approach for first class constraints, these classical equations become conditions that determine the physical states of the model:
\begin{equation}
\hat{\mc{H}}^\mu (\Lambda) | \psi_{phys} \rangle=0 \ ,
\end{equation}
in a way dependent on the value of the cosmological constant and spatial coordinates, through the dependence of fields and momenta on them. This same condition can be seen as a way to determine the value of the cosmological constant required for a given quantum state to be physical, as shown in the one-dimensional case. Once a basis is chosen in Hilbert space, one can solve the set of equations:
\begin{equation}\label{cceq}
\langle  \psi_{phys} | \hat{\mc{H}}^\mu (\Lambda) | \psi_{phys} \rangle=0
\end{equation}
in the parameter space spanned by the the cosmological constant itself, the (complex) coordinates which cover the specific (sub)space of states we are testing and possibly the additional undetermined parameters of the model.  In order to do so the dependence on spatial coordinates has to be integrated out, for instance looking at the Fourier modes of each equation. For each $\mu$ this reduces the number of equations from $d-1$ non countable infinities (an equation per spatial point) to $d-1$ countable ones (one equation per each Fourier mode). Moreover the commutation relations among the different modes might drastically reduce the number of independent equations. A notable example is the Weyl symmetry in string theory, which requires only three modes of the Virasoro generators to annihilate physical states.\\
This is true in any number of dimensions, as long as diffeomorphism invariance holds. In particular, due to the specific form of the cosmological constant term in the gravitational action, $\Lambda$ will appear linearly in the constraints. There will then be at most one value of the cosmological constant that allows a given quantum state to be physical.\\
To turn the formal equations \eqref{cceq} into something able to provide an actual result for $\Lambda$ it is of course necessary to have fully quantized the theory, so that the explicit form and algebra of $\hat{\mc{H}}^\mu$ are known. Therefore we have to turn our attention to those models of gravity that we are able to quantize, also in the presence of additional fields.
\smallbreak
Summarizing, the purpose of this thesis is to investigate whether there exists a mechanism that determines the value of the cosmological constant in a quantum theory which includes both quantum matter and quantum gravity.\\
As discussed above, the symmetries of a diffeomorphism invariant classical theory, when replaced by quantum constraints, seem to provide such a mechanism, determining equations that can be solved for the cosmological constant. Therefore $\Lambda$ can be determined once specific quantum states are required to be physical, \emph{i.e.} annihilated by the quantum constraints.\\
The need of a fully quantized theory of gravity and matter and the one-dimensional example lead us to approach the issue in the case of two-dimensional models. While pure General Relativity in two dimensions is trivial, since the Einstein-Hilbert Lagrangian is a total derivative, a very rich framework in which the nature of the cosmological constant can be studied is two-dimensional dilaton gravity. We will couple this class of models to scalar matter and a vector field, quantize it in the canonical approach and finally determine the spectrum of the cosmological constant for the lowest excitations of the theory.
\cleardoublepage
\pagestyle{empty}
\vspace{-4cm}
\begin{flushright}
\textit{
Tempus item per se non est, sed rebus ab ipsis\\
consequitur sensus, transactum quid sit in aevo, \\
tum quae res instet, quid porro deinde sequatur;\\
nec per se quemquam tempus sentire fatendumst\\
semotum ab rerum motu placidaque quiete.}\\[.4cm]
\small{Il tempo ancor non \`e per s\`e in natura:\\
Ma dalle sole cose il senso cava\\
Il passato il presente ed il futuro;\\
N\`e pu\`o capirsi separato il tempo\\
Dal moto delle cose e dalla quiete.\\[.4cm]
Even time exists not of itself; but sense \\
Reads out of things what happened long ago, \\
What presses now, and what shall follow after: \\
No man, we must admit, feels time itself, \\
Disjoined from motion and repose of things.}\footnote{Titus Lucretius Carus \textit{de rerum natura}  (Liber I vv. 459-463)\\Translations: Alessandro Marchetti, William Ellery Leonard.}
%Non si può dire che alcuno avverta il tempo separato da movimento delle cose e da quiete tranquilla.
\end{flushright}
\chapter{Dilaton-Maxwell gravity in 1+1 dimensions}
	\label{ch:dmax}
	\pagestyle{fancy}
	\section{Why 1+1 dimensions?}
		The elegance and beauty of Einstein's General Relativity are matched only by the great difficulties that are encountered in the attempts of understanding its deepest implications, already at the classical level.\\
It is then important to determine, depending on the specificity of the goals to be achieved, an appropriate framework within General Relativity with the right balance between simplification of the dynamics and interesting physical content. A typical example is given by models of cosmological interest, as the famous Friedman-Lema\^ itre-Robertson-Walker metric ansatz mentioned in the Introduction.\\
In our case we will focus on a broad class of models in 1+1 dimensions, commonly dubbed Generalized Dilaton Theories (GDTs). They are described by a rather general action principle, which determines the dynamics of the single gravitational degree of freedom and a dilaton field, and is invariant under space-time diffeomorphisms.\\
There are different motivations to focus on two-dimensional models, and in particular on GDTs:
\begin{itemize}
\item The diffeomorphism invariance of higher dimensional GR is reduced to conformal invariance - to be distinguished from Weyl invariance \emph{per se} - if no further scale is introduced\footnote{we will see that even with the introduction of a Cosmological Constant conformal invariance is not manifest, but it is still present.}.
\item The Riemann tensor has a single non vanishing component.
\item The two dimensional Einstein-Hilbert lagrangian density is a total derivative (a coupling with the dilaton field keeps it from being just a topological term).
\item They can generally be considered as special cases of Poisson Sigma Models \cite{Schaller:1994es}.
\item There is always a conserved quantity which classifies all classical solutions in the absence of matter, while a modified conservation law is present in models with matter \cite{Kummer:1998yg}.
\item If one imposes spherical symmetry to the line element in 3+1 dimensional GR (Spherically Reduced Gravity), the effective action for the remaining degrees of freedom reduces to a sub-case of the GDT action (see for example \cite{Grumiller:2002nm}, and references therein).
\item They are strictly related to string theory models with a dynamical background.
\end{itemize}
In the last two decades two-dimensional GDTs have proven to be very useful in the understanding of classical and quantum gravity, allowing to face conceptual issues also relevant to higher dimensions.\\
In particular an abundance of models has been studied, for example describing black hole (BH) solutions, Hawking radiation and obtaining a full non-perturbative quantization of geometry in the path integral approach, with ``virtual'' black holes states in the scattering of matter fields. Most of these results are well summarized in \cite{Grumiller:2002nm,Grumiller:2006rc}.
	\section{Dilaton-Maxwell gravity in two dimensions}
		A general action for a two dimensional model of dilaton gravity coupled to a Maxwell gauge field may be taken in the form\footnote{A similar, higher dimensional, action of this form is often used in cosmology, for instance in models of Quintessence or k-essence, in low energy string theory and other dark energy models. For instance, the $G(X)$ potential is used to investigate possible variations of the fine structure constant. See \cite{Copeland:2006wr} and references therein.}:
\begin{equation}\label{dil-max_action}
 S_{DM}=\frac{1}{\kappa}\int_\mc{M} dx^{2}\sqrt{-g}\left(XR-U(X)X_{,\mu}X^{,\mu}-2V(X)-\frac{1}{4}G(X)F_{\mu\nu}F^{\mu\nu}\right)\ ,
\end{equation}
where $X$ is the dilaton, $U$, $V$ and $G$ are arbitrary functions of $X$, and $F_{\mu \nu}$ is the usual field strength for the vector gauge field $A_\mu$:
\begin{equation}
F_{\mu \nu}=\partial_\mu A_\nu - \partial_\nu A_\mu = A_{[\nu, \mu]}\ .
\end{equation}
The parameter $\kappa$ denotes an overall factor. As for now, space-time is considered to be a smooth manifold and we will drop all boundary terms (as for instance the ones coming from integration by parts) simply by requiring all fields to vanish at space-time infinity.\\
Since in two dimensions the space-time metric is conformally flat, it is always possible to consider a general Weyl redefinition of the metric, hence in particular an arbitrary dilaton-dependent transformation of the following form is feasible:
\begin{equation}\label{conf_transf}
g_{\mu \nu}\rightarrow e^{\chi(X)}g_{\mu \nu}\ .
\end{equation}
Furthermore the metric tensor may be parametrized in terms of three independent fields, so that the line element reads:
\begin{equation}\label{line_element}
dx^{2}=e^{\varphi}\left(-\lambda_{0}\lambda_{1}dt^{2}+(\lambda_{0}-\lambda_{1})dt\ ds+ds^{2}\right) \ .
\end{equation}
Note that the action \eqref{dil-max_action} is not explicitly Weyl invariant: the Ricci scalar term and the kinetic term for the dilaton are uncoupled to the conformal mode $\varphi$; as indeed:
\begin{equation}
R \sim X_{,\mu}X^{,\mu} \sim e^{-\varphi} \ ,
\end{equation}
so that the $e^\varphi$ factor given by the square root of the determinant of the metric cancels out. On the other hand the potential term and the $U(1)$ gauge field term do not allow this simplification, so that:
\begin{equation}
\sqrt{-g}V(X) \sim e^\varphi \ , \qquad \sqrt{-g}G(X)F_{\mu\nu}F^{\mu\nu} \sim e^{-\varphi} \ .
\end{equation}
\subsection{Gauge Fixing}\label{gfixing}
In what follows the conformal and Coulomb gauges will be chosen for the gravitational and Maxwell sectors, respectively. For the gravitational fields this amounts to the choice:
\begin{equation}
\lambda_{0}=\lambda_1=1 \ ,
\end{equation}
which yields:
\begin{equation}\label{line_element_cgauge}
dx^{2}=e^{\varphi}\left(-dt^{2}+ds^{2}\right) \ .
\end{equation}
For the gauge field, with that choice of the metric, the condition $A^\mu_{,\mu}=0$ is reduced to:
\begin{equation}
A_{0,t}=A_{1,s} \ ,
\end{equation}
so that by fixing $A_0=const$ it gives to $A_{1,s}=0$.\smallbreak
In the following we will use therefore the gauge fixing condition:
\begin{equation}\label{cgauge}
g_{01}=0\ ,\qquad g_{00}+g_{11}=0 \ .
\end{equation}
\subsection{Equations of motion}
In the following we will consider the action \eqref{dil-max_action} inclusive of a Weyl transformation which casts the line element in the form\footnote{Equivalently one can perform the variation before applying a Weyl transformation.}
\begin{equation}\label{line_element_Weyl}
dx^{2}=e^{\varphi+\chi(X)}\left(-\lambda_{0}\lambda_{1}dt^{2}+(\lambda_{0}-\lambda_{1})dt\ ds+ds^{2}\right) \ .
\end{equation}
Equations of motion easily follow from the variation of the action with respect to the fields followed by the imposition of the gauge fixing conditions \eqref{cgauge} (henceforth commas denoting derivatives are suppressed without ambiguities, while the subscript $t$ (resp., $s$) indicates a time (resp., space) derivative).
Varying with respect to the $\lambda$'s one finds:
\begin{equation} \begin{split}
-A_{1t}^{2}&G(X)e^{-\chi(X)-\varphi}-2\left(U(X)-\chi'(X)\right)\left(X_{s}\pm X_{t}\right){}^{2}+\\&+2\left(X_{s}\pm X_{t}\right)\left(\varphi_{t}\pm \varphi_{s}\right)-4\left(X_{s}\pm X_{t}\right)_{s}- 4V(X)e^{\chi(X)+\varphi}=0\ ,
\end{split}
\end{equation}
while variation with respect to the dilaton $X$ leads to:
\begin{equation}  \begin{split} \label{xEOM}
\left(X_{s}^{2}-X_{t}^{2}\right)&\left(U'(X)-\chi''(X)\right)-\varphi_{ss}+\varphi_{tt}+\\&+2\left(X_{ss}-X_{tt}\right)\left(U(X)-\chi'(X)\right)+\\&+\partial_{X}\left(\frac{1}{2}A_{1t}^{2}G(X)e^{-\chi(X)-\varphi}-2V(X)e^{\chi(X)+\varphi}\right)=0\ .
\end{split}\end{equation}
Furthermore, for the conformal mode $\varphi$ and the gauge field components one finds:
\begin{subequations}
\begin{align}
&-X_{ss}+X_{tt}-2V(X)e^{\chi(X)+\varphi}-\frac{1}{2}A_{1t}^{2}G(X)e^{-\chi(X)-\varphi}=0\label{cEOM} \ ,\\
&\partial_{s}\left(A_{1t}G(X)e^{\chi(X)+\varphi}\right)=0 \ ,\\
&\partial_{t}\left(A_{1t}G(X)e^{\chi(X)+\varphi}\right)=0 \ ,
\end{align}
\end{subequations}
where the last two equations determine a classical constant of motion for the system. Even though all classical solutions may be obtained in closed form for the present classes of models, quantum mechanically the non linear coupling of the dilaton field, $X$, and the conformal mode of the metric, $\varphi$, prevents one from pursuing a non-perturbative approach.
\smallbreak
It is thus desirable to possibly find out if and under which conditions the system may equivalently be described in a (partially) decoupled regime, in which different degrees of freedom could be quantized independently and non-perturbatively. It is possible to see that a subclass of the dilaton-Maxwell gravity models has a dual description in terms of Liouville fields \cite{Zonetti:2011ky}. This duality is a first original contribution of this thesis and will be thoroughly described in the next Section.
	\section{Duality with Liouville field theory}
		\subsection{Decoupling and Liouville fields}\label{sec_decoupling}
In what follows, for the sake of simplicity, all functions $U$, $V$ and $G$ are assumed to be non-vanishing. Whenever one or more of these functions vanishes the analysis proceeds along similar steps, and of course presents then a simpler structure.\\
In order to obtain a system in which the gravitational degrees of freedom are decoupled, one can combine \eqref{xEOM} to \eqref{cEOM}, by introducing an arbitrary function $F(X)$. In particular, looking at \eqref{xEOM} and the combination $\eqref{cEOM}+F'(X)\eqref{xEOM}$, one can isolate the factors multiplying the two exponential terms and impose a condition. We will choose these factors to be two arbitrary functions $\alpha(X),\gamma(X)$  times the potentials $G(X)$ and $V(X)$, in the attempt of separating the dynamics of the exponentials:
\begin{subequations}\begin{align}
-F'(X)G(X)+G'(X)-G(X)\chi'(X)=\gamma'(X)G(X) \ ,\\
-F'(X)V(X)-V'(X)-V(X)\chi'(X)=\alpha'(X)V(X)\ ,
\end{align}\end{subequations}
These equations may be solved for $G(X)$ and $\chi(X)$ by rewriting them as:
\begin{subequations}\begin{align}
\partial_X \ln G(X) = \partial_X \left (\chi(X)-F(X)+\gamma(X) \right ) \ ,\\
\partial_X \chi(X) = \partial_X \left (\alpha(X)+F(X)-\ln V(X) \right ) \ ,
\end{align}\end{subequations}
leading to:
\begin{subequations}\begin{align}
G(X)&=\frac{c_{1}e^{\gamma(X)-\alpha(X)}}{V(X)} \ ,\\
e^{\chi(X)}&=\frac{e^{-\alpha(X)-F(X)+c_0 }}{V(X)} \ , \label{ct_decoupling}
\end{align}\end{subequations}
where the quantities $c_0$ and $c_1$ are integration constants, and we assume $V(X)\neq0$ on $\mc{M}$. This assumption is quite restrictive on the form of $V$. We do not only assume that $V(X)$ is a non vanishing function of $X$, but also that when $X$ is expressed in terms of its classical solution $X(t,s)$ we have $V(X(t,s))\neq 0\ , \ \forall \ \{t,s\}$.\\
We can then introduce two newly defined fields:
\begin{subequations}\begin{align}
Z&=\varphi-F(X)-\alpha(X)+c_{0} \ ,\\
Y&=\varphi-F(X)-\gamma(X)+c_{0} \ ,
\end{align}\end{subequations}
so that \eqref{xEOM} reduces to\footnote{Reminder: the commas denoting ordinary derivatives are omitted, so that a subscript $t$ (resp. $s$) denotes a derivative with respect to the time (resp. space) coordinate.}:
\begin{equation}
-2e^{Z}-X_{ss}+X_{tt}-\frac{1}{2}e^{-Y}A_{1t}^{2}c_{1}=0 \ ,
\end{equation}
while the combination $\eqref{cEOM}+\left (U(X) +\partial_{X}\ln V(X) +\alpha'(X) \right ) \eqref{xEOM}$ is
\begin{equation}
\begin{split}
\left(\partial_{t}^{2}-\partial_{s}^{2}\right)&\left(Z-\ln(V(X))-\int_{1}^{X}U(y)dy\right)+\\&+2e^{Z}\left(-U(X)-\partial_{X}\ln(V(X))\right)-\\&-\frac{1}{2}e^{-Y}A_{1t}^{2}c_{1}\left(-\gamma'(X)+U(X)+\partial_{X}\ln(V(X))+\alpha'(X)\right) \ .
\end{split}
\end{equation}
Furthermore, by requiring the resulting factors of the exponentials to be constant, as is the case for the equation of motion of a Liouville field, one has to impose:
\begin{subequations}\begin{align}
U(X)+\partial_{X}\ln(V(X))&=\vartheta \ ,\\
-\gamma'(X)+U(X)+\partial_{X}\ln(V(X))+\alpha'(X)&=-\vartheta \ ,
\end{align}\end{subequations}
which are solved by:
\begin{subequations}\begin{align}
U(X)=&\vartheta-\partial_{X}\ln(V(X)) \ ,\\
e^{\gamma(X)}=&e^{\alpha(X)+2\vartheta X}+\frac{\vartheta_{G}}{\vartheta \ c_{1}} \ ,
\end{align}\end{subequations}
where the $\vartheta$'s are conveniently defined arbitrary constants, with the condition $\vartheta \neq 0$.\\This, with the conditions defined above, gives:
\begin{equation}
X=\frac{1}{2\vartheta}(Z-Y) \ .
\end{equation}
Finally, with the new fields expressed as:
\begin{subequations}\begin{align}\label{newfields}
Z&=\varphi-\bar{F}(X)+c_{0} \ ,\\
Y&=\varphi-\bar{F}(X)+c_{0}-2\vartheta X \ ,
\end{align}\end{subequations} 
where $\bar{F}(X)=\alpha(X)+F(X)$, the set of equations is reduced to:
\begin{subequations}\begin{align}\label{deom}
\begin{split}\left(Z_{t} \pm Z_{s}\right)^{2} &\mp 4 \left(Z_{t} \pm Z_{s}\right)_{s}-8e^{Z}\vartheta-\\&-\left(Y_{t} \pm Y_{s}\right)^{2} \pm 4 \left(Y_{t} \pm Y_{s}\right)_{s}-2\vartheta_{G}A_{1t}^{2}e^{-Y}=0\end{split} \ ,\\
&Y_{tt}-Y_{ss}+\vartheta_{G}A_{1t}^{2}e^{-Y}=0 \ ,\\
&Z_{tt}-Z_{ss}-4e^{Z}\vartheta=0 \ ,\\
&\partial_{s}\left(\vartheta_{G}A_{1t}e^{-Y}\right)=0 \ ,\\
&\partial_{t}\left(\vartheta_{G}A_{1t}e^{-Y}\right)=0 \ .
\end{align}\end{subequations}
It is clear that the gravitational system is completely decoupled, and is equivalent to two Liouville fields $Z$ and $Y$, which are constrained further by the first two equations of motion, as is indeed to be expected in a diffeomorphic invariant system in two dimensions.\\
Such a decoupled behaviour is of course particular to the specific choice made for the arbitrary functions contributing to the original action. The form of the function $G$ and, most importantly, of the function $U$ has been determined in the process, restricting the generality of the mechanism. On the other hand, as it is clear from \eqref{ct_decoupling}, the function $\chi$ entering the Weyl redefinition is left unconstrained, {\it i.e.,} no restrictions on the $F$ and $\alpha$ functions are required, thereby preserving the gauge symmetries of the model. In particular one requires:
\begin{subequations}\label{dconditions}\begin{align}
&U(X)=\vartheta-\partial_{X}\ln(V(X)) \ ,\\
&G(X)=\frac{\vartheta_{G}e^{2\vartheta X}}{\vartheta V(X)} \ .
\end{align}\end{subequations}
Comparing with \cite{Grumiller:2006rc,Grumiller:2002nm}, one may see that such a restriction allows still for enough freedom to cover some classes of dilaton gravity models. In particular one can easily recognize:
\begin{itemize}
\item A subset of the so-called \textit{ab-}family. Among other models it includes the Witten black hole and the CGHS models \cite{Witten:1991yr, Elitzur:1991cb, Mandal:1991tz, Callan:1992zr}, with
\begin{equation}
 U(X)=\vartheta-\frac{a}{X} \ , \quad V(X)=-\frac{B}{2}X^a \ , \quad G(X)=-\frac{2\vartheta_G}{\vartheta B}e^{2\vartheta X}X^{-a} \ ,
\end{equation}
where $a,B$ are arbitrary constants and the $\vartheta$ contribution to $U(X)$ may then be removed through a conformal transformation which is linear in $X$.
\item Liouville gravity \cite{Nakayama:2004vk}
\begin{equation}
U(X)=a  \ , \quad V(X)=b e^{(\vartheta - a) X} \ , \quad G(X)=\frac{\vartheta_{G}e^{(\vartheta+a) X}}{\vartheta b} \ ,
\end{equation}
where again $a,b$ are arbitrary constants.
\end{itemize}
\subsection{A dual action}
Given the new set of equations of motion \eqref{deom} and constraints \eqref{dconditions} obtained above, one can build a dual action involving two Liouville fields, a gauge vector field and the two constraints:
\begin{equation}\label{Laction}
\begin{split}
 S_{eom}=&\int d^2x
\frac{\xi^2\sqrt{-g_{\flat}}}{\kappa} \Bigl[\frac{1}{2}\left(Z_{\mu}Z^{\mu}-Y_{\mu}Y^{\mu}\right)-4\vartheta e^{Z} -\\&-\frac{e^{-Y}}{2}\vartheta_{G}F_{\mu\nu}F^{\mu\nu}+\left(Z-Y\right)R_\flat\Bigr] \ ,
\end{split}
\end{equation}
where the metric tensor has the same form as in \eqref{line_element} with $\varphi=0$, while $\xi^2$ is an overall factor which is irrelevant for the calculation of the equations of motion. Once again the commas denoting derivation have been omitted without risk of ambiguities.\\
In this formulation the gravitational sector is pure gauge, since the two $\lambda$'s behave like Lagrange multipliers and may always be chosen to give a flat Minkowski metric in the gauge fixing procedure. The role of the Ricci scalar $R_{\flat}$ is in fact just to ensure that the correct constraints are obtained when variation with respect to the $\lambda$'s is performed.\\
In order to fix the overall scale factor, and show that such an action is indeed a general result which is independent from the gauge choice made in the previous Sections, one may fix the arbitrary functions and constants appearing in \eqref{newfields}\footnote{Which as a matter of fact corresponds to fixing the arbitrary part of the Weyl redefinition $\chi(X)$.} and explicitly solve for $X,\varphi$. This is straightforward enough for all polynomial functions of $X$ and readily reproduces the form of \eqref{Laction}.\\
The factor $\xi$ can then be fixed by comparison, and it is easy to verify that $\xi^2=(2\vartheta)^{-1}$ is required for the two actions to coincide. By rescaling the fields as in
\begin{equation}
Z \rightarrow \xi^{-1} Z \ , \qquad Y \rightarrow \xi^{-1} Y \ ,
\end{equation} one can view $\xi$ as defining a coupling constant and define the Liouville action dual to dilaton-Maxwell gravity as:
\begin{equation} \label{Leff}
\begin{split}
 S_{dual}=&\int d^2x\frac{\sqrt{-g_{\flat}}}{\kappa} \Bigl[\frac{1}{2}\left(Z_{\mu}Z^{\mu}-Y_{\mu}Y^{\mu}\right)-2 e^{Z/\xi}-\\& -\frac{\xi^2\vartheta_{G}}{2}F_{\mu\nu}F^{\mu\nu}e^{-Y/\xi}+\xi\left(Z-Y\right)R_\flat\Bigr] \ .
\end{split}
\end{equation}
This last form of the action closely resembles that of the action quantized in \cite{Govaerts:2011p3916} and \cite{Curtright:1982}.
\subsection{The cosmological constant}
The $0th$ order in the power expansion of $V(X)$, in non vanishing, is in fact a cosmological constant term, modulo overall factors:
\begin{equation}
V(X)=\Lambda+\sum_{n=1}^{\infty}v_{n}X^{n}= \Lambda\left(1+\Lambda^{-1}\sum_{n=1}^{\infty}v_{n}X^{n}\right)= \Lambda\left(1+v(X)\right) \ .
\end{equation}
with this form for $V(X)$ the conditions on the functions and (rescaled) fields become
\begin{subequations}\label{final_decoupling}
\begin{align}
V(X) & =  \Lambda+\Lambda v(X) \ ,\\
U(X) & =  \vartheta-\partial_{X}\ln(V(X)) \ ,\\
G(X) & =  \frac{\vartheta_{G}e^{2\vartheta X}}{\vartheta\Lambda\left(1+v(X)\right)} \ ,\\
e^{\chi(X)}&=\frac{e^{-\alpha(X)-F(X)+c_0 }}{\Lambda\left (1+v(X)\right )} \ ,\\
Z & =  \xi \left (\varphi+\chi(X)+\ln(1+v(X))+\ln(\Lambda)\right ) \ ,\label{Zdef}\\
Y & =  \xi \left (\varphi+\chi(X)+\ln(1+v(X))+\ln(\Lambda)-2\vartheta X\right ) \ , \label{Ydef}
\end{align}
\end{subequations}
so that the cosmological constant is nothing else than a constant shift of the two rescaled Liouville fields $Z,Y$.
The cosmological constant can then be isolated by redefining $Z=\bar{Z}+\xi\ln(\Lambda)$ and $Y=\bar{Y}+\xi\ln(\Lambda)$. In the dual action a shift in the Liouville fields amounts to a rescaling of the two $\vartheta$'s:
\begin{equation}
\begin{split}
S_{dual}=&\int d^{2}x\frac{\sqrt{-g_{\flat}}}{\kappa}\Bigl[\frac{1}{2}\left(\bar{Z}_{\mu}\bar{Z}^{\mu}-\bar{Y}_{\mu}\bar{Y}^{\mu}\right)-2\Lambda e^{\bar{Z}/\xi}-\\&-e^{-\bar{Y}/\xi}\frac{\vartheta_{G}}{2\Lambda}F_{\mu\nu}F^{\mu\nu}+\xi\left(\bar{Z}-\bar{Y}\right)R_{\flat}\Bigr] \ .
\end{split}
\end{equation}
This final form for the action is the most convenient in our attempt to study the cosmological constant problem in 1+1 dimensional Dilaton-Maxwell gravity: the gravitational system is decoupled into two Liouville fields, nevertheless maintaining the symmetry content of the original theory while everything can be formulated on a static Minkowski background, and the cosmological constant $\Lambda$ appears explicitly.\\ In the following, for the sake of simplicity, we will drop the bars denoting the shifted fields without ambiguities.
\subsection{Additional fields}\label{sec_addfields}
It is clear that the duality between \eqref{dil-max_action} and \eqref{Leff} strongly relies on the presence of arbitrary dilaton couplings involving each of the dynamical terms in the action: the dilaton kinetic term is coupled to $U(X)$, the gauge field kinetic term to $G(X)$ and there is a dilaton potential $V(X)$ which couples to the space-time dynamics.\\
It is the interplay between these potentials and the conformal mode $\varphi$, manipulated as described in Section \ref{sec_decoupling}, which determines the conditions \eqref{dconditions} and the form of the two Liouville fields \eqref{newfields}. In order to consider additional fields, and keep the decoupling mechanism working without modifications, it is important that no additional conformal couplings, \emph{i.e.} terms proportional to powers of $e^\varphi$, are introduced, hence only terms which do not explicitly break Weyl invariance are allowed.\\
For example the standard kinetic term for scalar fields:
\begin{equation}
\sqrt{-g}\phi_{,\mu}\phi^{,\mu} \sim e^\varphi \times e^{-\varphi} = 1
\end{equation}
has the right behaviour, while a mass term:
\begin{equation}
\sqrt{-g} m^2 \phi^2 \sim e^\varphi
\end{equation}
has not. It is then possible to include massless scalar fields directly in \eqref{Leff} by adding:
\begin{equation}
\mathcal{L}_\phi = -\frac{1}{2}\sqrt{-g_{\flat}}\phi_{,\mu}\phi^{,\mu}
\end{equation}\\
A discussion on the possibility of including fermions is of sure interest, but it would require a complete reformulation of the problem within the First order formalism, and is left for future work.
\subsection{The Gibbons-Hawking-York boundary term}
In the presence of a boundary, the same procedure used to obtain the dual action \eqref{Laction} can be applied to the standard Gibbons-Hawking-York boundary term. In dilaton gravity this term has the form
\begin{equation}
S_{GHY}=-\frac{1}{2}\int_{\partial M}dx\sqrt{\gamma}XK \ ,
\end{equation}
where $K$ is the extrinsic curvature, which has the general expression:
\begin{equation}
K_{\mu\nu}:=\gamma_{\mu}^{\rho}\gamma_{\nu}^{\sigma}\nabla_{\rho}n_{\sigma}= \gamma_{\mu}^{\rho}\gamma_{\nu}^{\sigma}\frac{1}{2}\left(\mathcal{L}_{n}\gamma\right)_{\rho\sigma} \ . 
\end{equation}
Here $\gamma_{\mu\nu}$ is the induced metric on the boundary, $\gamma$ is its determinant and $n$ is the normal to the boundary.\\
We can then consider this term, inclusive of a general dilaton dependent Weyl transformation $\varphi\to\varphi+\chi(X)$, and replace the conformal mode and the dilaton with the new fields $Z,Y$. 
For every choice of the function $\chi$ one can see that the GHY term in the Liouville field theory side becomes (inclusive of an overall factor $\xi^{2}$ as done for the dual action above):
\begin{equation}
S_{GHY}=-\int_{\partial M}dx\ (Z-Y)\left(\xi K_{\flat}+\xi^{2}\left(1+8g_{\flat}\right)\partial_{t}Z\right)
\end{equation}
where again the $\flat$ indicates quantities calculated with respect to the conformally flat metric, i.e. $\varphi=0$. Note how the cosmological constant does not appear in this expression, as a shift in both $Z,Y$ is irrelevant.
	\section{Classical analysis}
		\subsection{Classical solutions with explicit cosmological constant}
Once again the equations of motion for the Liouville theory side follow readily from \eqref{Leff} by variation with respect to the different fields. The gauge fixing conditions $\lambda_{0}=\lambda_1=1$ and $A_0= A_{1s} = 0$ are imposed after variation, so that one has:
\begin{subequations}\label{eff_eom}
\begin{align}
-Y_{\text{ss}}+Y_{tt}+\frac{\vartheta_{G}\left(A_{1t}\right)^{2}}{\Lambda\xi}e^{-\frac{Y}{\xi}}&=0 \ ,\\
\frac{2e^{\frac{Z}{\xi}}\Lambda}{\xi}+Z_{ss}-Z_{tt}&=0 \ ,\\
e^{-\frac{Y}{\xi}}\xi^{-1}Y_{s}\left(A_{1}\right)_{t}&=0 \ , \label{A0eom}\\
e^{-\frac{Y}{\xi}}\left(-\xi^{-1}Y_{t}\left(A_{1t}\right)+\left(A_{1tt}\right)\right)&=0 \ ,
\end{align}
\end{subequations}
with two constraints
\begin{equation}
\begin{split}\label{eff_constraints}
\left(Y_{t} \pm Y_{s} \right)^{2}  &\mp 4\xi\left(Y_{t} \pm Y_{s}\right)_{s} + \frac{2\vartheta_{G}\left(A_{1t}\right)^{2}}{\Lambda}e^{-\frac{Y}{\xi}}-\\&
-\left( Z_{t} \pm Z_{s} \right)^{2}\pm4\xi\left(Z_{t} \pm Z_{s}\right)_{s}+4e^{\frac{Z}{\xi}}\Lambda=0 \ .
\end{split}
\end{equation}
The two equations stemming from the variation with respect to the gauge field are in fact Maxwell equations. Recalling that $A_{1,s}=0$ for our choice of gauge fixing we can rewrite them as:
\begin{equation}
\partial_{\mu}\left(A_{1t}e^{-Y/\xi}\right)=0 \ ,
\end{equation}
so that there is a classically conserved quantity $E=A_{1t}e^{-Y/\xi}=const$.\\
This allows to express the gauge field $A_{1}$, which is a time-only dependent field in term of the $Y$ field:
\begin{equation}
A_{1,t}(t)=Ee^{Y(t,s)/\xi} \ ,
\end{equation}
which forces the $Y$ field itself to be space-independent. After the imposition of the constraints, general solutions for $Z,Y$ are:
\begin{eqnarray}
\label{Zsol} Z \! &= \!& \xi \! \ln \! \left [ \frac{z_{3}x_{0}\xi^{2}}{\Lambda} \frac{S(t,s)^2}{T(t,s)^2}\right ] \quad \qquad\\
Y\! &= \!& 2\xi\ln\left(\text{sech}\left(\sqrt{x_{0}}(t-t_{0})\right)\right)+\xi\ln\left(\frac{2\xi^{2}x_{0}\Lambda}{E^{2}\vartheta_{G}}\right) \ ,\\
A_{1}\! &= \!& a_{0}+\frac{2\xi^{2}\sqrt{x_{0}}\Lambda}{E\vartheta_{G}}\tanh\left(\sqrt{x_{0}}(t-t_{0})\right) \ .
\end{eqnarray}
where:
\begin{equation*}
S(t,s) = \text{sech}\left(\frac{\sqrt{x_{0}}}{2}\left(t+s+z_{0}\right)\right)\text{sech}\left(\frac{\sqrt{x_{0}}}{2}\left(t-s+z_{1}\right)\right) \ ,
\end{equation*}
\begin{equation*}
T(t,s) = z_{3}\tanh\left(\frac{\sqrt{x_{0}}}{2}\left(t+s+z_{0}\right)\right)+\tanh\left(\frac{\sqrt{x_{0}}}{2}\left(t-s+z_{1}\right)\right)+z_{2}
\end{equation*}
and $z_0,z_1,z_2,x_0,t_o$ are integration constants. If $x_{0}$ is negative the hyperbolic functions are replaced by trigonometric ones. The two $tanh$ functions are replaced by $i\ tanh$, so that by choosing $z_2$ to be purely imaginary the overall $i$ factor in the denominator of \eqref{Zsol} simply gives a minus sign when squared. Such solutions however are highly singular, since they contain negative powers of $sin$ and $cos$. We can then restrict to the case in which $x_{0}>0$.\\
To recover the corresponding solutions in the dilaton-Maxwell gravity formulation, one can choose the Weyl transformation $\chi(X)$ in the most convenient way. In particular by choosing:
\begin{equation}\label{ctchoice}
\chi(X)=-\ln(1+v(X))
\end{equation}
for simplicity and leaving the cosmological constant explicit one gets the relations inverse to the last two of \eqref{Zdef} and \eqref{Ydef}:
\begin{subequations}
\begin{align}
\varphi & = \xi^{-1}Z \ ,\label{inverseZ}\\
X & = \xi\left(Z-Y\right) \ ,\label{inverseY}
\end{align}
\end{subequations}
so that, with the chiral coordinates $x^{+}=t+ s+z_{0},\ x^{-}=t- s+z_{1}$:
\begin{eqnarray*}
\varphi & = & 2\ln\left(\frac{\text{sech}\left(\frac{\sqrt{x_{0}}}{2}x^{+}\right)\text{sech}\left(\frac{\sqrt{x_{0}}}{2}x^{-}\right)}{z_{3}\tanh\left(\frac{\sqrt{x_{0}}}{2}x^{+}\right)+\tanh\left(\frac{\sqrt{x_{0}}}{2}x^{-}\right)+z_{2}}\right)-\ln\left(\frac{\Lambda}{z_{3}x_{0}\xi^{2}}\right) \ ,\\
X & = & 2\xi^{2}\ln\left(\frac{\text{sech}\left(\frac{\sqrt{x_{0}}}{2}x^{+}\right)\text{sech}\left(\frac{\sqrt{x_{0}}}{2}x^{-}\right)}{\left(z_{3}\tanh\left(\frac{\sqrt{x_{0}}}{2}x^{+}\right)+\tanh\left(\frac{\sqrt{x_{0}}}{2}x^{-}\right)+z_{2}\right)}\right)-\\
&&-2\xi^2 \ln \left (\text{sech}\left(\frac{\sqrt{x_{0}}}{2}(x^{+}+x^{-}+2t_{0}-z_{0}-z_{1})\right)\right )-\\&&-\xi^{2}\ln\left(\frac{2\Lambda^{2}}{z_{3}E^{2}\vartheta_{G}}\right) \ .
\end{eqnarray*}
If one requires the conformal factor to be real the sign of $x_{0}$
has to be the same as the sign of $\Lambda$ to guarantee the existence of the logarithm. But $\varphi$ itself
is not an observable and it appears either derived or exponentiated
in observable quantities, so this requirement can be lifted.\\
On the other hand the dilaton field $X$ contributes to the curvature
scalar directly, depending on the form of the potential $V(X)$, so
restrictions apply. In particular one requires $z_{3}\vartheta_{G}>0$.\\
Notice that a change in the sign of $z_{2}$ amounts to a parity transformation
on the chiral coordinates $x^{\pm}\rightarrow-x^{\pm}$. We can therefore
restrict to the case $z_{2}\ge0$ without loss of generality.\\
The integration constants can then be reorganized for an easier interpretation.
By choosing the case $\vartheta_{G}>0$, so that $z_{3}>0$, one can
define:
\begin{equation}
\sqrt{z_{3}}=\alpha \ , \quad z_{2}/\sqrt{z_{3}}=M\qquad\text{with }M\ge0,\ \alpha>0 \ ,
\end{equation}
and incorporate all the shifts of the origin as $\tau=2t_{0}-z_{0}-z_{1}$. 
The solutions can then be rewritten as:
\begin{equation}
\varphi = 2\ln\left(\frac{\text{sech}\left(\frac{\sqrt{x_{0}}}{2}x^{+}\right)\text{sech}\left(\frac{\sqrt{x_{0}}}{2}x^{-}\right)}{\alpha\tanh\left(\frac{\sqrt{x_{0}}}{2}x^{+}\right)+\alpha^{-1}\tanh\left(\frac{\sqrt{x_{0}}}{2}x^{-}\right)+M}\right)-\ln\left(\frac{\Lambda}{x_{0}\xi^{2}}\right)\label{csolution} \ ,
\end{equation}
\begin{equation}
\begin{split}
X& = 2\xi^{2}\ln\left(\frac{\text{sech}\left(\frac{\sqrt{x_{0}}}{2}x^{+}\right)\text{sech}\left(\frac{\sqrt{x_{0}}}{2}x^{-}\right)}{\left(\alpha\tanh\left(\frac{\sqrt{x_{0}}}{2}x^{+}\right)+\alpha^{-1}\tanh\left(\frac{\sqrt{x_{0}}}{2}x^{-}\right)+M\right)}\right)-\\
&\quad-2\xi^2 \ln \left (\text{sech}\left(\frac{\sqrt{x_{0}}}{2}(x^{+}+x^{-}+\tau)\right)\right ) -\xi^{2}\ln\left(\frac{2\Lambda^{2}}{E^{2}\vartheta_{G}}\right)\label{xsolution} \ ,
\end{split}
\end{equation}
where it is clear that $\alpha$ is a measure of an asymmetry between
the chiral coordinates, while $M$ determines the existence of singularities,
i.e. curves on which the solutions are divergent.
\subsubsection{Solutions on cylindrical space-time}
If one considers a compactified space dimension, so that $s\in[0,2\pi)$,
some of the arbitrary constants are necessarily fixed in order to
have the values of $Z$ to match at $s=0$ and $s=2\pi$.
In particular the argument of the logarithm in \eqref{Zsol} will have to match at the extremities of the interval, and this requires:
\begin{eqnarray*}
z_{1} & = & z_{0}+2\pi \ ,\\
\alpha & = & 1 \ .
\end{eqnarray*}
By shifting $t$ to include $z_{0}$:
\begin{eqnarray*}
Z(t,s) & = & 2\xi\ln\left(\frac{\text{sech}\left(\frac{\sqrt{x_{0}}}{2}\left(t+s\right)\right)\text{sech}\left(\frac{\sqrt{x_{0}}}{2}\left(t-s+2\pi\right)\right)}{\tanh\left(\frac{\sqrt{x_{0}}}{2}\left(t+s\right)\right)+\tanh\left(\frac{\sqrt{x_{0}}}{2}\left(t-s+2\pi\right)\right)+z_{2}}\right)-\\&&-\xi\ln\left(\frac{\Lambda}{x_{0}\xi^{2}}\right) \ ,\\
Y(t) & = & 2\xi\ln\left(\text{sech}\left(\sqrt{x_{0}}(t+t_{0})\right)\right)+\xi\ln\left(\frac{2\xi^{2}x_{0}\Lambda}{E^{2}\vartheta_{G}}\right) \ ,\\
A_{1}(t) & = & a_{0}+\frac{2\xi^{2}\sqrt{x_{0}}\Lambda}{E\vartheta_{G}}\tanh\left(\sqrt{x_{0}}(t+t_{0})\right) \ .
\end{eqnarray*}
By using again the inverse relations \eqref{inverseZ} and \eqref{inverseY} one obtains the solutions on the dilaton-Maxwell gravity side:
\begin{subequations}
\begin{align}
\varphi =& 2\ln\left(\frac{\text{sech}\left(\frac{\sqrt{x_{0}}}{2}\left(t+s\right)\right)\text{sech}\left(\frac{\sqrt{x_{0}}}{2}\left(t-s+2\pi\right)\right)}{\tanh\left(\frac{\sqrt{x_{0}}}{2}\left(t+s\right)\right)+\tanh\left(\frac{\sqrt{x_{0}}}{2}\left(t-s+2\pi\right)\right)+z_{2}}\right)-\ln(\frac{\Lambda}{x_{0}\xi^{2}}) \ , \label{compactcsol}\\
\begin{split}
X  = &2\xi^{2}\ln\left(\frac{\text{sech}\left(\frac{\sqrt{x_{0}}}{2}\left(t+s\right)\right)\text{sech}\left(\frac{\sqrt{x_{0}}}{2}\left(t-s+2\pi\right)\right)}{\left(\tanh\left(\frac{\sqrt{x_{0}}}{2}\left(t+s\right)\right)+\tanh\left(\frac{\sqrt{x_{0}}}{2}\left(t-s+2\pi\right)\right)+z_{2}\right)}\right)-\\&-2\xi^2\ln\left (\text{sech}\left(\sqrt{x_{0}}(t+t_{0})\right)\right )-\xi^{2}\ln\left(\frac{2\Lambda}{E^{2}\vartheta_{G}}\right) \ .
\end{split}\label{compactxsol}
\end{align}
\end{subequations}
\subsection{Killing vector(s)}\label{killingv}
For the sake of completeness we can also determine the equations for the Killing vectors in the model, to be applied in the investigation of the properties of classical solutions, e.g. black holes. However such studies are beyond the scope of this thesis and are left for future work. We leave this Section as a reference.\\
Killing vectors are generators of isometries, and are identified by requiring the Lie derivative of the metric with respect to them to vanish:
\begin{equation}
\left(\mathcal{L}_{K}g\right)_{\mu\nu}=g_{\mu\nu,\rho}K^{\rho}+2g_{\mu\rho}K_{,\nu}^{\rho}=g_{\mu\nu}^{(\flat)}e^{\varphi}\varphi_{,\rho}K^{\rho}+g_{\mu\rho}^{(\flat)}e^{\varphi}K_{,\nu}^{\rho}+g_{\nu\rho}^{(\flat)}e^{\varphi}K_{,\mu}^{\rho}=0 \ ,
\end{equation}
or in compact form (Killing's equation):
\begin{equation}
K_{(\mu;\nu)}=0 \ .
\end{equation}
For the different components of the metric tensor:
\begin{eqnarray*}
\left(\mathcal{L}_{K}g\right)_{00} & \propto & \varphi_{,t}K^{0}+\varphi_{,s}K^{1}+2V_{,t}^{0}=0 \ ,\\
\left(\mathcal{L}_{K}g\right)_{01} & \propto & -K_{,s}^{0}+K_{,t}^{1}=0\ ,\\
\left(\mathcal{L}_{K}g\right)_{11} & \propto & \varphi_{,t}K^{0}+\varphi_{,s}K^{1}+2V_{,s}^{1}=0 \ .
\end{eqnarray*}
By combining the equations and defining chiral derivatives $\partial _\pm = \partial_t \pm \partial_s$, one gets:
\begin{eqnarray*}
\partial_{-}\left(K^{0}+K^{1}\right) & = & 0 \ ,\\
\partial_{+}\left(K^{0}-K^{1}\right) & = & 0 \ ,\\
\varphi_{,t}K^{0}+\varphi_{,s}K^{1}+K_{,t}^{0}+K_{,s}^{1} & = & 0 \ ,
\end{eqnarray*}
and again, by defining $k_\pm = K_0 \pm K_1$:
\begin{eqnarray*}
K^{0} & = & k_{+}(x^{+})+k_{-}(x^{-}) \ ,\\
K^{1} & = & k_{+}(x^{+})-k_{-}(x^{-}) \ ,\\
k_{+}\partial_{+}\varphi+k_{-}\partial_{-}\varphi & = & -2k_{+}'(x^{+})-2k_{-}'(x^{-}) \ ,
\end{eqnarray*}
which have always a solution for a given $k_{-}(x^{-})$:
\begin{equation}
\begin{split}
k_{+}(x^{+})=&e^{-\frac{1}{2}\varphi(x^{+},x^{-})}\times \\& \times \left(c_{1}+\int_{1}^{x^{+}}dy\frac{1}{2}e^{\frac{1}{2}\varphi(y,x^{-})}\left(2k_{-}'(x^{-})-k_{-}(y)\partial_{-}\varphi(y,x^{-})\right)\right) \ ,
\end{split}
\end{equation}
so that there are infinitely many Killing vectors.
		\subsection{Curvature and singularities}
\begin{figure}
\includegraphics[scale=.8]{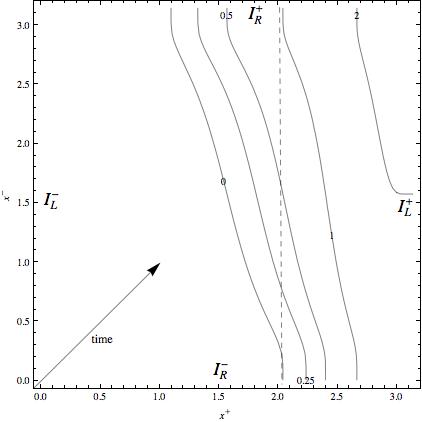}
\caption{Penrose-like diagram showing different $\gamma_\mathbb{R}$ singularity curves for different values of $M$ and with other parameters set to $\alpha=2,\tau=0,x_{0}=1$. The dashed line is the event horizon for the $M=1$ singularity associated with the right handed null rays.}\label{fig_singularities}
\end{figure}

Once the inverse relations \eqref{inverseZ} and \eqref{inverseY} are at hand, provided the choice \eqref{ctchoice} and the gauge fixing discussed in Section \eqref{gfixing}, it is quite straightforward to obtain the Ricci scalar, \emph{i.e.} the space-time scalar curvature, as a function of the Liouville fields $Z,Y$ and the dilaton $X$.\\
In particular from the dilaton-Maxwell side one has that:\begin{equation}
R = \frac{1}{2} e^{-\varphi-\chi(X)} \left(\partial_{tt}-\partial_{ss}\right) \left(\varphi+\chi(X)\right) \ ,
\end{equation}
which in the Liouville theory formulation corresponds to:
\begin{equation}
R=\frac{\left(1+v(X)\right)}{2}e^{-\xi^{-1}Z} \left(\partial_{tt}-\partial_{ss}\right) \left(\xi^{-1}Z-\ln\left(1+v(X)\right)\right) . 
\end{equation}
One can then employ the equations of motion for the Liouville fields \eqref{deom}
\begin{eqnarray*}
\frac{\vartheta_{G}E^{2}}{\Lambda}e^{\frac{Y}{\xi}} & = & -\xi Y_{tt} \ ,\\
Z_{tt}-Z_{ss} & = & 2\xi^{-1}\Lambda e^{Z/\xi} \ ,
\end{eqnarray*}
so that one can write $R$ as:
\begin{equation}\label{riccis}
\begin{split}
R = \left(1+v(X)\right)\xi^{-2}\Lambda&-v'(X)\left(\Lambda+\frac{\vartheta_{G}E^{2}}{2\Lambda}e^{-X\xi^{-2}}\right)+\\&+\left(\frac{v'(X)^{2}}{1+v(X)}-v''(X)\right)\frac{1}{2}\xi^{2}e^{-\varphi}\left(X_{t}^{2}-X_{s}^{2}\right) \ .
\end{split}
\end{equation}
In this form it is interesting to see how the model independent part
of the dynamics, given by the fields $(\varphi,X)$ or $(Z,Y)$,
is coupled to the model dependent one, here represented by the $v(X)$
potential, the only remaining free function from the original dilaton-Maxwell gravity action \eqref{dil-max_action}.\\
As one can easily see by comparing with the definition for $v(X)$ given in \eqref{final_decoupling} the first contribution to the scalar curvature is nothing more than $V(X)\xi^{-2}$ (where the factor $\xi^{-2}$ was introduced in \eqref{Laction}). The second contribution is due to the two Liouville potentials in \eqref{Leff}, as it is clear from the presence of the ``coupling constants'' $\Lambda, \vartheta_G$. Finally, the last term is connected to the dynamics of the dilaton field $X$.
\smallbreak
Singularities will then be easily identified by considering the form of $V(X)$, and consequently of $v(X)$. From the physical point of view it is a fair assumption to consider $v(X)$ to be a smooth function of $X$, so that a possible singular behaviour is only induced by the dynamics.\\
It is then possible to express $v(X)$ in terms of a polynomial of order $n$:
\begin{equation}
v(X)=\Lambda^{-1}\sum_{i>0}^{n}v_{i}X^{i} \ ,
\end{equation}
so that given the Ricci scalar \eqref{riccis} the only source of singularities are singularities in the fields $\varphi,X$. Looking then at the solutions \eqref{csolution}, \eqref{xsolution}, \eqref{compactcsol} and \eqref{compactxsol} it is clear that singularities arise on the space-time curve(s) which are solutions of the equations:
\begin{eqnarray}
\gamma_\mathbb{R}:&  & \alpha\tanh\left(\frac{\sqrt{x_{0}}}{2}x^{+}\right)+\alpha^{-1}\tanh\left(\frac{\sqrt{x_{0}}}{2}x^{-}\right)+M=0\label{gamma} \ ,\\
\gamma_{c}:  &  & \tanh\left(\frac{\sqrt{x_{0}}}{2}\left(t+s\right)\right)+\tanh\left(\frac{\sqrt{x_{0}}}{2}\left(t-s+2\pi\right)\right)+M=0\label{gammap} \ ,
\end{eqnarray}
for infinite and compactified space topologies respectively.\\
The presence of the $sech$ functions in the solutions ensures the regularity of the other factors inside the logarithms. At the same time this guarantees that
the metric is never degenerate at finite times. For $s\in \mathbb{R}$, we can define the quantity:
\begin{equation}
\Gamma_\mathbb{R}(\gamma)=-2\ln\left[\alpha\tanh\left(\frac{\sqrt{x_{0}}}{2}x^{+}\right)+\alpha^{-1}\tanh\left(\frac{\sqrt{x_{0}}}{2}x^{-}\right)+M\right]_{\gamma} \ ,
\end{equation}
where the subscript $\gamma$ indicates that it is calculated on some curve. In the limit in which $\gamma\to\gamma_\mathbb{R}$ we have $\Gamma_\mathbb{R}(\gamma) \gg 0$, while the Ricci scalar will behave as:
\begin{eqnarray*}
R_{\gamma} & \simeq & \xi^{-2}v_{n}\Gamma_\mathbb{R}^{n}-nv_{n}\Gamma_\mathbb{R}^{n-1}\left(\Lambda+\frac{\vartheta_{G}E^{2}}{2\Lambda}e^{-\Gamma_\mathbb{R}}\right)+\\&&+\Lambda^{-1}nv_{n}\Gamma_\mathbb{R}^{n-2}\frac{1}{2}\xi^{2}e^{-\Gamma_\mathbb{R}}x(t,s)e^{\Gamma_\mathbb{R}}=\\
 & \simeq & \xi^{-2}v_{n}\Gamma_\mathbb{R}^{n} \ ,
\end{eqnarray*}
where $x(t,s)$ the regular part (the numerator) of $X_{t}^{2}-X_{s}^{2}$ in \eqref{riccis}.
There are then curvature singularities in models with $n>0$, \emph{i.e.} $v(X)\neq0$, whose sign depends on the coefficient $v_{n}$. The same conclusion can be reached for the compactified case. In this formulation all singularities are extended one dimensional objects.\\
Let us focus on the case of $s \in \mathbb{R}$. Since the function $tanh$ is limited in the interval $(-1,1)$, singularities will appear only for:
\begin{equation}
|M|\le\frac{1+\alpha^{2}}{\alpha} \ .
\end{equation}
To visualize the location of the singularities, \emph{i.e.} the $\gamma$ curves, in space time one can map the real line(s) into finite intervals by replacing $x^\pm \rightarrow tan\left (x^\pm -\pi / 2\right )$, with $x^\pm\in[0,\pi]$, in a sort of Penrose-like diagram.\\
By way of example we can pick a choice for the parameters and visualize different $\gamma$ curves for different values of the $M$ parameter in Figure \ref{fig_singularities}.
As stated before the $\alpha$ parameter determines an asymmetry between the chiral coordinates $x^\pm$, as it is manifest in the diagram where $\alpha = 2$ was chosen. This has as consequence the presence of different ``black hole'' horizons associated with different directions for the null rays.
In Figure \ref{fig_singularities} null rays would be represented by lines of constant $x^\pm$ depending on their direction. Let us pick for example the case $M=1$.\\
All null rays originated at $\mc{I}^-_L$ with $x^-=const$ are bound to reach the singularity, so that the boundary at $\mc{I}^-_L$ itself is an event horizon, and no null geodesics reaches $\mc{I}^+_L$.\\
On the other hand null rays originated on $\mc{I}^-_R$ with $x^+\in [0,\sim 2]$ are able to escape to $\mc{I}^+_R$, while only null rays with $x^+ \in [\sim 2, ~2.65]$ fall into the singularity. There is then an event horizon, represented with a dashed line, for right handed null rays, located at $x^+\sim2.65$.\\ The rightmost portion of space-time is causally disconnected from the patch that is on the left of the singularity line.\\
As we mentioned earlier curvature singularities are in fact identified with singularities in the dilaton field and the conformal mode $\varphi$. This in turn means that the line element is divergent too, so that all space-like geodesics with endpoints on the $\gamma_\mathbb{R}$ curve reach the singularity only at infinite proper time.
\paragraph{Asymptotics}
Given the form of the Ricci scalar \eqref{riccis} it is possible to check its behaviour at the boundary of space-time by looking at the boundary limits of each of its terms.
The boundary quantities that enter its expression are simply the conformal mode $\varphi$ (which also provides information on the boundary behaviour of the line element \eqref{line_element_cgauge}), the dilaton field $X$ and the combination $X^2_t-X^2_s$, as described by the classical solutions \eqref{csolution}\eqref{xsolution}.\\
For the conformal mode, due to the presence of the two $sech$ functions, we can easily conclude that in each of the four different limits $x^\pm \rightarrow \mc{I}^\pm_{R,L}$\footnote{see Fig. \eqref{fig_singularities}} one has: 
\begin{equation}
\varphi \rightarrow - \infty \ .
\end{equation}
This is true also in the presence of the singularities described above. By taking the limits to the boundary and to the singularity at the same time it is clear that while the $sech$ contribute exponentially to the numerator inside the logarithm, the denominator goes to zero linearly. In fact, looking at \eqref{csolution}:
\begin{equation} \label{b_cph}
 \lim_{|x^\pm| \to +\infty}\varphi \approx 2 \ln \left ( e^{-\frac{\sqrt{x_0}}{2}|x^\pm|} \right ) = -\sqrt{x_0}|x^\pm| \ .
 \end{equation}
The dilaton field at the boundaries is given by:
\begin{subequations}
\begin{align}
\mc{I}^\pm_R:\quad X=X^\pm_R&\approx\xi ^2 \ln\left(\frac{E^2 \vartheta_G }{2 \Lambda ^2}\frac{\alpha ^2 \left(\text{tanh}\left(\frac{\sqrt{x_0} x^+}{2}\right)\pm1\right)^2}{\left(M \alpha +\alpha ^2 \text{tanh}\left(\frac{\sqrt{x_0} x^+}{2}\right)\pm1\right)^2}\right) \ ,\\
\mc{I}^\pm_L:\quad X=X^\pm_L&\approx\xi ^2 \ln\left(\frac{E^2 \vartheta_G }{2 \Lambda ^2}\frac{\alpha ^2 \left(\text{tanh}\left(\frac{\sqrt{x_0} x^-}{2}\right)\pm1\right)^2}{\left(M \alpha + \text{tanh}\left(\frac{\sqrt{x_0} x^-}{2}\right)\pm\alpha ^2\right)^2}\right) \ .
\end{align}
\end{subequations}
The singularities related to the parameters $\alpha,M$ are in this case still relevant as the rest of the factors inside the logarithm are constants or regular functions of the boundary variable.\\
The combination $X^2_t-X^2_s$, when approaching the boundaries behaves as:
\begin{equation}\label{b_XX}
 \lim_{|x^\pm| \to +\infty} \left (X^2_t-X^2_s\right ) \approx \text{sech}^4\left (\frac{\sqrt{x_0}}{2}x^\pm\right ) \text{cosh} \left ( \sqrt{x_0} x^\pm \right ) \approx e^{\sqrt{x_0} |x^\pm|}  \ .
\end{equation}
We can now look at the boundary behaviour of the Ricci scalar
\begin{equation}
\begin{split}
R = \left(1+v(X)\right)\xi^{-2}\Lambda&-v'(X)\left(\Lambda+\frac{\vartheta_{G}E^{2}}{2\Lambda}e^{-X\xi^{-2}}\right)+\\&+\left(\frac{v'(X)^{2}}{1+v(X)}-v''(X)\right)\frac{1}{2}\xi^{2}e^{-\varphi}\left(X_{t}^{2}-X_{s}^{2}\right)
\end{split}
\end{equation}
term by term:
\begin{itemize}
\item The first term will simply behave as a polynomial of the boundary dilaton field $X_{R,L}^\pm$, exhibiting singularities in correspondence to the ``bulk'' singularities described above.
\item The second term will be:
\begin{subequations}
\begin{align}
\mc{I}^\pm_R:  \quad & \approx -v'(X)\left(\Lambda+\Lambda\frac{\alpha^{2}\left(\text{tanh}\left(\frac{\sqrt{x_{0}}x^{+}}{2}\right)\pm 1\right)^{2}}{\left(M\alpha+\alpha^{2}\text{tanh}\left(\frac{\sqrt{x_{0}}x^{+}}{2}\right)\pm 1\right)^{2}}\right) \ ,\\
\mc{I}^\pm_L: \quad & \approx - v'(X)\left(\Lambda+\Lambda\frac{\alpha^{2}\left(\text{tanh}\left(\frac{\sqrt{x_{0}}x^{-}}{2}\right)\pm 1\right)^{2}}{\left(M\alpha+\text{tanh}\left(\frac{\sqrt{x_{0}}x^{-}}{2}\right)\pm \alpha^{2}\right)^{2}}\right) \ ,
\end{align}
\end{subequations}
again singular only in correspondence to the ``bulk'' singularities.
\item The third term contains a first factor that by the assumptions made above is simply a polynomial in $X$, so that it is as regular as $X$ itself. The combination $e^{-\varphi}\left (X^2_t-X^2_s\right )$ is of order unity, as the two divergencies \eqref{b_XX} and \eqref{b_cph} cancel out. With some lines of calculations it is possible to see that when approaching the boundary this factor reduces to:
\begin{eqnarray*}
\mc{I}_R^-:\quad &-\frac{\left(\alpha^{2}+M\alpha-1\right)\Lambda\xi^{2}}{\alpha^{2}}\left(\left(\alpha^{2}-M\alpha-1\right) \right. - \\& \left . -2e^{\sqrt{x_{0}}x^{+}}\left(1+\alpha^{2}\right)+e^{2\sqrt{x_{0}}x^{+}}\left(\alpha^{2}+M\alpha-1\right)\right) \ ,\\
\mc{I}_R^+:\quad &-\frac{e^{-2\sqrt{x_{0}}x^{+}}\left(\alpha^{2}-M\alpha-1\right)\Lambda\xi^{2}}{\alpha^{2}}\left(\left(\alpha^{2}-M\alpha-1\right)\right. - \\& \left . -2e^{\sqrt{x_{0}}x^{+}}\left(1+\alpha^{2}\right)+e^{2\sqrt{x_{0}}x^{+}}\left(\alpha^{2}+M\alpha-1\right)\right) \ ,\\
\mc{I}_L^-:\quad &-\frac{\left(\alpha^{2}-M\alpha-1\right)\Lambda\xi^{2}}{\alpha^{2}}\left(\left(\alpha^{2}+M\alpha-1\right)\right. + \\& \left .+2e^{\sqrt{x_{0}}x^{-}}\left(1+\alpha^{2}\right)+e^{2\sqrt{x_{0}}x^{-}}\left(\alpha^{2}-M\alpha-1\right)\right) \ ,\\
\mc{I}_L^+:\quad &-\frac{e^{-2\sqrt{x_{0}}x^{-}}\left(\alpha^{2}+M\alpha-1\right)\Lambda\xi^{2}}{\alpha^{2}}\left(\left(\alpha^{2}+M\alpha-1\right)\right. + \\& \left .+2e^{\sqrt{x_{0}}x^{-}}\left(1+\alpha^{2}\right)+e^{2\sqrt{x_{0}}x^{-}}\left(\alpha^{2}-M\alpha-1\right)\right) \ ,
\end{eqnarray*}
which is regular everywhere on the boundary.
\end{itemize}
Concluding the analysis of the asymptotics of the Ricci scalar we can look at the values of curvature at asymptotic values of the time variable, \emph{i.e.} the limits $x^\pm\rightarrow + \infty$ and $x^\pm\rightarrow - \infty$, dubbed ``distant future'' and ``distant past'' respectively in the context of Penrose diagrams.\\
In these limits the dilaton field goes to a constant value, in particular
\begin{equation}
\lim_{\to\ DF,\ DP} X = \xi ^2 \left(\ln\left[\frac{2 E^2 \alpha ^2}{\left(1\pm M \alpha +\alpha ^2\right)^2 \Lambda ^2}\right]+\ln\left [\vartheta_G \right ]\right) \ ,
\end{equation}
while the factor $e^{-\varphi}\left (X^2_t-X^2_s\right )$ goes to
\begin{equation}
\lim_{\to\ DF,\ DP} e^{-\varphi}\left (X^2_t-X^2_s\right ) = -\frac{\left(1-\left(2+M^2\right) \alpha ^2+\alpha ^4\right) \Lambda  \xi ^2}{\alpha ^2} \ .
\end{equation}
We can then conclude that the scalar curvature is constant in both limits.\smallbreak
The line element, on the other hand, has a diverging negative conformal factor on the whole boundary of space-time, so that the metric tensor is degenerate in these limits.
	\section{Hamiltonian and BRST Formulation}
		Having in mind the idea of applying the Canonical Quantization procedure to the classes of models of dilaton-Maxwell gravity subject to the restriction \eqref{dconditions}, the next necessary step is the analysis of their dual formulation \eqref{Leff} within the Hamiltonian formalism.\\
In order to take advantage of the formal equivalence of field modes and quantum harmonic oscillators, which will allow to express quantum field operators in terms of creation and annihilation operators on a suitably defined Fock space of quantum states, we will focus our attention on the case of a space-time with a cylindrical topology.\\
In particular we will consider:\begin{equation}
\mc{M} = \mathbb{R} \otimes S^1\ ,
\end{equation}
where the time coordinate $t$ takes values on the real line and the space coordinate $s$ is replaced by an angular coordinate limited to the interval $[0,2\pi)$. In this process we are implicitly introducing a length scale $\ell_c$ characteristic of the size of our compactification. In the following we will work in units which give $\ell_c=1$. We are entitled to this choice in our two dimensional case: the ``natural'' units are usually chosen by fixing $c=\hbar=G=1$. In two dimensions, however, Newton's constant is dimensionless, so that instead of fixing $G$ we can fix our compactification scale $\ell_c$. Therefore we can simply replace our spatial coordinate $s \in \mathbb{R}$ with $s\in [0,2\pi)$ with no ambiguities.\\
All fields will be then required to be periodic in the space coordinate, so that for any field $f(t,s=0)=f(t,s=2\pi)$. This is analogous to the so-called ``box quantization'' procedure employed in the canonical approach to Quantum Field Theory.\\
Due to the presence of gauge symmetries we will employ the Dirac approach to constrained dynamics\cite{Henneaux:1994,Govaerts:1991}. This will result in a set of constraints for phase space variables, with well defined algebraic properties, which in turn  determine a submanifold in phase space on which the physically relevant dynamics, namely the dynamics modulo gauge transformations, takes place.\\
\subsection{Hamiltonian formulation}
Since the symmetry content of the original action \eqref{dil-max_action} is preserved in the Liouville theory formulation \eqref{Leff}, it is expected to obtain three primary constraints, two related to the diffeomorphism invariance and one to the $U(1)$ gauge invariance.\\
This is indeed the case, as it is clear from the definition of the conjugate momenta, three of which are constrained to vanish:
\begin{subequations}
\begin{align}
\lambda_0:\quad&P_{0}=0\ ,\\\lambda_1:\quad&P_{1}=0\ ,\\
Y:\quad&P_{Y}=-\frac{1}{\left(\lambda_{0}+\lambda_{1}\right)} \left[\left(\lambda_{0}-\lambda_{1}\right)Y_{s} -2\left(Y_{t}-\xi\left(\lambda_{0}-\lambda_{1}\right)_{s}\right)\right]\ ,\\
Z:\quad&P_{Z}=\frac{1}{\left(\lambda_{0}+\lambda_{1}\right)} \left[\left(\lambda_{0}-\lambda_{1}\right)Z_{s} -2\left(Z_{t}-\xi\left(\lambda_{0}-\lambda_{1}\right)_{s}\right)\right]\ ,\\
A_0:\quad&\Pi_{0}=0\ ,\\
A_1:\quad&\Pi_{1}=-\frac{4\vartheta_{G}}{\left(\lambda_{0}+\lambda_{1}\right)}e^{-Y/\xi}\left(A_{0s}-A_{1t}\right)\ .
\end{align}
\end{subequations}
Three primary constraints are then present: $L^1=P_0,\ L^2 = P_1\ ,L^3 = \Pi_0$, and the primary Hamiltonian is defined as:
\begin{equation}
H_{p} = \int ds \left (\sum_\text{\tiny phase space} \! p_i \dot{q}^i - \mc{L}(p,q) + \sum_{k=1}^3\ell_k L^k \right )\ ,
\end{equation}
where $\ell_i$ are Lagrange multipliers.
Consistency conditions have to be imposed on the $L^k$'s, requiring their Poisson brackets with the primary Hamiltonian to be vanishing. This ensures that primary constraints are preserved under the classical time evolution, \emph{i.e.}:
\begin{equation}
\dot{L}^k= \left \{ L^k,H \right \} \approx 0\ ,
\end{equation}
where $\approx$ here stands for a weak equality, \emph{i.e.} an equality valid on the constraint hypersurface in phase space.\\
This, in turn, produces a set of secondary constraints:
\begin{subequations}\label{secondaryconstraints}
\begin{align}
\begin{split}
L^\pm=&-\frac{1}{4}\left(P_{Z}\mp Z_{s}\right)^{2}\mp\xi\left(P_{Z}\mp Z_{s}\right)_{s}+\Lambda e^{Z/\xi}+\\&\quad+\frac{1}{4}\left(P_{Y}\pm Y_{s}\right)^{2}\mp\xi\left(P_{Y}\pm Y_{s}\right)_{s}+\frac{\Lambda}{8\vartheta_{G}}e^{Y/\xi}\Pi_{1}^{2} \ ,\end{split}\\
L^\emptyset=&\ \Pi_{1s}\ .
\end{align}
\end{subequations}
Note again the two similiar Liouville sectors, one of which is coupled to the conjugate momentum of the gauge field component $A_1$.\\
The complete set of constraints is, as expected, first-class, with two of these constraints being the generators of space-time diffeomorphisms and a third one being Gauss' law. The only non-identically vanishing brackets\footnote{Smeared over suitable test functions, denoted here by $f$ and $g$.} reproduce the classical Virasoro algebra, extended to include the contributions of the gauge field:
\begin{subequations}\label{generators}
\begin{align}
\{L^\pm (f),L^\pm (g)\}&=\pm L^\pm(fg'-f'g)\approx0 \ ,\\
\{L^+(f),L^-(g)\}&=-\frac{1}{4\vartheta_{G}}\left (e^{Y/\xi}\Pi_{1}L^\emptyset\right )\left (fg\right )\approx0 \ .
\end{align}
\end{subequations}
Consequently no further constraints arise. The Hamiltonian density itself is a linear combination of the first-class constraints:
\begin{equation}
\mc{H}=\lambda_0 L^+ + \lambda_1 L^- + A_0 L^\emptyset
\end{equation}
and is therefore vanishing on the constraint hypersurface as required by the invariance under time-reparametrization.
\subsection{Additional fields: massless scalar fields}
If additional fields are present, the restrictions described in Section \ref{sec_addfields} guarantee that the classical constraint analysis follows the steps described above, with additional terms to be included in the definitions of the generators of gauge transformations \eqref{generators}.\\
The case of a massless scalar field $\phi$ is the simplest. By adding in \eqref{Leff} a kinetic term in the form\footnote{Mind the suppressed comma denoting derivatives.}:
\begin{equation}
\mc{L}_\phi = -\frac{1}{2}\sqrt{-g_{\flat}} \phi_\mu \phi^\mu \ ,
\end{equation}
and by defining the conjugate momentum:
\begin{equation}
\pi_\phi = -\frac{1}{\left(\lambda_{0}+\lambda_{1}\right)} \left[\left(\lambda_{0}-\lambda_{1}\right)\phi_{s}-2\phi_t\right]\ ,
\end{equation}
there will be an extra contribution to the $L^\pm$:
\begin{equation}\label{Lscalar}
L^{\pm,\phi} = \frac{1}{4}\left(\pi_\phi \pm \phi_{s}\right)^{2} \ ,
\end{equation}
without modifications of the algebra \eqref{generators}.
%\subsection{St\"uckelberg Mechanism}\label{std_stueck_ham}
%If we include in the effective action \eqref{Leff} the Standard St\"uckelberg (effective) Lagrangian \eqref{std_stueck_eff}, the momentum conjugate to $B$ will be simply defined by:
%\begin{equation}
% P_{B}=\frac{m}{\left(\lambda_{0}+\lambda_{1}\right)}\left[2\left(A_{0}-m^{-1}B_{t}\right)-\left(A_{1}-m^{-1}B_{s}\right)\left(\lambda_{0}-\lambda_{1}\right)\right] \ .
%\end{equation}
%We can label three different additional terms in the $L^\pm$:
%\begin{subequations}\label{std_stueck_Ls}
%\begin{align}
%L^{\pm,A} & = -\frac{1}{4}m^{2}A_{1}^{2} \ ,\\
%L^{\pm,B} & = -\frac{1}{4}\left(P_{B}\mp B_{s}\right)^{2} \ ,\\
%L^{\pm,I} & = \mp\frac{1}{2}\left(P_{B}\mp B_{s}\right)mA_{1} \ ,
%\end{align}
%\end{subequations}
%which are a mass term for the gauge field, a kinetic term for the St\"uckelberg field and an interaction term, respectively. The only non vanishing brackets of these new terms are:
%\begin{subequations}
%\begin{align}
%\{L^{\pm,B} (f),L^{\pm,B} (g)\}&=\pm L^{\pm,B}(fg'-f'g) \ ,\\
%\{L^{\pm,I} (f),L^{\pm,I} (g)\}&=\pm L^{\pm,A}(fg'-f'g) \ ,\\
%\{L^{\pm,B} (f),L^{\pm,I} (g)\}+\{L^{\pm,I} (f),L^{\pm,B} (g)\}&=\pm L^{\pm,I}(fg'-f'g) \ .
%\end{align}
%\end{subequations}
%In this case also the Gauss' constraints acquires an extra term, and it reads:
%\begin{equation}
%L^{\emptyset}=mP_{B}-\Pi_{1s} \ .
%\end{equation}
%It is easy to check that despite the modification to the constraints, the algebra \eqref{generators} is preserved.
		\subsection{BRST formulation}
In order to quantize a classical theory in the presence of constraints exhibiting a non-abelian algebra, it is very useful to consider a formulation in the BRST formalism, originally developed in \cite{Becchi:1974md,Tyutin:1975qk} and nowadays a fundamental tool in quantum physics.\\
Gauge symmetries are often difficult to implement quantum mechanically due to the appearance of ghost degrees of freedom and the risk of Gribov problems upon gauge fixing. A BRST formulation allows to keep things under control. The phase space is extended, with the addition of anti-commuting ghosts degrees of freedom, and gauge symmetries are replaced by symmetry under BRST transformations: this new symmetry is built in a way that makes always possible to fix the gauge and eliminate the ghosts from the dynamics. In the following we assume the reader to be familiar with the BRST formalism. For a textbook treatment see e.g. \cite{Govaerts:1991,Henneaux:1994}.
\subsubsection{BRST extensions}
Given the set of six constraints $C_1=L^1$, $C_2=L^2$, $C_3=L^3$, $C_{4,5}=L^\pm $ and $C_6=L^\emptyset$ one may introduce six pairs of anti-commuting canonically conjugate \emph{BRST ghosts} $c^a, {p}_a$, with $a$ taking the values $a=1,\dots,6$. Let us stress that this procedure is independent of the specific content of the model, as long as the constraints obey the algebra obtained above, so we will not specify whether we include scalar fields, St\"uckelberg mechanisms \emph{et cetera}.\\
The \emph{ghost number} for any expression on space-time is defined to be:
\begin{equation}
g_n(f) = \{ f , Q_c\}
\end{equation}
where the charge $Q_c$ is defined as $Q_c =\int d s \left (i \, \sum c^a {p}_a\right )$, so that for the original set of canonical variables $g_n = 0$, while the ghost degrees of freedom $c^a, {p}_a$ have respectively $g_n=1$ and $g_n=-1$ \cite{Govaerts:1991}. In contradistinction to this \emph{ghost sector}, the set of the $A_\mu,Y,Z$ fields and possibly additional matter fields will be referred to as the \emph{bosonic sector}.
The BRST charge $Q_B$ is defined to be real, of ghost number $g_c = 1$, Grassmanian odd, nilpotent and such that $\frac{\partial}{\partial c^a} Q_B | _{c^a = {p}_a = 0} = C^a$.
One can easily see that the expression:
\begin{equation}\label{QBRST}
Q_B = \int d \sigma \left (\sum_{a=1}^6  c^a C_a - c^4 c^4_s {p}_4 + c^5 c^5_s {p}_5 \right)\ ,
\end{equation}
meets all these properties.
Through Poisson brackets, the action of the BRST charge on the ghost variables ${p}_4$ and ${p}_5$ gives the BRST extension of the constraints $C_4$ and $C_5$,
\begin{subequations}
\begin{align}
C_{4}^{BRST}&=C_{4}-c^{4}p_{4s}-2c^{4}_s p_{4} \ ,\\
C_{5}^{BRST}&=C_{5}+c^{5}p_{5s}+2c^{5}_s p_{5}\ .
\end{align}
\end{subequations}
One can directly check that both expressions fulfil the requirements for BRST extended observables, and exhibit the same algebra as the original constraints constructed out of the original fields only, $L^\pm$.\\
No extension is obtained for the last abelian constraint:
\begin{equation}
C_{6}^{BRST}=C_{6}\ .
\end{equation}

The BRST extension of the Hamiltonian density can be obtained using an arbitrary function $\Psi$ on extended phase space, of odd Grassmann parity, of ghost number $g_c = -1$, and which is anti-hermitian. The complete BRST Hamiltonian density then reads,
\begin{equation}\label{BRSThamiltonian}
\mc{H}^{BRST}  =- \{ \Psi, Q_B \}\ .
\end{equation}
Suitable boundary conditions have to be considered for the ghost sector. In the case of a cylindrical space-time it will be required for all these extended degrees of freedom to be periodic in the $s$ spatial coordinate, and to be vanishing at $t$-infinity.

\subsubsection{BRST Gauge fixing}
In order to proceed to the quantization of the theory we partially fix the gauge freedom in both the gravitational and the Maxwell sectors. To implement the gauge choice of the conformal gauge and Coulomb gauge respectively, as described in Section \ref{gfixing}, one can fix a specific form for the $\Psi$ function \cite{Govaerts:1991},
\begin{equation}
\Psi=\frac{{p}_{1}\left(\lambda_{0}-1\right)}{\beta}+\frac{{p}_{2}\left(\lambda_{1}-1\right)}{\beta}+\frac{{p}_{3}\left(A_{0}-\alpha\right)}{\beta}+\lambda_{0}{p}_{4}+\lambda_{1}{p}_{5}+A_{0}{p}_{6}\ ,
\end{equation}
where $\beta$ is a free real parameter that will be taken to vanish later on and $\alpha$ is a constant that can be set to 0 to agree with \eqref{cgauge}. The BRST extended Hamiltonian (density)  \eqref{BRSThamiltonian} is the given by:
\begin{equation}\label{BRSThamgauge}
\begin{split}
\mc{H}^{BRST}=\frac{\lambda_0 - 1}{\beta} P_0 + \frac{\lambda_1 - 1}{\beta} P_1 + \frac{A_0 - \alpha}{\beta} \Pi_0 + \frac{c^1p_1 + c^2 p_2 + c^3 p_3}{\beta} + \\
+ c^1 p_4 + c^2 p_5 + c^3 p_6 + \lambda_0 C_4 ^{BRST}+ \lambda_1 C_5 ^{BRST}+ A_0 C_6
\end{split}
\end{equation}
By computing the equations of motion for the phase space variables, and then rescaling the fields in order to absorb the factor $\beta$:
\begin{equation}\label{ghostrescale}
{p}_{1,2,3}\rightarrow \frac{i\beta}{2 \pi} b_{4,5,6}\quad P_{0,1}\rightarrow\beta P_{0,1}\quad\Pi_{0}\rightarrow\beta\Pi_{0}\ ,
\end{equation}
one obtains, in a compact notation for the indices:
\begin{subequations}
\begin{align}
\lambda_{(0,1)t}\beta&=1-\lambda_{(0,1)}\ ,\\
A_{0,t}\beta&=\alpha-A_{0}\ ,\\
-\beta P_{(0,1)t}&=P_{(0,1)}+C_{(4,5)}^{BRST}\ ,\\
-\beta\Pi_{0t}&=\Pi_{0}+C_{6}^{BRST}\ ,\\
\beta c_{(1,2,3)t}&=c_{1,2,3}\ ,\\
\beta b_{(4,5,6)t}&=b_{(4,5,6)}- 2 \pi i \ {p}_{(4,5,6)}\ .
\end{align}
\end{subequations}
By taking the limit $\beta \rightarrow 0$ all the l.h.s. terms vanish, reducing the equations of motion to the gauge fixing conditions: 
\begin{subequations} \label{brstgaugefix}
\begin{align}
\lambda_0 &= \lambda_1 = 1\ ,\\
A_0 & = \alpha\ ,\\
P_{(0,1)}&= - C_{(4,5)}^{BRST} \ ,\\
\Pi_{0}& = -C_{6}^{BRST}\ ,\\
c_{1,2,3}&=0\ ,\\
{p}_{(4,5,6)} &= -\frac{i}{2\pi} \ b_{(4,5,6)}\ .
\end{align}
\end{subequations}
By imposing these on-shell conditions on the BRST Hamiltonian density \eqref{BRSThamiltonian} one gets:
\begin{equation}
\mc{H}^{BRST}=C_{4}^{BRST}+C_{5}^{BRST}+\alpha C_{6}^{BRST}\ .
\end{equation}
Since the ghosts $(c^{(4,5,6)},b_{(4,5,6)})$ are canonically conjugate,
\begin{equation}\label{ghostpsb}
\{c^a(s),b_a(s')\}^+=-2i\pi\delta_{2\pi}(s -s')\ ,
\end{equation}
with $\delta_{2\pi}(s-s')$ being the $2\pi$-periodic Dirac $\delta$ distribution on the unit circle, one may check that $Q_B$ is still nilpotent, {\it i.e.}, it has a vanishing Poisson bracket with itself.\\
For the BRST extended constraints, after gauge fixing, one has:
\begin{subequations}
\begin{align}
C_4^{BRST}=&L^{+, BRST} = L^+ + \frac{i}{2\pi}\left ( c^{4}b_{4s}+2c^{4}_s b_{4} \right ) =  L^+ + L^{+, g}\ ,\\
C_{5}^{BRST}=&L^{-, BRST} =L^- -\frac{i}{2\pi}\left (c^{5}b_{5s}+2c^{5}_s b_{5}\right ) =  L^- + L^{-, g}\ .
\end{align}
\end{subequations}
Note that, given also the form of \eqref{secondaryconstraints}, the BRST extended Virasoro generators are in fact a sum of terms from the different (uncoupled) sectors of the model:
\begin{equation}\label{BRSTgenerators}
L^{\pm, BRST} =L^{\pm, Z} + L^{\pm, Y} + L^{\pm, g} + \sum_\text{\tiny add. fields} L^\pm_{a.f.} \ ,
\end{equation}
where the sum indicates possible contributions from additional fields, as for example free massless scalars. This feature is of paramount importance in the quantization procedure, since it allows to quantize each sector separately.\\
For the sake of a uniform and simple notation, in relation with the indices $\pm, \emptyset$ used in the Lagrangian and Hamiltonian analysis, we can now replace the indices introduced in this section as in:
\begin{equation*}
c^4 \rightarrow c^+ \quad c^5 \rightarrow c^- \quad c^6 \rightarrow c^\emptyset \ .
\end{equation*}
\subsection{The constraint algebra}
It is straightforward to see that the $C_4^{BRST}=L^{+, BRST}$,  $C_5^{BRST}=L^{-, BRST}$ and $C_6^{BRST}=L^{\emptyset, BRST}$ extended constraints obey the smeared algebra:
\begin{eqnarray}
&\{ L^{\pm,BRST} (f) , L^{\pm,BRST} (g) \} = \pm L^{\pm,BRST} (g_s f- f_s g)\ ,\\
&\{ L^{\pm,BRST} (f) , L^{\emptyset,BRST} (g) \} = 0\ ,
\end{eqnarray}
hence the equations of motion for $L^{\pm,BRST}$, computed with the gauge fixed Hamiltonian, are:
\begin{equation}
L^{\pm,BRST}_t = \pm L^{\pm,BRST}_s\ ,
\end{equation}
which admit as solutions the mode expansions:
\begin{equation}\label{constraintsexpansion}
L^{\pm,BRST} = \sum_{n \in \mathbb{Z}} \frac{1}{2\pi}L^{\pm,BRST}_n \exp{-i n (t \pm s)}\ .
\end{equation}
In this way it is immediate to compute the algebra for the modes $L^{\pm,BRST}_n$ through a Fourier transformation,
\begin{equation}
\{ L^{\pm,BRST}_n, L^{\pm,BRST}_m \} = - i (n-m) L^{\pm,BRST}_{n+m}\ .
\end{equation}
For each chiral sector this is the celebrated Virasoro algebra, {\it i.e.}, the partially gauge fixed classical theory is a \emph{conformal invariant theory}. In particular, note how the two chiral sectors do commute with one another.
\cleardoublepage
\pagestyle{empty}
\vspace{-4cm}
\begin{flushright}
\textit{Nunc et seminibus si tanta est copia quantam\\
enumerare aetas animantum non queat omnis,\\
visque eadem et natura manet quae semina rerum\\
conicere in loca quaeque queat simili ratione\\
atque huc sunt coniecta, necesse est confiteare\\
esse alios aliis terrarum in partibus orbis\\
et varias hominum gentis et saecla ferarum.}\\[.4cm]
\small{Or; se dunque de' semi \`e tanto grande\\
La copia quanto a numerar bastevole\\
Non \`e degli animai l'etade intera,\\
E la forza medesma e la natura\\
Ritengono i principii atta a vibrarli\\
In tutti i luoghi nella stessa guisa\\
Ch'\'e fur lanciati; in questo egli \'e pur d'uopo\\
Confessar ch'altre terre in altre parti\\
Trovinsi, et altre genti ed altre specie\\
D'uomini e d'animai vivano in esse.\\[.4cm]
And now, if store of seeds there is \\
So great that not whole life-times of the living \\
Can count the tale... \\
And if their force and nature abide the same, \\
Able to throw the seeds of things together \\
Into their places, even as here are thrown \\
The seeds together in this world of ours, \\
'Tmust be confessed in other realms there are \\
Still other worlds, still other breeds of men, \\
And other generations of the wild.}
\footnote{Titus Lucretius Carus \textit{de rerum natura} (Liber II vv. 1070-1076)\\ Translations: Alessandro Marchetti, William Ellery Leonard.}
\end{flushright}		
\chapter{Quantum Theory}	
	\pagestyle{fancy}\label{ch:qt}
	\section{Quantization generalities}
Having obtained the formulation of our model in the BRST formalism and having explicitly calculated the algebra of the BRST extended constraints, we can proceed towards the quantization of the theory.\\
As mentioned earlier we will apply the so-called ``canonical quantization'' scheme in the case of a space-time with a cylindrical topology, in which the space dimension is compactified on a circle of length $2\pi$.\\ The classical fields, their conjugate momenta and all functions of them are to be replaced by quantum operators. This requires us to choose a particular ordering convention, following the introduction of non commutative products, e.g. between a field and its conjugate momentum. We adopt normal ordering, placing the annihilation operators to the right of the creation ones.\\
The classical Poisson brackets will be replaced by commutators or anti-commutators, as the case may be, inclusive of the extra factor $i\hbar$ multiplying the values of the corresponding classical brackets.\\
In this procedure quantum anomalies might appear, due to the ordering issues mentioned above. In order to maintain the classical gauge symmetries at the quantum level these anomalies will have to cancelled, implying specific restrictions. As we are quantizing a conformal field theory on a cylinder we know from ordinary string theory that we will have to deal with central extensions of the Virasoro algebra.\\
In the following we will work in natural units $c=\hbar=1$. However we will keep track of the $\hbar$ factors to distinguish classical from quantum contributions and possibly discuss the classical limit $\hbar \to 0$ in specific cases. We will denote these dimensionless $\hbar$ factors as $\HB$.\\
Thanks to the specific form of the BRST generators \eqref{BRSTgenerators} we can apply the quantization procedure in each sector separately.
\section{Quantization of the ghost sector}
After BRST gauge fixing \eqref{brstgaugefix} only three of the six ghost pairs are left, $(c^a,b_a)$ with $a=\pm, \emptyset$.\\
The equations of motion for $c^\pm$ and $b_\pm$ are also readily calculated from \eqref{BRSThamgauge} and read:
\begin{equation}
c^\pm _t = \pm c^\pm_s \qquad b_{\pm t} = \pm b_{\pm s} \ .
\end{equation}
Solutions can be found with the method of characteristics, and are:
\begin{subequations}\label{ghostexpansions}
\begin{align}
c^\pm (t,s)& = \sum_{n \in \mathbb{Z}} c^\pm _n e^{-i n (t \pm s)} \ ,\\
b_\pm (t,s)& = \sum_{n \in \mathbb{Z}} b_\pm ^n e^{-i n (t \pm s)} \ .
\end{align}
\end{subequations}
Given the Poisson brackets \eqref{ghostpsb} one can compute the algebra for the modes $c_n$ and $b^n$:
\begin{equation}\label{ghostmodespb}
\{ c^\pm_n , b^m_\pm \} = i \delta_{n+m} \ .
\end{equation}
Given the anti-commutation relations coming from \eqref{ghostmodespb}, $[ c^\pm_n, b^\pm_m ]^+ = \HB \delta_{n+m}$ it is natural to adopt the Fock space quantization of the ghost sector, remembering that since the classical ghost fields $c^a$ and $b_a$ are Grassmann odd variables, the quantized theory obeys the Fermi-Dirac statistics.\\
For the ghost operators, omitting the hat emphasizing the operator character of observables, one defines the following mode expansions, at time $t = 0$ in the Schr\"odinger picture,
\begin{equation}\label{ghostsexpansions}
c^\pm (s) = \sum_{n \in \mathbb{Z}} c^\pm _n \exp{\mp i n s}, \qquad 
b_\pm (s) = \sum_{n \in \mathbb{Z}} b_\pm ^n \exp{\mp i n s}, \ ,
\end{equation}
with $\{c^\pm_n,b^\pm_m\}=\HB \delta_{n+m,0}$, ${c^\pm_n}^\dagger=c^\pm_{-n}$ and ${b^n_\pm}^\dagger=b_\pm^{-n}$.
The vacuum of the theory, denoted as $|\Omega \rangle$, is the tensor product of all vacua for all modes $n\in\mathbb{Z}$.
The ghost modes operators are defined in a way that when acting on Fock states one has:
\begin{equation}
\text{creation op.}\begin{cases}
c_{n} & n<0\\
b^{n} & n\le0
\end{cases}\ ,\qquad\text{annihilation op.}\begin{cases}
c_{n} & n\ge0\\
b^{n} & n>0
\end{cases} \ .
\end{equation} 
Therefore one has a countable infinity of ghost/anti-ghost pairs of operators, one such pair for each $n\in\mathbb{Z}$ in \eqref{ghostsexpansions}.\\
A similar expansion in Fock operators can be also formally employed in the $\emptyset$-ghost sector, with ghost modes $c^\emptyset_n$ and $b_\emptyset^n$.\\
Given the decoupling of the two chiral sectors of the ghost variables in the conformal gauge, both for the canonically conjugate pairs of ghost degrees of freedom as well as their contributions to the constraints, an efficient way to compute the quantum ghost Virasoro algebra is through \emph{radial quantization} \cite{Difrancesco:1999, Ginsparg:1988ui, Blumenhagen:2009zz}. \\
Considering the expansions \eqref{ghostexpansions} and \eqref{constraintsexpansion} for $t = 0$, one can define the complex variable $z = e^{i s}$ and its complex conjugate $\bar{z}$, which can be considered as an independent variable. In this sense the mapping to the complex plane is overcomplete.\\
Holomorphic (\emph{i.e.} involving the $z$ variable) mode expansions will take the form:
\begin{equation}
f(z) = \sum_{n \in \mathbb{Z}} f_n z^{-n-h} \ ,
\end{equation}
where $h$ is the conformal weight. Analogously anti-holomorphic modes will be defined with a conformal weight $\bar{h}$.\\
The conformal weight for $b$'s and $c$'s can be inferred on dimensional considerations, the BRST charge and the charge $Q_C$ being both scalars:
\begin{subequations}
\begin{align}
[Q_B] = 1 &= [\ell] [c^2] [\ell^{-1}] [b] \ ,\\
[Q_C] = 1 &= [\ell] [c] [b] \ ,
\end{align}
\end{subequations}
so that $[c] = \ell$ and $[b] = \ell^{-2}$. This also gives $[L^\pm] = \ell^{-2}$.
One has then the complex plane expressions:
\begin{equation}
\begin{array}{c c}
L^{+,g} (z) = L^g(z)= \sum \frac{1}{\pi}L^g_n z^{-n-2} \ , & L^{-,g} (\bar{z}) =\bar{L}^ g(\bar{z})= \sum \frac{1}{\pi} \bar{L}^g_n \bar{z}^{-n-2} \ ,\\
\\
b(z) =  \sum b_n z^{-n-2} \ , & \bar{b}(\bar{z}) =  \sum \bar{b}_n \bar{z}^{-n-2}  \ , \\
c(z) =  \sum c_n z^{-n+1} \ ,  & \bar{c}(\bar{z}) =  \sum \bar{c}_n \bar{z}^{-n+1}  \ , 
\end{array}
\end{equation}
where the sums are over the integers $\mathbb{Z}$.\\ 
In this formulation the single modes can be extracted with contour integrals, exploiting the residue theorem. The derivative with respect to $s$ becomes a derivative with respect to $z$ (resp., $\bar{z}$) for holomorphic (resp., antiholomorphic) functions, taking a factor $i$ ($-i$). Again the derivative will be written as $\partial_z f(z) = f(z)_z = f_z$. In this way the contributions of the ghost sector to the constraints \eqref{BRSTgenerators} read:
\begin{subequations}\label{LLcomplex}
\begin{align}
L^ g(z) &= -\frac{1}{2\pi} \left ( b_z(z) c(z) + 2 b(z) c_z(z) \right ) \ ,\\
L^ g(\bar{z}) &= -\frac{1}{2\pi} \left ( \bar{b}_{\bar{z}} (\bar{z}) \bar{c} (\bar{z}) + 2 \bar{b} (\bar{z}) \bar{c}_{\bar{z}} (\bar{z}) \right ) \ .
\end{align}
\end{subequations}
In order to compute the quantum commutator involving the modes of these quantities, one needs to compute the singular part of the radial ordered function $R(L^g(z) L^g(w))$, to be integrated in $z$ and $w$ on suitable contours (an analogous procedure has to be applied for the antiholomorphic functions). Using the Wick theorem, via the contraction between two $b$'s or $c$'s, it is possible to reduce the radial ordered function to\footnote{Redundant specification of variable dependence is omitted, \emph{i.e.} $f_z(z) = f_z$.}:
\begin{equation}\label{RorderedLL}
\begin{split}
R\left(L^g(z) L^g(w) \right)= \left ( \frac{1}{2 \pi i} \right )^2  R \Bigl[b_z c(z) b_w c(w) + 2 b_z c(z) b(w) c_w \\+ 2 b(z) c_z b_w c(w) + 4 b_z c(z) b_w c(w) \Bigr] \ ,
\end{split}
\end{equation}
where one has four terms in the form $R( b_1 c_1 b_2 c_2)$. Given the anti-commutation relations, one obtains the general result:
\begin{equation}
R( b_1 c_1 b_2 c_2) = \underline{ b_1 c_2} :c_1 b_2: + \underline{c_1 b_2} :b_1 c_2: + \underline{b_1 c_2} \: \underline{c_1 b_2} + \text{reg.terms} \ ,
\end{equation}
where underlining denotes contractions between operators. Only contractions involving different complex variables are non-zero, and contractions between two $c$'s or $b$'s are vanishing. In this way everything is reduced to the calculation of the contraction $\underline{c(w) b(z)} = \underline{b(w) c(z)}$ and its derivatives.\\
We can then determine $R ( c(w) b(z) )$, in agreement with \cite{Govaerts1989186} and checking that the anti-commutator $[\hat{c}_n,\hat{b}^m] = \HB \delta _{n+m}$ is reproduced:
\begin{equation}
R\bigl( c(w) b(z) \bigr ) = -\frac{\HB}{2}\frac{w}{z^2} \frac{z+w}{z-w} +\text{reg.terms} \ .
\end{equation}
It is now straightforward to compute the needed R-product \eqref{RorderedLL}, using the definition of $L^g$ on the complex plane \eqref{LLcomplex}:
\begin{equation}
\begin{split}
R\left(L^g(w) L^g(z) \right) =&\HB \left ( \frac{1}{2 \pi i} \right )^2 \Bigl[ - 2\pi \left( \frac{L^g_w}{z-w} + \frac{2L^g(w)}{(z-w)^2}\right)+\\&+\HB\left (\frac{13}{(z-w)^4} -\frac{2w^{-2}}{(z-w)^2} -\frac{2w^{-3}}{z-w}\right ) \Bigl] \ .
\end{split}
\end{equation}
The quantum algebra for the modes is now simply obtained by integrating on the complex plane:
\begin{subequations}
\begin{align}
[ \hat{L}^g_n,  \hat{L}^g_m ] &= (n-m)\HB \hat{L}^g_{m+n} -\HB^{2}\delta_{n+m}\left(\frac{13}{6}n^{3}-\frac{1}{6}n\right) \ ,\\
[ \hat{\bar{L}}^g_n,  \hat{\bar{L}}^g_m ] &= (n-m) \HB\hat{\bar{L}}^g_{m+n}-\HB^{2}\delta_{n+m}\left(\frac{13}{6}n^{3}-\frac{1}{6}n\right) \ ,\\
[ \hat{L}^g_n,  \hat{\bar{L}}^g_m ] &= 0 \ .
\end{align}
\end{subequations}
Hence, as it is well known  \cite{Polchinski:1998rq, Green:1987sp,Govaerts1989186}, the Virasoro algebra in the ghost sector acquires a quantum central extension, namely a quantum anomaly which breaks the conformal symmetry of the classical ghost sector.
\smallbreak
To cross check the consistency with the techniques applied in the Liouville sector, we can show that the same result is easily obtained by explicitly calculating the commutator of the modes of the Virasoro operators with an exponential regularization.\\
For simplicity we will work in the $c^+,\ b_+$ sector. We consider the regularized expansions at time $t=0$:
\begin{eqnarray}
c^+(s)&=&\sum_{n}c_{n}e^{-ins}e^{-\epsilon|n|}\ ,  \\
b_+(s)&=&\sum_{n}b^{n}e^{-ins}e^{-\epsilon|n|} \ ,
\end{eqnarray}
where the sum runs over the integers $n\in \mathbb{Z}$.
The Virasoro operators in the $+$ ghost sectors then read:
\begin{equation}
L^{+,g}(s)=\frac{1}{2\pi}\sum_{n}\sum_{m}\left(m+2n\right):c_{n}b^{m}:e^{-i\left(n+m\right)s}e^{-\epsilon\left(|n|+|m|\right)} \ ,
\end{equation}
where normal ordering implies that annihilation operators are always on the right of creation ones.\\ The modes of these operators are readily extracted with a Fourier transform:
\begin{equation}
L^{+,g}_r=\sum_{n}\left(r+n\right):c_{n}b^{r-n}:e^{-\epsilon\left(|n|+|r-n|\right)} \ .
\end{equation}
We can then calculate the commutator:
\begin{eqnarray*}
\left[L_{r}^{+},L_{q}^{+}\right] &=&\sum_{n,m}\left(q+m\right)\left(r+n\right)\left(c_{n}b^{r-n}c_{m}b^{q-m}-c_{m}b^{q-m}c_{n}b^{r-n}\right)\times \\ &&\times e^{-\epsilon\left(|n|+|r-n|+|m|+|q-m|\right)}
\end{eqnarray*}
and, carefully treating the regulating exponential factors, we obtain once again:
\begin{equation}
\left[L_{r}^{+},L_{q}^{+}\right]=(r-q)\HB L_{r+q}^{+,g}-\HB^{2}\delta_{r+q}\left(\frac{13}{6}r^{3}-\frac{1}{6}r\right) \ .
\end{equation}

\section{Quantization of Liouville field theory}
Starting from the classical constraints $L^\pm$ \eqref{secondaryconstraints}, in order to ensure the closure of their quantum algebra the possibility of quantum corrections to the coupling constant $\xi$ needs to be considered \cite{Curtright:1982}, in a manner dependent on the fields. As a matter of fact, only terms involving the fields linearly need to be corrected, namely the $Z$ and $Y$ fields only with the replacements $\xi \rightarrow \xi_Z = \xi + \delta_Z$ and $\xi \rightarrow \xi_Y = \xi + \delta_Y$ for the corresponding couplings, respectively. The factor $\xi$ appearing in the exponential Liouville term contributions to $L^\pm$ remains unchanged:
\begin{equation}\label{quantumL}
\begin{split}
L^{\pm}=&\left[-\frac{1}{4}\left(P_{Z}\mp Z_{s}\right)^{2}\mp\xi_{Z}\left(P_{Z}\mp Z_{s}\right)_{s}+\Lambda e^{Z/\xi}\right]+\\&+\left[\frac{1}{4}\left(P_{Y}\pm Y_{s}\right)^{2}\mp\xi_{Y}\left(P_{Y}\pm Y_{s}\right)_{s}+\frac{\Lambda}{8\vartheta_{G}}e^{Y/\xi}\Pi_{1}^{2}\right] \ ,
\end{split}
\end{equation}
As may be seen from \eqref{secondaryconstraints} and \eqref{Lscalar}, terms associated to different fields have a similar form. Hence the computation of the quantum algebra for the $L^{\pm,Z}$ contributions provides the general result which may be particularized to all other fields. However, because of the Liouville exponential term involving the $Z$ field, radial quantization can no longer be used: even if $Z$ is expressed as the sum of holomorphic and antiholomorphic contributions, the exponential coupling between these two sectors through the Liouville potential does not allow to separate the two complex variables. Consequently one has to consider Fourier mode expansions of the fields and compute directly the commutators for these modes.\\
Following \cite{Curtright:1982} the field $Z$ and its conjugate momentum $P_Z$ are expressed in terms of a creation/annihilation zero-mode pair $(a_0, a^\dagger_0)$ and two chiral sets of  non-zero mode Fock operators, $a_n, \bar{a}_n$ for $n \neq 0$, $n\in\mathbb{Z}$, where positive (resp., negative) $n$'s correspond to annihilation (resp., creation) operators. Given the singularities that arise from local products of operators at the same spatial point, a regularization procedure needs to be introduced to define infinite sums over field modes. For convenience of computation, we have opted for a simple exponential damping regularization factor $e^{-\varepsilon |n|}$, with $\varepsilon\rightarrow 0^+$, to be included in all field mode expansions,
\begin{subequations}\label{Zmodes}
\begin{align}
Z(s) &= \frac{i}{2 \sqrt{\pi} } \left [ a_0 - a^\dagger_0 + {\sum_n}' \frac{1}{n}\left ( a_n e^{-ins} + \bar{a}_ne^{ins} \right ) e^{-\varepsilon |n|} \right ]\ ,\\
P_Z(s) &= \frac{1}{2 \sqrt{\pi}} \left [ a_0 + a^\dagger_0 + {\sum_n}' \left ( a_n e^{-ins} + \bar{a}_ne^{ins} \right ) e^{-\varepsilon |n|} \right ]\ ,
\end{align}
\end{subequations}
where the primed sum, $\sum'_n$, stands for a sum over all non zero modes, $n\ne 0$, $n\in\mathbb{Z}$. The given mode operators obey the following algebra of commutation relations,
\begin{equation}
[a_n , a_m ] = [\bar{a}_n, \bar{a}_m ] = n\HB \ \delta^n_{-m}, \ a^\dagger_n=a_{-n},\
\bar{a}^\dagger_n=\bar{a}_{-n}, \ [a_0, a^\dagger_0] = \HB.
\end{equation}
In terms of the fields $Z$ and $P_Z$, these commutation relations translate to the required Heisenberg algebra, once the limit $\varepsilon\rightarrow 0^+$ is applied.\\
In order to keep a compact notation, let us define $\chi_Z^\pm = P_Z \pm Z_s$.
The mode expansions \eqref{Zmodes} give:
\begin{equation}
\chi_Z^\pm = \frac{1}{\sqrt{\pi}}\left ( \frac{1}{2} ( a_0 + a^\dagger_0 )+ {\sum_n}' \left ( \begin{array}{c} a_n \\ \bar{a}_n \end{array} \right ) e^{-\varepsilon |n| }e^{\mp i n s} \right ),
\end{equation}
so that one can easily compute:
\begin{subequations}
\begin{align}
\chi_Z^\pm (s) \chi_Z^\pm (s') & = : \chi_Z^\pm (s) \chi_Z^\pm (s') : + \frac{\HB}{\pi} \sum_{n>0} n e^{- \varepsilon n } e^{-i n (s - s')}, \\
\left [ \chi_Z^\pm (s) \chi_Z^\pm (s') \right ] & = \pm i \HB (\partial_s - \partial_{s'} ) \Delta (s - s') = \pm 2 i \partial_s \Delta (s - s'),
\end{align}
\end{subequations}
where $\Delta$ is the regularized Dirac $\delta$ function on the circle,
\begin{equation}
\Delta (s - s') = \frac{1}{2 \pi} \sum_n e^{- \varepsilon |n|} e ^{-i n (s - s')}.
\end{equation}
Handling the sums with care one can compute the commutators:
\begin{subequations}
\begin{align}
\begin{split}
&\int d s d s' \left [ \chi^\mp_Z (s)^2  ,  \chi^\mp_Z (s')^2 \right ] e ^{\pm i r s} e ^{\pm i q s'} = \\ & \qquad = - 4 \HB (r-q) \int d s   \chi^\mp_Z(s)^2  e ^{\pm i (r+q) s} - \frac{4}{3}\left (r^3 -r \right )\HB^2\delta_{r+q}\ ,
\end{split}\\ 
&\int d s d s' \left [ \chi^\mp_{Zs}  ,  \chi^\mp_{Zs'} \right ] e ^{\pm i r s} e ^{\pm i q s'} = - 4 \HB \pi r^3 \delta_{r+q}\ ,\\
\begin{split}
& \int d s d s' \left ( \left [ \chi^\mp_{Z} (s)^2 ,  \chi^\mp_{Zs'} \right ] + \left [ \chi^\mp_{Zs}  ,  \chi^\mp_Z (s')^2 \right ]  \right ) e ^{\pm i r s} e ^{\pm i q s'} =\\ &\quad= - 4 \HB (r-q) \int d s \chi^\mp_{Zs} e ^{\pm i (r+q) s}\ ,
\end{split}
\end{align}
\end{subequations}
where normal ordering is always implied when needed. To deal with the exponential term $K(s) = \exp{Z(s)/ \xi } $, which commutes with itself, one first needs to compute:
\begin{subequations}
\begin{align}
&\chi^\pm_Z (s) K(s') = :\chi^\pm_Z (s) K(s') : - \frac{i\HB}{2 \pi \xi} \sum_{n>0} e^{- \varepsilon n} e ^{\mp i n (s - s')} K(s'),\\
&\left [\chi^\pm_Z (s), K(s') \right ]= - \frac{i\HB}{ \xi} K(s') \Delta (s - s'),
\end{align}
\end{subequations}
which are the building blocks for the terms involved in the algebra:
\begin{subequations}
\begin{align}
\begin{split}
 &\int d s d s' \left [ \chi^\mp_Z (s)^2  ,  K(s') \right ] e ^{\pm i r s} e ^{\pm i q s'} = -\frac{i\HB}{\xi} \int d s  :\chi^\mp_Z (s) K(s) :e ^{\pm i (r+q) s} +  \\
  & \qquad +\frac{\HB^2}{2 \pi \xi^2} \sum_{n>0} \left ( e^{- 2 \varepsilon ( n + |n+r| )} -e^{- 2 \varepsilon ( n + |n-r| ) } \right ) K_{\pm(r+q)},
\end{split}\\
&\int\! d s d s' \left ( \left [ \chi^\mp_{Zs}  ,  K (s') \right ] + \left [ K (s)  ,  \chi^\mp_{Zs'}\right ]  \right ) e ^{\pm i r s} e ^{\pm i q s'} = \mp \frac{\HB}{\xi} (r-q) K_{\pm (r+q)}.
\end{align}
\end{subequations}
At this point we can finally put all terms together and write the quantum algebra for the $Z$ sector:
\begin{equation}
\begin{split}
 \left [ L_r^{Z\pm} , L_q^{Z\pm} \right ] =& (r-q)\HB \Biggl [ \int d s \left ( -\frac{1}{4} : \chi^\mp_Z (s)^2 : \mp \xi_Z \chi^\mp_{Z s} \right ) e^{\pm i (r-q) s} +\\  &+2\Lambda K_{\pm(r+q)} \left (\frac{\xi_Z}{\xi}+\frac{\HB}{8 \pi \xi^2} \right ) \Biggr ] -\\& -\HB\left (\frac{\HB}{12}\left (r^3 +2r\right )+4 \pi \xi_Z^2 r^3 \right ) \delta_{r+q} \ .
\end{split}
\end{equation}
As we can see the factor on the second line multiplying $K$ has to be equal to $1$ if we want the algebra to close. We can then exploit the possibility of a quantum correction in $\xi_Z$ and fix it to get a Virasoro algebra with a central extension. By choosing $\xi_Z = \xi - \frac{\HB}{8 \pi \xi}$
we finally get:
\begin{equation}\label{Zvirasoro}
 \left [ L_r^{\pm,Z} , L_q^{\pm,Z} \right ] = (r-q)\HB L^{\pm,Z}_{r+q} - \HB \left [ \left ( \frac{\HB}{12} + 4 \pi \left ( \xi - \frac{\HB}{8 \pi \xi} \right )^2 \right ) r^3 +\HB \frac{1}{6}r \right ] \delta_{r+q}\ .
\end{equation}
Using the same procedure, with some attention to the different signs, one can compute the quantum algebra of the $Y$ sector:
\begin{equation}\label{Yvirasoro}
 \left [ L_r^{\pm,Y} , L_q^{\pm,Y} \right ] = (r-q)\HB L^{\pm,Y}_{r+q} + \HB \left [ \left ( \frac{\HB}{12} + 4 \pi \left ( \xi - \frac{\HB}{8 \pi \xi} \right )^2 \right ) r^3 +\HB \frac{1}{6}r \right ] \delta_{r+q}\ ,
\end{equation}
where once again the quantum correction to the coupling constant $\xi_Y$ appearing in \eqref{quantumL} is fixed by the requirement of a closed Virasoro algebra for the $Y$ sector:
\begin{equation}
\xi_Y = \xi - \frac{\HB}{8 \pi \xi} \ .
\end{equation}
Note how the central charges in the $Z$ and $Y$ sector are identical in absolute value but appear with opposite signs.
\section{Quantization of additional fields}
\subsection{Free massless scalar fields}
Given the form of \eqref{Lscalar} and the expansions \eqref{Zmodes}, the contribution of a massless scalar field $\phi$ to the total central extension of the Virasoro algebra is readily established:
\begin{equation}
\left [ L_r^{\pm,\phi} , L_q^{\pm,\phi} \right ] = (r-q)\HB L^{\pm, \phi}_{r+q} + \HB^2\left ( \frac{1}{12}  r^3 + \frac{1}{6}r \right ) \delta_{r+q} \ ,
\end{equation}
in agreement with known results \cite{Polchinski:1998rq, Green:1987sp}.
\subsection{The $Y$ field in the absence of the Maxwell field }\label{YnoMax}
It is interesting to see what is the effect of considering the subclass of models in which the Maxwell field has no dynamics, \emph{i.e. $A_\mu = const$}.\\
This choice, in terms of \eqref{Leff}, obviously amounts to discarding the Maxwell field strength term, which in turns, in terms of \eqref{quantumL}, would simply eliminate the Liouville potential term for $Y$.\\
As it is clear from \eqref{Leff} however, the $Y$ field maintains a coupling to the (flat) Ricci scalar $R_{\flat}$, which is the reason why the term linear in $Y$ in \eqref{quantumL} is still present in this case. We will denote the $Y$ field as $\Y$ in this subclass of models.\\
Having preserved the coupling with $R_{\flat}$, the coupling constant $\xi_\Y$ is still allowed to acquire a quantum correction that, due to the absence of the Liouville potential, will not require to be fixed to a specific value for the algebra to close. The quantum Virasoro algebra for the $\Y$ sector, in the absence or for constant Maxwell field will then be:
\begin{equation}
\left [ L_r^{\pm,\Y} , L_q^{\pm,\Y} \right ] = (r-q) \HB L^{\pm,\Y}_{r+q}+ \HB \left [ \left ( \frac{\HB}{12} + 4 \pi \xi_\Y^2\right ) r^3 +\HB \frac{1}{6}r \right ] \delta_{r+q} \ .
\end{equation}
This additional freedom, which in some sense is surprising since we removed rather than added degrees of freedom, can be exploited for example to eliminate part of the total central charge later on.
\section{Quantum Virasoro Algebra}
When the quantum Virasoro algebra has been determined in all sectors it is possible to sum up all contributions to the generators $L^{\pm, BRST}$, inclusive of the Liouville, gauge, ghost fields and possibly a collection of a number $D$ of free massless scalars $\phi_i$:
\begin{equation}\label{Lassum}
L^{\pm,BRST} = L^{\pm,Z} + L^{\pm,Y} + L^{\pm,g} + \sum_{i=1}^D L^{\pm,\phi_i} \ .
\end{equation}
The central charge contributions from the $Z$ and $Y$ sectors cancel out exactly, by virtue of the opposite sign in \eqref{Zvirasoro} and \eqref{Yvirasoro}. The only contributions are then given by the ghosts and the scalars, so that for the modes of the total Virasoro generators:
\begin{equation}
\left [ L_r^{\pm,BRST} , L_q^{\pm,BRST} \right ] = (r-q)\HB L^{\pm, BRST}_{r+q} + c \ \delta_{r+q} \ ,
\end{equation}
with:
\begin{equation}
c=\HB^{2}\left(\frac{D}{12}\left(r^{3}+2r\right)-\left(\frac{13}{6}r^{3}-\frac{1}{6}r\right)\right)=\frac{\HB^2}{12}\left (D-26\right )r^3 + \HB^2\frac{D+1}{6} r \ .
\end{equation}
In this form the central charge $c$ breaks the Virasoro algebra at the quantum level. However, it is possible to eliminate the $r^3$ term in $c$ by tuning the number of free scalars to $D=26$, in the same way as the number of space-time dimensions is tuned in bosonic string theory \cite{Polchinski:1998rq, Green:1987sp}. In this way the cubic term of the central charge is cancelled and the remaining linear term can be reabsorbed with a shift of the zero modes $L^\pm_0 \to L^\pm_0 -\HB 27/12$ so that the quantum Virasoro algebra is finally closed:
\begin{equation}
\left [ L_r^{\pm,BRST} , L_q^{\pm,BRST} \right ] = (r-q)\HB L^{\pm, BRST}_{r+q}\ .
\end{equation}
\subsection{The case with no Maxwell field}
As mentioned in Section \ref{YnoMax}, if the Maxwell field is not present at the classical level one is left with the possibility of an arbitrary quantum correction to the coupling constant $\xi_Y$.
The total central charge of the quantum Virasoro algebra of the quantum operators:
\begin{equation}
L^{\pm,BRST} = L^{\pm,Z} + L^{\pm,\Y} + L^{\pm,g} + \sum_{i=1}^D L^{\pm,\phi_i} \ ,
\end{equation}
will then be:
\begin{equation}
c\left (\xi_\Y\right ) =\frac{1+D}{6} r \HB ^2 + \left ( 4\pi\left(\xi_{\Y}^{2}-\xi^{2}\right)+\frac{D-14}{12}\HB-\frac{\HB^{2}}{16\pi\xi^{2}} \right ) r^3 \HB  \ .
\end{equation}
Using the freedom to fix the quantum correction in $\xi_\Y = \xi + \delta_\Y$ we may cancel the $r^3$ term in the central extension with the choice:
\begin{equation}
\delta_\Y =-\xi+\mbox{sgn}(\xi)\sqrt{\xi ^2-\frac{(D-14) \HB }{48 \pi }+\frac{\HB ^2}{64 \pi ^2 \xi ^2}} \ ,
\end{equation}
which recovers the right limit when $\HB$ is taken to zero. Reality conditions on the terms in the square root can be seen as a constraint on the number of scalar fields $D$:
\begin{equation}
D \leq 14+\frac{48 \pi  \xi ^2}{\HB }+\frac{3 \HB }{4 \pi  \xi ^2} \ .
\end{equation}
As we can see in the classical limit, \emph{i.e.} $\HB\to0$, no restriction on the number of scalar field is present. Using that freedom we are again left with a central charge that affects only the zero modes of the $L$'s, which may be redefined as :
\begin{equation}\label{shift_nomax}
L^{\pm,BRST}_0 \Rightarrow L^{\pm,BRST}_0 - \HB(D +1)/12 \ ,
\end{equation} hence giving an algebra which is finally free of central extensions.
	\section{Quantum constraints and the choice of basis}
		\subsection{Quantum constraints}
Once the quantum Virasoro algebra is obtained, it is possible to find the quantum realization of the constraints on Hilbert space, following the usual Dirac prescription that physical states have to be annihilated by the constraints.
As a matter of fact the cosmological constant $\Lambda$ (and the coupling constant $\xi$) are still free parameters: by requiring certain quantum states to be physical, {\it e.g.}, the Fock vacuum, $\Lambda$ will be constrained to take a specific value.\\
The presence of the Liouville potentials involving the $Z$ and, depending on whether one includes the Maxwell field, the $Y$ fields prevents one from following the most direct approach, {\it i.e.}, extracting the modes $L_n^\pm$ of the quantum constraints with a discrete Fourier transform and looking for the states that satisfy $L^\pm_n |\psi \rangle=0$ with $n = 0,1,2,\ldots$, as in ordinary String Theory \cite{Green:1987sp, Polchinski:1998rq}. For our purpose, however, it is sufficient to use the weaker condition:
\begin{equation}\label{quantumconstraints}
\langle \psi | L^\pm(\sigma) | \psi \rangle = 0,
\end{equation}
under the hypothesis that $|\psi\rangle$ is physical.\\ This is because we are not looking to determine the set of physical states of the model, but rather to determine how the cosmological constant is constrained at the quantum level by the requirement that a given quantum state of the Universe is physical.\\
The space-coordinate dependence will have to be carried through and in some cases traded for a mode expansion {\it via} a Fourier transformation once the matrix elements between suitable states spanning the Hilbert space have been calculated. In particular, considering linear combinations of the shifted Virasoro generators, the quantum constraints for an arbitrary quantum physical state will be:
\begin{equation}
\langle L^+ + L^- \rangle  =0\ , \qquad \langle L^+ - L^- \rangle = 0\ , \qquad \langle L^\emptyset \rangle = 0 \ ,
\end{equation}
with:
\begin{equation}\label{Lshifted}
L^{\pm} = L^{\pm,Z} + L^{\pm,Y} + \sum_{i=1}^D L^{\pm,\phi_i} - c/2 \ ,
\end{equation}
where the ghost sector is omitted by taking advantage of the BRST invariance and $c$ is the shift of the zero modes which cancels out the central charge, as previously determined in \eqref{shift_nomax}.\\
As the cosmological constant enters the expressions for $ L^{\pm}$ only through the Liouville potential terms, which are identical for the $+$ and $-$ cases, only $\langle L^+ + L^- \rangle$ will depend on $\Lambda$. For the general case:
\begin{subequations}
\begin{align}
\begin{split}
\langle L^+ + L^- \rangle =& -\frac{1}{2}\langle P_{Z}^2+ Z^2_{s}\rangle + 2 \xi_Z \langle Z_{ss} \rangle + \frac{1}{2}\langle P_{Y}^2+ Y_{s}^2\rangle-2\xi_Y\langle Y_{ss} \rangle + \\&+ \sum \frac{1}{2}\langle P_{i}^2+ \phi_{i,s}^2\rangle+\Lambda \left [  2 \langle e^{Z/\xi} \rangle + \frac{1}{4\vartheta_{G}}\langle e^{Y/\xi}\Pi_{1}^{2}\rangle \right ]  - c\ ,\end{split}\\
\langle L^+- L^- \rangle =& \langle P_{Z}Z_{s}\rangle -2\xi_{Z}\langle P_{Z,s}\rangle+\langle P_{Y}Y_{s}\rangle-2\xi_{Y}\langle P_{Y,s}\rangle + \sum \langle P_{i}\phi_{i,s}\rangle \ ,\\
\langle L^\emptyset \rangle =& \langle \Pi_{1s}\rangle \ .
\end{align}
\end{subequations}
By solving the first constraint for $\Lambda$ a first result is established: the Liouville field $Z$ contributes with an opposite sign to the cosmological constant value as compared to the $Y$ field and to the scalar fields $\phi_i$, possibly addressing the cosmological constant problem: the presence of quantum fluctuations of the dynamical gravitational degrees of freedom, {\it i.e.}, the conformal mode, partially compensates the positive contributions to the value of the cosmological constant stemming from the quantum fluctuations of the scalar (matter) fields and the dilaton field. This will be shown explicitly for a particular set of states later on.

\subsection{Representation of the Hilbert space and the quantum constraints}
In order to illustrate in detail how the quantum constraints provide a mechanism to constrain the cosmological constant $\Lambda$ to a specific value, we will now focus our attention on the sole case in which no classical Maxwell field is present.\\
While this choice simplifies the discussion, by reducing the Liouville field $Y$ to a scalar field $\Y$ non minimally coupled to the flat Ricci scalar $R_{\flat}$, there is no loss in generality and the same analysis can be carried out with no sensible differences but a more lengthy and involved description of the spectrum for the cosmological constant.\\
As a basis for the Hilbert space two possibilities are at hand: coherent states, being eigenstates of the annihilation operators, have the advantage of providing rather simple expressions for the quantum constraints, and therefore seem to be the most obvious choice. On the other hand, since our first goal is to obtain values for $\Lambda$ which follow from the requirement for the lower excitations of the spectrum of the theory to be physical, a Fock basis is the best option.\\
However if the quantum constraints are expressed in terms of creation and annihilation operators the exponential terms of the Liouville potentials in $L^{\pm}$ would spread every Fock excitation of the field $Z$ over the entire spectrum, making the calculation of the matrix elements quite problematic. To avoid this it is possible to use a diagonal representation for the constraint operators in the coherent state (overcomplete) basis \cite{Klauder:1968dq}. This has the advantage of turning all the matrix elements calculations into Gaussian integrals over complex variables. By writing a general state as a tensor product of linear combinations of Fock excitations of the Fock vacua we will be able to obtain two constraint equations involving the cosmological constant $\Lambda$.\\
To simplify the picture, we can reorganize the Fock operators defined in the expansions \eqref{Zmodes} and used in the quantization of every field:
\begin{equation}
\mathbf{a}_n = \left \{
\begin{array}{l l}
\hat{a}_0 & :\  n = 0\ ,\\ \\
\frac{1}{\sqrt{n}} \hat{a}_n &:\  n > 0\ ,\\ \\
\frac{1}{\sqrt{|n|}} \hat{\bar{a}}_{|n|} &:\   n < 0\ ,\\
\end{array} \right.\quad ,
\qquad 
\mathbf{a}_n^\dagger = \left \{
\begin{array}{l l}
\hat{a}_0^\dagger &\ : n = 0\ ,\\ \\
\frac{1}{\sqrt{n}} \hat{a}_n^\dagger &\ : n > 0\ ,\\ \\
\frac{1}{\sqrt{|n|}} \hat{\bar{a}}_{|n|}^\dagger &\ :  n < 0\ ,\\
\end{array} 
\right. 
\end{equation}
so that $ \left[ \mathbf{a}_n , \mathbf{a}_m^\dagger \right] = \HB \ \delta^n_m$.
\paragraph{Diagonal coherent states representation for quantum operators}
Given the quantum operators $L^\pm(s)$ defined in \eqref{Lshifted}, following the procedure described in Appendix \ref{kernelrep}, we get:
\begin{equation} \label{kernelrepconstraints}
\begin{split}
 L^\pm(s) =  \int \prod_m \left[ \frac{d z_m d \bar{z}_m}{2 \pi} \right]  |\ul{z} \rangle \Biggl ( & \Lslash^{\pm,Z} (s, z, \bar{z}) + \Lslash^{\pm,\Y} (s, z, \bar{z}) + \\ & + \sum_{i=1}^{D-1} \Lslash^{\pm,\phi_i} (s, z, \bar{z}) -\HB\frac{D+1}{12} \Biggr )\langle \ul{z} |\ ,
\end{split}
\end{equation}
where $m$ runs over all the modes of creation and annihilation operators. Coherent states are defined as:
\begin{equation}\label{coherentstate}
 | \ul{z} \rangle = \bigotimes_{f}^{ fields}\left ( \bigotimes_n |z_n^f \rangle \right )\ ,
\end{equation} 
where the first tensor product is over the bosonic fields, excluding the ghosts, by virtue of the BRST symmetry established above.\\
The $\Lslash^{\pm,f}$ are the kernels:
\begin{equation}
\begin{split}
 \Lslash^{\pm,Z} (s, z, \bar{z}) =&   2\Lambda \ \exp{-\frac{1}{4 \pi \xi^2} \left (1 + 2\sum_{n>0}\frac{1}{n} \right )} \langle e^{Z/\xi} \rangle -\\  & - \frac{1}{4}\langle\left(\chi_Z^\mp\right)^2 \rangle \mp \xi_Z  \langle\left( \chi_Z^{\mp}\right)_s \rangle + \frac{1}{2\pi}\left (\frac{1}{4}+\sum_{n>0} n \right )\ ,
\end{split}
\end{equation}
\begin{equation}
\Lslash^{\pm,\Y} (s, z, \bar{z}) = \frac{1}{4}  \langle \left( \chi_\Y^\pm \right)^2 \rangle \pm  \xi_\Y \langle \left(\chi_\Y^\pm \right)_s \rangle - \frac{1}{2\pi}\left (\frac{1}{4}+\sum_{n>0} n \right )\ ,
\end{equation}
\begin{equation}
\Lslash^{\pm,\phi_i} (s, z, \bar{z}) =  \frac{1}{4} \langle \left(\chi_{\phi_i}^\pm\right)^2 \rangle - \frac{1}{2\pi}\left (\frac{1}{4}+\sum_{n>0} n \right )\ ,
\end{equation}
where $\langle \mc{O}\rangle$ denotes a diagonal matrix element between two coherent states in the form of \eqref{coherentstate} and $\chi_f^\pm = P_f \pm f_s$ for a generic field $f$. Once again normal ordering is implied everywhere needed. 
The infinite sums\footnote{regularized as specified in \eqref{kernel}} appearing in these functions are absorbed in the calculation of matrix elements of \eqref{kernelrepconstraints} that will be performed later on.

\paragraph{Physical States in the Fock basis}
As said before, because of the decoupling between the fields of the bosonic sector the Hilbert space is a direct product of the Hilbert spaces for each field. Furthermore each field is described by a mode expansion, so that its Hilbert space is itself a tensor product of independent Hilbert spaces, one for each $n$ labelling the modes. For a single field $f$ a completely general state may be written as:
\begin{equation}\label{physicalstatef}
| \psi^f(d) \rangle = \bigotimes_{n \in \mathbb{Z}} \left [ \sum_{\mu \geq 0} d_\mu^f(n) | \mu_n^f \rangle \right ]\ ,
\end{equation}
where $n$ labels the modes, $\mu^f_n$ is the occupation number of the mode $n$ of the field $f$, and the $d$'s are complex coefficients. 
Considering then the whole set of fields in the model, any state in the complete Hilbert space (inclusive of the quantum fields $Z,Y$ and $D$ free scalar fields $\phi_i$) can be written then as a sum of factorized states in the form of \eqref{physicalstatef}:
\begin{equation}\label{physicalstate}
|\psi\rangle = \sum_{\{d^Z, d^\Y, d^i\}} | \psi^Z(d^Z) \rangle | \psi^\Y(d^\Y) \rangle \bigotimes_{i=1}^{D}| \psi^i(d^i) \rangle \ ,
\end{equation}
where the sum is over an arbitrary number of sets of $d$ coefficients.\\
This choice is not the most intuitive but it has the advantage of providing us with complete control on the single coefficients of every field, so that specific quantum states are easily selected for the purpose of a spectrum analysis; in the simple example of two decoupled systems, $A$ and $B$, with an Hilbert space basis $|n\rangle$ and $|m\rangle$ respectively, the easiest way to write a general state has the form $|\psi_g\rangle = \sum_{n,m} \psi(n,m)|n\rangle|m\rangle$. Considering factorized states $|\psi_f (a,b)\rangle = \sum_{n} a(n)|n\rangle\otimes \sum_mb(m)|m\rangle$, a sum over different sets of coefficients $\{a,b\}$ reproduces the general state given the identification $\psi(n,m) = \sum_{a,b} a(n)b(m)$, as in a series expansion.\smallbreak
To simplify the picture, and take advantage of the decoupling, without loosing insight in the mechanism that constrains the cosmological constant, we will consider a subset of the Hilbert space, in which the quantum states \eqref{physicalstate} are defined with a single set of $d$ coefficients, so that the sum is dropped. In this way the quantum state is completely factorized, and we can work in each sector separately.\\
Such a state will give, when contracted with coherent states \eqref{coherentstate}:
\begin{equation}
\begin{split}
|\langle \ul{z}|\psi\rangle|^2= | \psi (\ul{z}) |^2 = \prod_{f \in \text{\tiny fields}}| \psi^f (\ul{z}) |^2=\\= \prod_{f \in \text{\tiny fields}} \prod_n \left [\sum_{\mu, \nu \geq 0} d^f_\mu(n) \bar{d}^{f}_\nu (n) \bar{z}^\mu_n z^\nu_n e^{-|z_n|^2} \right ]\ .
\end{split}
\end{equation}
\subsection{Matrix elements of the quantum constraints}
By using the factorization the constraint equations \eqref{quantumconstraints} reduce to a sum of independent integrals over complex variables:
\begin{equation}
\begin{split}
 \langle L^\pm(s) \rangle = &\int \prod_m \left[ \frac{d z_m d \bar{z}_m}{2 \pi} \right] \Lslash^{\pm,Z} (s, z, \bar{z})  | \psi^Z (\ul{z}) |^2+\\  &+ \int \prod_m \left[ \frac{d z_m d \bar{z}_m}{2 \pi} \right] \Lslash^{\pm,\Y} (s, z, \bar{z})  | \psi^\Y (\ul{z}) |^2 +\\  &+ \sum_i^{D} \int \prod_m \left[ \frac{d z_m d \bar{z}_m}{2 \pi} \right]\Lslash^{\pm,i} (s, z, \bar{z})| \psi^i (\ul{z}) |^2+ \\&- \HB\frac{D+1}{12}\ .
\end{split}
\end{equation}
The integrals are all Gaussian in the $z$'s, since $|\psi|^2$ carries a Gaussian factor for each mode. Taking again an orthogonal combination of the constraints, we can finally obtain the equations:

\begin{equation}
\begin{split}\label{qc1}
 \langle L^+ + L^- \rangle =   \frac{2}{\sqrt{\pi}} \sum'_n |n|^{3/2} \left [ \frac{\xi_\Y}{\varpi_n^\Y}\Im \left (\omega^{(1)\Y}_n e^{- i n s}  \right ) - \frac{\xi_Z}{\varpi_n^Z}\Im\left (\omega^{(1)Z}_n e^{- i n s}  \right ) \right] - \\
  - \frac{1}{4\pi} \sum_{f}^{\mbox{\tiny fields}} \beta(f) \Biggl \{ \Bigl ( \sum_{n,m \geq 0} + \sum_{n,m \leq 0} \Bigr )_{n \neq m} \frac{4 \sqrt{n^* m^*}}{\varpi^f_n \varpi^f_m} \Re \left ( \omega^{(1)f}_n e^{- i n s}  \right ) \Re \left ( \omega^{(1)f}_m e^{- i m s}  \right ) +\\
 + \Bigl ( \sum'_n + 2\delta^n_0 \Bigr ) \left [ \frac{2 n^*}{\varpi^f_n} \left ( \Re \left ( \omega^{(2)f}_n e^{- i 2 n s}  \right ) + \tilde{\omega}^f_n - 1 \right ) \right ] \Biggr \} - \HB\frac{D+1}{6} + 2 \Lambda \prod_\ell \T_\ell ,
\end{split}
\end{equation}
\begin{equation}
\begin{split}\label{qc2}
 \langle L^+ - L^- \rangle = \frac{2}{\sqrt{\pi}} \sum'_n n |n|^{1/2} \left [ \frac{\xi_\Y}{\varpi_n^\Y}\Im \left (\omega^{(1)\Y}_n e^{- i n s}  \right ) - \frac{\xi_Z}{\varpi_n^Z}\Im\left (\omega^{(1)Z}_n e^{- i n s}  \right ) \right] - \\
  -\frac{1}{4\pi} \sum_{f}^{\mbox{\tiny fields}} \beta{f} \Biggl \{ \Bigl ( \sum_{n,m \geq 0} - \sum_{n,m \leq 0} \Bigr )_{n \neq m}  \frac{4 \sqrt{n^* m^*}}{\varpi^f_n \varpi^f_m}  \Re \left ( \omega^{(1)f}_n e^{- i n s}  \right ) \Re \left ( \omega^{(1)f}_m e^{- i m s}  \right ) +\\
 + \sum'_n \left [ \frac{2 n}{\varpi^f_n} \left ( \Re \left ( \omega^{(2)f}_n e^{- i 2 n s}  \right ) + \tilde{\omega}^f_n  \right ) \right ] \Biggr \} ,
\end{split}
\end{equation}
where: $$n^* = \Biggl \{ \begin{array}{c c} \frac{1}{4}&\ : n = 0\ ,\\ |n|&\ : n\neq 0\ .\end{array}$$
The sum over the fields, with the factor $\beta(f)$, means that there is one such contribution from each field in the model, with $\beta=-1$ for the $Z$ field and $\beta=1$ for all the others. $\ \T_\ell$ is the term coming from the integration of the Liouville potential:
\begin{equation}\label{thorn}
\begin{split}
 \T_\ell = & \sum_{\mu, \nu \geq 0} \sum_{\alpha = 0}^\mu \sum_{\beta = 0}^\nu \binom{\mu}{\alpha} \binom{\nu}{\beta} d^Z_\mu(\ell) \bar{d}^Z_\nu(\ell) \ i^{\mu - \alpha - \nu +\beta} \times \\  & \times \left [ \sum_\gamma^{\alpha+\beta} \binom{\alpha + \beta}{\gamma} \boldsymbol{\mathsf{S}}_\ell^{\alpha + \beta - \gamma} \int_{-\infty }^{+\infty } \! d x \ x^{\gamma }e^{-x^2}  \right] \times \\
 & \times \left [ \sum_\delta^{\mu + \nu - \alpha- \beta} \binom{\mu + \nu - \alpha- \beta}{\delta} \boldsymbol{\mathsf{C}}_\ell^{\mu + \nu - \alpha- \beta - \delta} \int_{-\infty }^{+\infty } \! d x \ x^{\delta }e^{-x^2} \right] \ ,
\end{split}
\end{equation}
with:
\begin{equation}
\boldsymbol{\mathsf{C}}_\ell = \Biggl \{ \begin{array}{l c} 0&\ell = 0\\ \frac{cos(\ell s)}{\sqrt{\pi |\ell|} 2 \xi}& \ell \neq 0 \end{array}\ , \qquad \quad
\boldsymbol{\mathsf{S}}_\ell = \Biggl \{ \begin{array}{l c} \frac{i}{2 \sqrt{\pi} \xi}&\ell = 0\\ \frac{sin(\ell s)}{\sqrt{\pi |\ell|} 2 \xi}& \ell \neq 0 \end{array}\ ,
\end{equation}
and the omega's are combinations of the $d$ coefficients which define the quantum state of the field $f$ they refer to:
\begin{subequations}\label{omegas}
\begin{align}
\varpi^f_n =& \sum_{\mu \geq 0} |d^f_\mu(n) |^2 \mu!\ ,\\
\omega^{(1)f}_n = & \sum_{\mu \geq 0} d^f_\mu(n) \bar{d}^f_{\mu+1}(n)(\mu + 1)!\ ,\\
\omega^{(2)f}_n = & \sum_{\mu \geq 0} d^f_\mu(n) \bar{d}^f_{\mu+2}(n)(\mu + 2)!\ ,\\
\tilde{\omega}^f_n =& \sum_{\mu \geq 0} |d^f_\mu(n) |^2 (\mu+1)!\ .
\end{align}
\end{subequations}
A second important result is explicit in these equations: while the second of \eqref{qc2} provides nothing more than a constraint on the coefficients $d$, the first one may be solved for the cosmological constant, $\Lambda$, for a given physical state, and determines its value as a function of the coupling constant $\xi$, and the $d$ coefficients themselves. Hence the requirement for a specific quantum state to be physical, {\it i.e.}, to be annihilated by the quantum constraints, can be realized only for a specific value of $\Lambda$.\\
Furthermore as mentioned before the inclusion of the gravitational sector together with the matter ones generates the possibility of compensating positive contributions to $\Lambda$ with a negative one, as the factor $\beta(f)$ clearly displays.
	\section{Spectrum analysis}
		Having obtained the expressions for the quantum constraints in terms of the coefficients which identify quantum states in Hilbert space, in an illustration of what kind of restrictions may arise for $\Lambda$, it is interesting to look into the spectrum of values that the cosmological constant takes when lower excitations of the model are required to be physical states.
\subsection{The Vacuum}
It seems a reasonable assumption for the Fock vacuum of the theory to be a physical state. Moreover in the model we are considering this choice corresponds to a static Minkowski solution.
Since this state is simply the tensor product of all Fock vacua, for every field $f$ and for every mode, the quantities defined in \eqref{thorn} and \eqref{omegas} will be:
\begin{equation}
\varpi^f_n = \tilde{\omega}^f_n = \T_n = 1, \quad \omega^{(1)f}_n = \omega^{(2)f}_n = 0,  \quad \forall n \in \mathbb{Z}, \forall f,
\end{equation}
so that \eqref{qc2} is identically vanishing, while \eqref{qc1} gives:
\begin{equation}
\Lambda = \frac{D+1}{12} \HB = \Lambda_\Omega.
\end{equation}
Hence the cosmological constant is forced by the quantum constraints to take a specific value, which is nothing else that the reabsorbed (linear) central charge which appears in the quantum Virasoro algebra \eqref{shift_nomax}. It is then just in an indirect way, {\it i.e.}, via the requirement of a conformal symmetry at the quantum level, that the coupling of scalar degrees of freedom to gravity induces a (generally) non vanishing cosmological constant in the vacuum, in contrast with the classical requirement $\Lambda=0$ for this solution. It is also worth to note that in spite of the dependence on $D$, which relates the cosmological constant to the matter content of the model, $\Lambda_\Omega$ is independent from the coupling constant $\xi$.
%\\
%Such a result is not surprising. In ordinary bosonic string theory the special case $D=26$ ensures a vanishing cosmological constant on the world-sheet, cancelling the anomaly arising from the quantization of the ghosts. In our model we have opposite contributions from the $Z$ and $Y$ fields, or equivalently from the dynamical gravitational degrees of freedom $X$ and $\varphi$, so that the consistency with string theory has to be expected. 
\subsection{First level excitations}
\begin{figure}
\includegraphics[width=\textwidth]{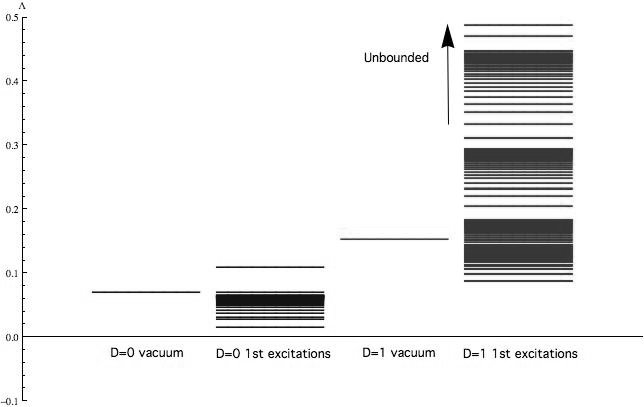}
\caption{Visualization of the spectrum for the cosmological constant with the following choice of parameters: $\HB = 1$, $\xi = 0.2$. In the cases in which one of the $d$ complex coefficients was unfixed by the constraints, we chose $d=1$.}\label{fig_spectrum}
\end{figure}

To go further, we will now show that imposing the same condition on a subset of the first level excitations of the fields\footnote{Namely states with occupation number at most 1 for each field.} will provide a spectrum of values for $\Lambda$, depending on the coefficients which define the quantum state and the coupling constant $\xi$.\\
Considering, for each field $f$, an excited state in the mode $n^f$ and the vacuum in all other modes:
\begin{equation}
|1\rangle = \bigotimes_f^{fields} |1^f \rangle = \bigotimes_f^{fields} \biggl [ \bigotimes_{n \neq n^f}  |\Omega\rangle \biggr ] \otimes \biggl [ d^f_{0} (n^f)| \Omega \rangle + d^f_{1} (n^f) |1^f_{n^f} \rangle \biggr ],
\end{equation}
we can calculate the quantum constraints \eqref{qc1} and \eqref{qc2} and apply a Fourier transform, so as to eliminate the space dependence. For the zero modes this leads to:
 \begin{equation}\label{1stzeromodes+}
\begin{split}
\langle L^+_0 + L^-_0 \rangle &= -\HB \frac{D+1}{6} + 2\Lambda \left ( 1 + \frac{|d^Z_{1}(n^Z)|^2}{4 \pi \xi^2 |n^Z|}\biggr | _{n^Z \neq 0} \right ) -\\ & - \frac{1}{4 \pi} \sum_f^{\mbox{\tiny fields}} \biggl [ \beta(f) \left ( 2 |n^f| + \delta^{n^f}_0 \right ) |d^f_{1}(n^f)|^2 \biggr ],
\end{split}
\end{equation}
\begin{equation}\label{1stzeromodes-}
\langle L^+_0 - L^-_0 \rangle \propto \sum_f^{\mbox{\tiny fields}} \biggl [ \beta (f) \ n^f |d^f_{1}(n^f)|^2 \biggr ],
\end{equation}
while for the other modes:
\begin{equation}\label{1stnmodes+}
\langle L^+_n + L^-_{-n} \rangle \propto 2\Lambda \left (\bar{d}^Z_0 (n) d^Z_1(n) + d^Z_0 (-n) \bar{d}^Z_1(-n) \right ),
\end{equation}
\begin{equation}\label{1stnmodes-}
\begin{split}
\langle L^+_n - L^-_{-n} \rangle  = &\xi_\Y \biggl (\bar{d}^\Y_0 (n) d^\Y_1(n) - d^\Y_0 (-n) \bar{d}^\Y_1(-n) \biggr ) -\\ &- \xi_Z \biggl ( \bar{d}^Z_0 (n) d^Z_1(n) - d^Z_0 (-n) \bar{d}^Z_1(-n) \biggr ),
\end{split}
\end{equation}
The analysis of such equations is rather complicated with an arbitrary number of scalar fields. We can then consider in detail the simplest (and more strictly constrained) cases, with $D=0$ and $D=1$.
\paragraph{1st excited level with no scalar fields} In this case, with only the $Z$ and $\Y$ fields present in the model, the results are quite straightforward: first, the excited fields have to be in a \emph{pure excited level}, {\it i.e.}, $|\psi^f_{n^f}\rangle \propto |1_{n^f} \rangle$, hence $d^f_0(n^f) = 0$. This follows directly from \eqref{1stnmodes+} and \eqref{1stnmodes-} when we exclude the solution $\Lambda=0$, which makes \eqref{1stzeromodes+} inconsistent.\\
The only possible solutions are then:\smallbreak
\begin{enumerate}
\item Both fields are excited\\
$$n^Z = n^\Y = N,$$
\begin{equation}\label{lambdadense}
 \Lambda =\HB\frac{1}{12} \left ( 1 + \frac{1}{4 \pi \xi^2 |n^Z|}\biggr | _{n^Z \neq 0} \right )^{-1} = 
\Biggl \{ \begin{array}{l l}
\HB\frac{1}{6}\frac{2\pi \xi^2 |N|}{1 + 4\pi \xi^2 |N|}&\ : N \neq 0\ ,\\ \\
\HB\frac{1}{12}&\ : N = 0\ .
\end{array}
\end{equation}
\item Only the $Z$ field is excited:
\begin{equation}\label{lambdanegative}
n^Z = 0: \qquad \qquad \Lambda = \HB\frac{1}{12}-\frac{1}{8\pi}\ .
\end{equation}
\item Only the $\Y$ field is excited:
\begin{equation}
n^\Y = 0: \qquad \qquad \Lambda = \HB\frac{1}{12}+\frac{1}{8\pi}\ .
\end{equation}
\end{enumerate}
As one can see there are strict constraints on the values that $n^Z$ and $n^\Y$ can take: in particular the fields can be excited in non-zero modes only together and in the same mode (which is reminiscent of level-matching conditions in string theory). If on the other hand one of $Z$ or $\Y$ is in its ground state, only the zero mode of the other field can be excited.\\
The spectrum of values that the cosmological constant $\Lambda$ is allowed to take is bounded and discrete.\\
As one can see from the equations in the case (1) it also becomes infinitely dense to the left of $\Lambda=\HB\frac{1}{6}$ for large $N$.
Furthermore if both fields are excited in their zero modes the cosmological constant will take the same value as obtained in the vacuum, namely $\Lambda_\Omega$.
\paragraph{1st excited level with $D=1$ scalar fields} Increasing the number of scalar fields loosens the restrictions imposed by the constraints. In fact already with one scalar field equations \eqref{1stzeromodes+}, \eqref{1stzeromodes-}, \eqref{1stnmodes+} and \eqref{1stnmodes-}, when the additional scalar field $\phi$ is excited in its zero mode, do not fix one of the $d$ coefficients, resulting in a $d$-dependent cosmological constant, {\it i.e.}, a finite part of the spectrum of $\Lambda$ is continuous. Of course when only pure excitations are considered, and all the $d$'s are fixed, as is usually done in string theory, the spectrum is discrete. \\
Furthermore the presence of a third field allows for much more freedom for which modes may be excited, removing (or reducing to inequalities) the constraints on the $n^f$'s obtained above. Again the value $\Lambda=0$ is excluded by the quantum constraints.\smallbreak
If all fields are excited we are forced to have pure excitations of the $Z$ and $\Y$ fields, {\it i.e.}, $d_0^Z(N) = d_0^\Y (N) = 0$:
\begin{enumerate}
\item $n^Z = n^\Y = N, \qquad n^\phi = 0$,
\begin{equation}\nonumber
\Lambda = \left (\HB\frac{1}{6} + \frac{|d_\phi^1(0)|^2}{8 \pi} \right ) \left ( 1 + \frac{1}{4 \pi \xi^2 |N|}\biggr | _{N \neq 0} \right )^{-1} .
\end{equation}
the spectrum for $\Lambda$ is positive, bounded on both sides and continuous.
\item $n^Z = -n^\Y = N \neq 0, \qquad n^\phi \neq 0, \qquad |d^\phi_1(n^1)|^2 = 2\frac{N}{n^\phi}$,
\begin{equation}\nonumber
\Lambda = \left ( \HB\frac{1}{6} + \frac{|N|}{2\pi} \right ) \left ( 1 + \frac{1}{4 \pi \xi^2 |N|} \right )^{-1} .
\end{equation}
the spectrum for $\Lambda$ is bounded from below, positive and discrete.
\item $ |n^Z| \neq |n^\Y|, \qquad n^\phi \neq 0, \qquad |d^\phi_1(n^\phi)|^2 = 2\frac{n^Z - n^\Y}{n^\phi}$,
\begin{itemize}
\item $ n^Z = 0, \qquad n^\Y \neq 0 $\\ $ \Lambda = \HB\frac{1}{3}+\frac{4|n^\Y| - 1}{8\pi}$,\\
$\Lambda$ is bounded from below and discrete.
\item $ n^Z \neq 0, \qquad n^\Y = 0$ \\ $ \Lambda = \left (\HB\frac{1}{6} + \frac{1}{8\pi} \right ) \left ( 1 + \frac{1}{4 \pi \xi^2 |n^Z|} \right )^{-1}$,\\
$\Lambda$ is bounded on both sides and discrete.
\item $ n^Z \neq 0, \qquad n^\Y \neq 0 $ \\ $\Lambda = \left ( \HB\frac{1}{6} + \frac{1}{4\pi}\left (|n^\Y| - |n^Z| + | n^\Y - n^Z|\right )\right ) \left ( 1 + \frac{1}{4 \pi \xi^2 |n^Z|} \right )^{-1}$,\\
$\Lambda$ is bounded from below, discrete and highly degenerate.
\end{itemize}
\end{enumerate}
\smallbreak
Two fields out of three are excited:
\begin{enumerate}
\item $Z$ and $\phi$ excited, $\Y$ in ground state. $Z$ is forced to be purely excited.
\begin{itemize}
\item $n^Z = n^\phi = 0 \qquad \longrightarrow \qquad \Lambda =\HB\frac{1}{6} + \frac{1}{8\pi} (|d_1^\phi(0)|^2 - 1 )$,\\
$\Lambda$ is bounded on both sides and continuous.
\item $n^Z \neq 0, \qquad n^\phi \neq 0, \qquad |d_1^\phi(n^\phi)|^2 = \HB\frac{n^Z}{n^\phi}$ \\ $ \Lambda = \frac{1}{6} \HB \left ( 1 + \frac{1}{4 \pi \xi^2 |n^Z|} \right )^{-1}$,\\
$\Lambda$ is negative, bounded on both sides and discrete.
\end{itemize} 
\item $\Y$ and $\phi$ excited, $Z$ in ground state. $\Y$  is forced to be purely excited.\\
%\begin{itemize}
%\item
$n^\Y = n^\phi = 0 \qquad \longrightarrow \qquad \Lambda = \HB\frac{1}{6} + \frac{1}{8\pi} (1+|d_1^\phi(0)|^2  )$,\\
$\Lambda$ is bounded on both sides and continuous.
%\item $n^\Y \neq 0 \qquad n^1 \neq 0 \qquad |d_1^1(n^1)|^2 = -\frac{n^\Y}{n^1} \longrightarrow \Lambda = \frac{25}{12} + \frac{|n^\Y|}{\pi}$\\
%$\Lambda$ is positive, unbounded and discrete.
%\end{itemize}
\item $Z$ and $\Y$ excited, $\phi$ in ground state. Both fields are forced to be purely excited, while $n^Z = n^\Y = N$,\\
\begin{equation}\nonumber
\Lambda = \HB\frac{1}{6} \left ( 1 + \frac{1}{4 \pi \xi^2 |N|}\biggr | _{N \neq 0} \right )^{-1} .
\end{equation}
\end{enumerate}
\smallbreak
Only one field is excited:
\begin{enumerate}
\item Field $Z$ purely excited in the zero mode $\longrightarrow \Lambda = \HB\frac{1}{6}-\frac{1}{8\pi}$.\\
\item Field $\Y$ purely excited in the zero mode $\longrightarrow \Lambda = \HB\frac{1}{6}+\frac{1}{8\pi}$.
\item Field $\phi$ excited in the zero mode $\longrightarrow \Lambda = \HB\frac{1}{6}-\frac{1}{8\pi}|d^\phi_1(0)|^2$.\\
$\Lambda$ is negative, bounded and continuous.
\end{enumerate}
Summarizing, the cosmological constant takes values from the minimum between $(\HB\frac{1}{6}-\frac{1}{8\pi})$ and $(\HB\frac{2}{3}\frac{\pi \xi^2 |N|}{1 + 4 \pi \xi^2 |N|})$ (depending again on the value of the coupling constant $\xi$), and is unbounded from above. It is everywhere discrete except for values within the range between $(\HB\frac{2}{3}\frac{\pi \xi^2 |N|}{1 + 4 \pi \xi^2 |N|})$ and $(\HB\frac{1}{6}+\frac{1}{8\pi})$, where countable infinities of the continuous bands mentioned above appear. These bands overlap differently depending on the value of the coupling constant $\xi$.
Again it is possible to have $\Lambda_1^{(D=1)}=\Lambda_\Omega$, with the necessary, but not sufficient, condition to have an $Z$ field excited in its zero mode.\smallbreak
Generalizing the $D=1$ case we can expect the spectrum for $\Lambda$ to consist of an infinite, countable, discrete set of values, in which some continuous bands appear, reflecting the presence of the unconstrained continuous coefficients $d$.
\smallbreak
An visualization of the spectrum just discussed is given in Figure ~\ref{fig_spectrum} in the case with $\HB = 1$, $\xi = 0.2$ and $d=1$ whenever one of the $d$ coefficients was not fixed by the constraints. In this specific case the spectrum for $\Lambda$ is unbounded towards positive infinity. This is due to the presence of the level matching conditions for the $Z$ and $\Y$ fields, which allow only for a partial cancellation between the negative contributions of the $Z$ field and the positive contributions of the $\Y$ and matter fields. These condition are loosened in the presence of additional matter fields, which introduce additional $d$ coefficients, \emph{i.e.} more free parameters, leaving the $Z$ field free to be excited arbitrarily high modes and provide greater negative contributions.\\ Furthermore no negative values of $\Lambda$ are admitted in this case. This is also due to the choice $\HB=1$. In a classical limit in which $\HB \to 0$ it is clear that \eqref{lambdanegative} gives $\Lambda = -1/8\pi <0$. Note also how regions of the spectrum are particularly dense when approaching some specific value of the cosmological constant from below.

\subsubsection{Higher excitations}
Due to the highly non-linear form of the quantum constraints \eqref{qc1}\eqref{qc2} a general analysis of higher excitations of the model is not easy to perform.
Besides the great number of states which would have to be considered, a main issue is given by the $s$-dependence in the Liouville potential term, which in general prevents us from performing a Fourier transform and work with a countable set of quantum constraints.\\
It is anyway worth pointing out, that it is only in higher (non purely) excited states that the quantum corrected coupling constants $\xi_Z, \xi_\Y$ play a role in determining the value of the cosmological constant, adding more quantum contributions to $\Lambda$ .
%	\section{Discussion and outlook}
%		\input{files/discussion}
\chapter{Conclusions and perspectives}
	\label{ch:end}
	\section{Summary}
Chapter \ref{ch:cc} was dedicated to an introduction to the cosmological constant problem. We described the historical background and recalled how the cosmological constant was conceived as a modification of Newtonian dynamics first and General Relativity later. Numbers, the relation between the abstract world of theory and the concreteness of the real world, found some space in the discussion of the astronomical observations that determine the value of $\Lambda$ that we see realized in Nature.\\
We have then examined the relation of the cosmological constant with the vacuum, finding a point of contact between General Relativity and Quantum Field Theory, which is the very core of the cosmological constant problem: how can we predict, account for or even control the tiny value of $\Lambda$? Why does the large vacuum energy predicted by Quantum Field Theory does not produce a large effect on the expansion of the Universe?\\
We have discussed also the coincidence problem and the solutions proposed by the landscape of string theory. We have reviewed how the case of one-dimensional gravity suggests to us that in a quantum theory of gravity a solution might be hidden and awaiting discovery, and we have tried to generalize this line of thought.

In Chapter \ref{ch:dmax} we have dealt with the classical theory of dilaton-Maxwell gravity in $1+1$ dimensions. We have argued that a two-dimensional model, despite the many simplifications with respect to a four-dimensional theory, is a very interesting and rich testing ground for new ideas on the path towards a quantum theory of gravity.\\
We have discussed how a subclass of the many models included in dilaton-Maxwell gravity can be reformulated in terms of a partially decoupled dual field theory. This is a first original contribution of this thesis: we have shown that the system of dilaton gravity and a non-minimally coupled gauge field, with a specific choice of the potentials, is in fact equivalent to two decoupled Liouville fields living on a flat space-time. Additionally the Liouville Field Theory exhibits the same symmetries of the original model, namely diffeomorphisms and $U(1)$ gauge invariances. We have then made explicit the presence of a cosmological constant in the dual theory, and discussed the possibility of adding additional fields.\\
We have also performed an analysis of the classical solutions of the dual theory, discussing the behaviour of space-time curvature and its singularities.\\
Paving the way to the quantum theory, we have described the formulation of the model in the Hamiltonian and BRST formalisms, determining the algebra of constraints.

Chapter \ref{ch:qt} contains the main original contributions of this work: we have devoted ourselves to the quantization and to the investigation of the mechanism that quantizes the cosmological constant. All the different sectors of the model were quantized, including possible additional fields, and the total quantum algebra of the Virasoro generators was obtained, inclusive of all the quantum corrections. The total central extension, in different cases, was calculated and cancelled.\\
We have then discussed how the quantum constraints can be implemented, explicitly deriving the form of their matrix elements after a suitable choice of their representation and of the basis in Hilbert space.\\
This allowed us to obtain that the cosmological constant is indeed quantized in two dimensions and its value is fixed once we require a quantum state to be physical. \\
Contributions from the gravitational degrees of freedom appear with a sign opposite to that of the contributions from the matter sectors, suggesting the possibility of a mechanism that could compensate the large vacuum contributions of the Standard Model of particle physics with excitations in the gravitational sector. Additional quantum corrections are also generated in the quantization of gravity, suggesting an important role of quantum gravity.\\
Finally, we have analysed the spectrum of the cosmological constant for the lower excitations of the model, with a brief discussion on what we expect from higher excitations.
\section{Discussion and outlook}
We have seen that in the case of two-dimensional dilaton-Maxwell gravity, and in particular for the subclass of model which admit a dual description in terms of Liouville fields, the realization of the classical symmetry at the quantum level provides a mechanism that fixes the cosmological constant to a specific value once a particular quantum state is required to be physical. In this approach $\Lambda$ is considered a free parameter which can be fixed by consistency conditions determined by quantum gravity.\\
The value of $\Lambda$ includes classical and quantum contributions and it is determined in a fully non-perturbative quantum theory of gravity and scalar matter. In this respect this result takes full advantage of the lower dimensional setting: the usual computation of the value of the cosmological constant in four dimensions is performed only considering one-loop contributions of matter and gauge fields in perturbation theory in a flat background geometry.\\
Turning the physicality conditions for quantum states, \emph{i.e.} annihilation by the action of the quantum constraints, into equations for $\Lambda$ we can account for the quantum vacuum fluctuations and the excitations of the quantum fields. In particular we can see that, in contrast with the positive contributions of the matter fields, one of the gravitational degrees of freedom, namely the $Z$ field, provides a negative term, therefore allowing for partial cancellations and small values of $\Lambda$ even in the presence of (excited) matter fields.\\
Despite the technical simplifications of the lower dimensional setting, however, it is not easy to freely study the case of arbitrarily high excitations and/or arbitrarily many matter fieds: an expression for $\Lambda$ in terms of occupation numbers and complex coefficients of the quantum states of the fields has to be obtained from imposing the Fourier modes of \eqref{qc1} and \eqref{qc2} to vanish. This keeps us from determining a general formula for the cosmological constant and we are forced to do it on a case by case basis. For simplicity we limited our study to the lower excitations only.\\
In our approach, considering the way we exploited the realization of the quantum constraints, we are able to determine, for a given quantum state of the model, which is the value that the cosmological constant has to take. In this way it is natural to study the spectrum of values of $\Lambda$ for particular subsets of quantum states. We can then see a parallel with the ideas behind the landscape of string theory. In both cases there is a large set of configurations (the $10^{500}$ vacua VS. the infinitely many quantum states) and for each of them there is a specific value for the cosmological constant. In neither cases we know a priori which one is the configuration we are in. Let us stress however that in our case, albeit two-dimensional, we are not introducing new physics but simply quantizing a theory of gravity and matter.\\
While the techniques applied in this work are especially two-dimensional, the results have a more general significance. As discussed in Section \ref{sec:ccinqg} the classical constaints which appear in the canonical formulation of any diffeomorphism invariant theory can be exploited in the quantum theory to determine the cosmological constant required for a given quantum state to be physical. We can therefore expect similar results in dimensions higher than two: in particular in the limit in which quantum gravity reduces to General Relativity we should be able to see the cosmological constant as made up by classical contributions and quantum corrections, as it is the case in our discussion.\\
A brief remark can be done on Spherically Reduced Gravity. First of all it has to be said that SRG is not one of the models which allows for a dual description in terms of decoupled Liouville fields. If no alternative decoupled formulation is accessible the only possibility is to proceed perturbatively, possibly employing the path integral approach, as developed in the literature (\cite{Grumiller:2002nm} and references therein). The dilaton field $X$ in SRG is related to the anglular dynamics of four-dimensional General Relativity. On the other hand in our decoupled model the $Z$ and $Y$ fields can be roughly related to the conformal mode $\varphi$ and the dilaton $X$, respectively. We can then say that it is the conformal mode to contribute negatively to $\Lambda$, while the dilaton contributes in a way similiar to a matter field. We can therefore speculate that in a four-dimensional (roughly spherically symmetric) scenario excitations of the conformal mode might provide cancellations that would allow for a smaller value of the cosmological constant in the same way that it happens in two-dimensions.
\section{Perspectives}
Different avenues are at hand for the further development of this work:
\begin{itemize}
\item \textit{Quantum physical states and the cosmological constant}\\
As briefly discussed above, it would be interesting to investigate further how the different quantum fields contribute to the cosmological constant, particularly for higher excitations of the $Z$ field, which is the only one contributing with a negative term to the cosmological constant.
\item \textit{Spherically reduced gravity}\\
Spherically Reduced Gravity, despite not being included in the class of models with a dual Liouville description, is the one model more directly related to higher dimensions. With substancial technical differences the cosmological constant problem could be investigated in this setting, providing more insight in possible solutions in higher dimensions.
\item \textit{Adding fermions}\\
The inclusion of fermionic matter fields would provide a more complete understanding of the different contributions to the cosmological constant. In order to treat half-integer spin fields a reformulation in terms of Cartan variables is required \cite{Meyer:2005lr} and additional second class constraints are bound to appear as a consequence of the $SU(2)$ symmetry.
\item \textit{Adding charged scalars}\\
The addition of charged complex scalar fields would allow us to test the model in the presence of a non-gravitational interaction, which however will have to be treated perturbatively.
\item \textit{Adding a mass to the vector field}\\
The vector field included in this work can be made massive via the St\"uckelberg mechanism \cite{Ruegg:2003ps}. However, in contrast with the case of a fixed background geometry, the interaction term between the St\"uckelberg field and the vector field cannot be eliminated with the addition of a gauge fixing term, generating a divergent contribution to the central charge of the quantum Virasoro algebra. A modified St\"uckelberg mechanism has to be introduced, with a modified dynamics that would exhibit a vanishing central charge. Work is in progress in this direction.
\item \textit{The cosmological constant problem in 2+1 dimensions}\\
An investigation following the same line of though for the case of 2+1 dimensional gravity is of course very interesting. However there are profound differences between the two- and three-dimensional theories and such a project will require a completely different approach.
\end{itemize}
\appendix
\chapter{Kernel representation for quantum operators}\label{kernelrep}
	Given a countable collection of Fock spaces, labelled with $n \in \mathbb{Z}$, one can build a coherent state as the tensor product:
\begin{equation}
|\ul{z} \rangle = \bigotimes_n |\ul{z}_n \rangle = \prod_n e^{-\frac{1}{2} |z_n|^2} e^{z_n \hat{a}_n^\dagger} |\Omega \rangle \ ,
\end{equation}
where $|\Omega \rangle$ is the Fock vacuum, while the Fock operators obey the usual commutator $[\hat{a}_n, \hat{a}^\dagger_m] = \delta^n_m$.
The action of the annihilation operators on coherent states gives:
\begin{equation}
\hat{a}_n |\ul{z} \rangle = z_n |\ul{z} \rangle\ , \qquad \qquad e^{\alpha \hat{a}_n}|\ul{z} \rangle = e^{\alpha z_n}|\ul{z} \rangle \ .
\end{equation}
For a generic quantum operator $\hat{\mc{O}}$, with $\langle \ul{z} | \hat{\mc{O}} | \ul{z}  \rangle = \mc{O}(z,\bar{z})$, one can define:
\begin{equation}
\mc{O}(z,\bar{z}) = \int \prod_m \left[ d^2 y_m e^{i y_m z_m} e^{i \bar{y}_m \bar{z}_m} \right] \tilde{\mc{O}}(y, \bar{y})\ ,
\end{equation}
and equivalently:
\begin{equation}
\tilde{\mc{O}}(y, \bar{y})= \int \prod_m \left[ \frac{d^2 z_m}{(2 \pi)^2} e^{-i y_m z_m} e^{-i \bar{y}_m \bar{z}_m} \right] \mc{O}(z,\bar{z})\ ,
\end{equation}
where $d^2 y = d y d \bar{y}$. Considering that :
\begin{equation}\nonumber
e^{-i y_m z_m} e^{-i \bar{y}_m \bar{z}_m} = \langle \ul{z} |e^{-i \bar{y}_m \hat{a}_m^\dagger} e^{-i y_m \hat{a}_m} |\ul{z} \rangle\ ,
\end{equation}
we have:
\begin{equation}
\mc{O}(z,\bar{z}) = \int \langle \ul{z} |\prod_m \left[ d^2 y_m  e^{-i \bar{y}_m \hat{a}_m^\dagger} e^{-i y_m \hat{a}_m} \right] |\ul{z} \rangle \tilde{\mc{O}}(y, \bar{y})\ .
\end{equation}
In this way the operator can be written as:
\begin{equation}
\hat{\mc{O}} = \int \prod_m \left[ d^2 y_m   e^{-i y_m \hat{a}_m} e^{-i \bar{y}_m \hat{a}_m^\dagger} e^{-y_m \bar{y}_m}\right] \tilde{\mc{O}}(y, \bar{y})\ ,
\end{equation}
where the extra exponential term comes from the swapping of the two exponential operators. By inserting the identity between them:
\begin{equation}
\begin{split}
\hat{\mc{O}} =& \int \prod_m \left[ \frac{d^2 y_m d^2 z_m}{2\pi}  e^{-i y_m z_m} e^{-i \bar{y}_m \bar{z}_m} e^{-y_m \bar{y}_m}\right] \tilde{\mc{O}}(y, \bar{y}) |\ul{z} \rangle\langle \ul{z} | = \\
=& \int \prod_m \left[ \frac{d^2 z_m}{2 \pi} \right]  |\ul{z} \rangle \Oslash (z, \bar{z})\langle \ul{z} |\ ,
\end{split}
\end{equation}
where $\Oslash (z, \bar{z})$, the integral kernel of the operator, is expressed as:
\begin{equation}\label{kernel}
\begin{split}
\Oslash (z, \bar{z}) &= \int \prod_m \left[ d^2 y_m  e^{i y_m z_m} e^{i \bar{y}_m \bar{z}_m} e^{-y_m \bar{y}_m}\right] \tilde{\mc{O}}(y, \bar{y}) =\\
&= \mbox{exp}\left( \sum_m\partial_{z_m} \partial_{\bar{z}_m} \right) \mc{O} (z, \bar{z})\ ,
\end{split}
\end{equation}
and the sum has to be considered regulated, e.g. with an dampening exponential $e^{-\epsilon|m|}$, with $\epsilon > 0$.
\chapter{Notation}
	\begin{tabular}{c|l}
%\hline
$G$&Newton's constant\\
$l_P$&Planck's length\\
$R,R_{\mu\nu}$&Ricci scalar and Ricci tensor\\
$T_{(\mu, \nu)}=T_{\mu \nu}+T_{\nu \mu}$&Symmetric tensor indices\\
$T_{[\mu, \nu]}=T_{\mu \nu}-T_{\nu \mu}$&Antisymmetric tensor indices\\
$\partial_\mu F = F_{,\mu}$&Partial derivation\\
$D_\mu F = F_{;\mu}$&Metric compatible covariant derivation\\
$\mc{L}\ , \mc{H}$&Langrangian and Hamiltonian densities\\
$L\ ,H$&Lagrangian and Hamiltonian functions\\
$F\approx 0$& Weak inequality (constrained dynamics)\\
$x\sim 10^3\  , F(X)\sim X$ & ``behaves as'' \\
%\hline
\end{tabular}
\chapter{Additional Research}
	\section*{The problem of time}
It is well known that in the canonical formulation of GR, employing the ADM decomposition, the resulting theory contains first-class constraints, which are generators of space-time diffeomorphisms. Most importantly the Hamiltonian function itself is a linear combination of such constraints, and is therefore vanishing on the constraints surface. It is then straightforward to conclude that no time evolution is possible for the physical states of a canonical quantum theory of gravity, as it is clearly shown by the famous Wheeler-DeWitt equation.\\
One of the possibilities in approaching this issue is the introduction of matter reference frames. In particular, the Brown-Kucha\v{r} mechanism \cite{kucharbrown:1995} consists in coupling GR to a specific matter model and, by manipulating the constraints, it make possible to obtain a so-called physical Hamiltonian, which would describe the (gravitational) dynamics with one of the matter variables playing the role of time.\\
In \cite{Cianfrani:2008p574} we applied the BK mechanism first to a generalized scalar field fluid and then to Schutz' model for a perfect (baryonic) fluid \cite{schutz:1970,schutz:1971}. The latter in particular is a much more realistic description of the cosmological fluid as compared to the dust model presented in the original work by Brown and Kucha\v{r}, especially in the early stages of the expansion of the Universe, where a non vanishing pressure better accounts for thermal energy, which is much greater than the rest mass of the particles themselves. A physical Hamiltonian was obtained, as a function of the gravitational (spatial) variables and the fluid entropy, successfully recovering a dynamical picture for the system. Moreover by choosing the frame co-moving with the fluid the logarithm of the entropy per baryon itself is interpreted as the time variable of the system: it is interesting to see that the entropy, an intrinsically future pointing variable, naturally plays the role of time in this framework, without it being foreseeable from the model itself.
\section*{Graviton confinement}
The long lasting interest in the existence of extra-dimensions has produced quite a variety of ways to accommodate their presence alongside the observed 4d Universe, taking into account the present experimental bounds. In particular Standard Model particles might be confined to a D3-brane, while gravity might be ``leaking'' in the extra-dimensions, at scales which have not been tested yet. In \cite{Murray:2010p3917} we investigated the case of self-gravitating hypermonopoles of any dimensions, showing that graviton confinement to the 4d bulk by curvature effects is present for any asymptotically flat space-time of more than 6d. For light enough resonances gravity is four-dimensional in an intermediate range, as required.
\section*{Affine quantization of gravity}
The initial singularity problem is one of the unresolved issues in modern cosmology. It is believed that quantum gravity might provide an answer, and many approaches have been put forward. An interesting proposal consists in the Affine quantization of gravity \cite{Klauder:1999ba,Klauder:2011ue}. In \cite{Fanuel:2012fk} we applied the Coherent States Affine Quantization program first to a toy model of FRLW cosmology and then to a reparametrization invariant minisuperspace model coupled to a scalar field. We found that the quantization procedure introduces additional dynamics in both cases, avoiding the initial singularity and resulting in a bouncing Universe. 
\cleardoublepage
\phantomsection
\addcontentsline{toc}{chapter}{Bibliography}
\bibliographystyle{hunsrt}
\bibliography{references}

\begin{thebibliography}{10}

\bibitem{Weinberg:1989}
Steven Weinberg.
\newblock The cosmological constant problem.
\newblock {\em Rev. Mod. Phys.}, 61(1):1--23, Jan 1989.

\bibitem{Weinberg:2000}
Steven Weinberg.
\newblock {The Cosmological constant problems}.
\newblock pages 18--26, 2000, astro-ph/0005265.

\bibitem{Carroll:1992}
S.~M. Carroll, W.~H. Press, and E.~L. Turner.
\newblock The cosmological constant.
\newblock {\em Ann. Rev. Astron. Astrophys.}, 30:499--542, 1992.

\bibitem{Peebles:2002gy}
P.J.E. Peebles and Bharat Ratra.
\newblock {The Cosmological constant and dark energy}.
\newblock {\em Rev.Mod.Phys.}, 75:559--606, 2003, astro-ph/0207347.

\bibitem{Carroll:2000fy}
Sean~M. Carroll.
\newblock {The Cosmological constant}.
\newblock {\em Living Rev.Rel.}, 4:1, 2001, astro-ph/0004075.

\bibitem{Bousso:2007gp}
Raphael Bousso.
\newblock {TASI Lectures on the Cosmological Constant}.
\newblock {\em Gen.Rel.Grav.}, 40:607--637, 2008, 0708.4231.

\bibitem{Rugh:2002}
S.~E. Rugh and H.~Zinkernagel.
\newblock The quantum vacuum and the cosmological constant problem.
\newblock {\em Studies In History and Philosophy of Science Part B: Studies In
  History and Philosophy of Modern Physics}, pages 663--705, 2002.

\bibitem{Bousso:2012dk}
Raphael Bousso.
\newblock {The Cosmological Constant Problem, Dark Energy, and the Landscape of
  String Theory}.
\newblock 2012, 1203.0307.

\bibitem{Seeliger:1895fk}
H.~Seeliger.
\newblock {\em Astronom. Nachr.}, 137:129, 1895.

\bibitem{Neumann:1896lr}
C.~Neumann.
\newblock {\em Uber das Newtonische Prinzip der Fernwirkung}, volume S. 1 u. 2.
\newblock Leipzig, 1896.

\bibitem{Norton:1999lr}
J.~D. Norton.
\newblock The cosmological woes of newtonian gravitation theory.
\newblock In H.~Goenner, J.~Renn, J.~Ritter, and T.~Sauer, editors, {\em The
  Expanding Worlds of General Relativity: Einstein Studies}, volume~7, pages
  271--322. Birkh\"{\"a}user (Boston), 1999.

\bibitem{Einstein:1917fk}
A.~Einstein.
\newblock {\em Sitz. Preuss. Akad. d. Wiss. Phys.-Math}, 142, 1917.

\bibitem{Einstein:1952fj}
A.~Einstein.
\newblock {\em The Principle of Relativity}, chapter Cosmological
  considerations on the general theory of relativity.
\newblock Dover Publ. (New York), 1952.

\bibitem{hawking:1973}
S.~W. Hawking and G.~F.~R. Ellis.
\newblock {\em The Large Scale Structure of {Space-Time}}.
\newblock Cambridge University Press, February 1975.

\bibitem{Gibbons:1977ceh}
G.~W. Gibbons and S.~W. Hawking.
\newblock Cosmological event horizons, thermodynamics, and particle creation.
\newblock {\em Phys. Rev. D}, 15:2738--2751, May 1977.

\bibitem{Edwards:1972lr}
D.~Edwards.
\newblock Exact expressions for the properties of the zero-pressure friedmann
  models.
\newblock {\em M.N.R.A.S.}, 159, 51, 1972.

\bibitem{Perlmutter:1998np}
S.~Perlmutter et~al.
\newblock {Measurements of Omega and Lambda from 42 high redshift supernovae}.
\newblock {\em Astrophys.J.}, 517:565--586, 1999, astro-ph/9812133.

\bibitem{Riess:1998cb}
Adam~G. Riess et~al.
\newblock {Observational evidence from supernovae for an accelerating universe
  and a cosmological constant}.
\newblock {\em Astron.J.}, 116:1009--1038, 1998, astro-ph/9805201.

\bibitem{tegmark:2004}
Max Tegmark and et~al.
\newblock Cosmological parameters from sdss and wmap.
\newblock {\em Phys. Rev. D}, 69(10):103501, May 2004.

\bibitem{Guth:1980zm}
Alan~H. Guth.
\newblock {The Inflationary Universe: A Possible Solution to the Horizon and
  Flatness Problems}.
\newblock {\em Phys.Rev.}, D23:347--356, 1981.

\bibitem{Rindler:1969kx}
W.~Rindler.
\newblock {\em Essential Relativity}.
\newblock Van Nostrand, New York, 1969.

\bibitem{Zeldovich:1971fk}
Ya.~B. Zel'dovich and I.~D. Novikov.
\newblock {\em Teoriya Tyagoteniya i evolyutsii zvezd (Theory of Gravitation
  and Stellar Evolution)}.
\newblock Nauka, 1971.

\bibitem{Linde:1974at}
Andrei~D. Linde.
\newblock {Is the Lee constant a cosmological constant?}
\newblock {\em JETP Lett.}, 19:183, 1974.

\bibitem{Veltman:1974au}
M.J.G. Veltman.
\newblock {Cosmology and the Higgs Mechanism}.
\newblock {\em Phys.Rev.Lett.}, 34:777, 1975.

\bibitem{weinberg:qtf2}
Steven Weinberg.
\newblock {\em The Quantum Theory of Fields, Volume {II}: Modern Applications}.
\newblock Cambridge University Press, May 2005.

\bibitem{Kolb:1990vq}
Edward~W. Kolb and Michael~S. Turner.
\newblock {The Early universe}.
\newblock {\em Front.Phys.}, 69:1--547, 1990.

\bibitem{Sato:1980yn}
K.~Sato.
\newblock {First Order Phase Transition of a Vacuum and Expansion of the
  Universe}.
\newblock {\em Mon.Not.Roy.Astron.Soc.}, 195:467--479, 1981.

\bibitem{Rugh:1999rt}
S.E. Rugh, H.~Zinkernagel, and T.Y. Cao.
\newblock The casimir effect and the interpretation of the vacuum.
\newblock {\em Studies in History and Philosophy of Modern Physics},
  30(1):111--139, 1999.

\bibitem{Bohr:1983ys}
N.~Bohr and L.~Rosenfeld.
\newblock {\em Quantum Theory and Measurement}, pages 465 -- 522.
\newblock Princeton University Press, 1983.

\bibitem{Martin:2012p5779}
Jerome Martin.
\newblock Everything you always wanted to know about the cosmological constant
  problem (but were afraid to ask).
\newblock May 2012, 1205.3365.

\bibitem{Dicke:1957zz}
R.H. Dicke.
\newblock {Gravitation without a Principle of Equivalence}.
\newblock {\em Rev.Mod.Phys.}, 29:363--376, 1957.

\bibitem{Barrow:1986lr}
J.D. Barrow and F.J. Tipler.
\newblock {\em The Anthropic Cosmological Principle,}.
\newblock Oxford University Press, 1986.

\bibitem{Kimpton:2012rv}
Ian Kimpton and Antonio Padilla.
\newblock {Cleaning up the cosmological constant}.
\newblock 2012, 1203.1040.

\bibitem{Polchinski:2006gy}
Joseph Polchinski.
\newblock {The Cosmological Constant and the String Landscape}.
\newblock pages 216--236, 2006, hep-th/0603249.

\bibitem{Bousso:2006ev}
Raphael Bousso.
\newblock {Holographic probabilities in eternal inflation}.
\newblock {\em Phys.Rev.Lett.}, 97:191302, 2006, hep-th/0605263.

\bibitem{Bousso:2007kq}
Raphael Bousso, Roni Harnik, Graham~D. Kribs, and Gilad Perez.
\newblock {Predicting the Cosmological Constant from the Causal Entropic
  Principle}.
\newblock {\em Phys.Rev.}, D76:043513, 2007, hep-th/0702115.

\bibitem{Bousso:2010vi}
Raphael Bousso and Roni Harnik.
\newblock {The Entropic Landscape}.
\newblock {\em Phys.Rev.}, D82:123523, 2010, 1001.1155.

\bibitem{Bousso:2010zi}
Raphael Bousso, Ben Freivogel, Stefan Leichenauer, and Vladimir Rosenhaus.
\newblock {A geometric solution to the coincidence problem, and the size of the
  landscape as the origin of hierarchy}.
\newblock {\em Phys.Rev.Lett.}, 106:101301, 2011, 1011.0714.

\bibitem{Govaerts:2004ba}
Jan Govaerts.
\newblock The cosmological constant of one-dimensional matter coupled quantum
  gravity is quantised.
\newblock In {\em Proc. 3rd International Workshop on Contemporary Problems in
  Mathematical Physics, 1-7 November 2003, Cotonou (Benin), eds. J. Govaerts,
  M.N. Hounkonnou and A.Z. Msezane}, pages 244--272. World Scientific
  (Singapore), 2004, hep-th/0408022.

\bibitem{Schaller:1994es}
Peter Schaller and Thomas Strobl.
\newblock {Poisson structure induced (topological) field theories}.
\newblock {\em Mod.Phys.Lett.}, A9:3129--3136, 1994, hep-th/9405110.

\bibitem{Kummer:1998yg}
W.~Kummer and G.~Tieber.
\newblock {Universal conservation law and modified Noether symmetry in 2-D
  models of gravity with matter}.
\newblock {\em Phys.Rev.}, D59:044001, 1999, hep-th/9807122.

\bibitem{Grumiller:2002nm}
D.~Grumiller, W.~Kummer, and D.~V. Vassilevich.
\newblock {Dilaton gravity in two dimensions}.
\newblock {\em Phys. Rept.}, 369:327--430, 2002, hep-th/0204253.

\bibitem{Grumiller:2006rc}
Daniel Grumiller and Rene Meyer.
\newblock {Ramifications of lineland}.
\newblock {\em Turk.J.Phys.}, 30:349--378, 2006, hep-th/0604049.

\bibitem{Copeland:2006wr}
Edmund~J. Copeland, M.~Sami, and Shinji Tsujikawa.
\newblock {Dynamics of dark energy}.
\newblock {\em Int.J.Mod.Phys.}, D15:1753--1936, 2006, hep-th/0603057.

\bibitem{Zonetti:2011ky}
Simone Zonetti and Jan Govaerts.
\newblock {Duality between 1+1 dimensional Maxwell-Dilaton gravity and
  Liouville field theory}.
\newblock {\em J. Phys. A: Math. Theor. 45 (2012) 042001}, 2012, 1111.1612.

\bibitem{Witten:1991yr}
Edward Witten.
\newblock {On string theory and black holes}.
\newblock {\em Phys.Rev.}, D44:314--324, 1991.

\bibitem{Elitzur:1991cb}
S.~Elitzur, A.~Forge, and E.~Rabinovici.
\newblock {Some global aspects of string compactifications}.
\newblock {\em Nucl.Phys.}, B359:581--610, 1991.

\bibitem{Mandal:1991tz}
Gautam Mandal, Anirvan~M. Sengupta, and Spenta~R. Wadia.
\newblock {Classical solutions of two-dimensional string theory}.
\newblock {\em Mod.Phys.Lett.}, A6:1685--1692, 1991.

\bibitem{Callan:1992zr}
Curtis Callan, Steven Giddings, Jeffrey Harvey, and Andrew Strominger.
\newblock Evanescent black holes.
\newblock {\em Physical Review D}, 45(4):R1005, 1992.

\bibitem{Nakayama:2004vk}
Yu~Nakayama.
\newblock {Liouville field theory: A Decade after the revolution}.
\newblock {\em Int.J.Mod.Phys.}, A19:2771--2930, 2004, hep-th/0402009.

\bibitem{Govaerts:2011p3916}
Jan Govaerts and Simone Zonetti.
\newblock Quantized cosmological constant in 1+1 dimensional quantum gravity
  with coupled scalar matter.
\newblock {\em Class. Quantum Grav.}, 28:185001, 2011, 1102.4957v1.

\bibitem{Curtright:1982}
Thomas~L. Curtright and Charles~B. Thorn.
\newblock Conformally invariant quantization of the liouville theory.
\newblock {\em Phys. Rev. Lett.}, 48(19):1309--1313, May 1982.

\bibitem{Henneaux:1994}
Marc Henneaux and Claudio Teitelboim.
\newblock {\em Quantization of Gauge Systems}.
\newblock {Princeton University Press}, August 1994.

\bibitem{Govaerts:1991}
Jan Govaerts.
\newblock {\em Hamiltonian Quantisation and Constrained Dynamics}, volume~4 of
  {\em Leuven Notes in Mathematical and Theoretical Physics}.
\newblock Leuven University Press, 1991.

\bibitem{Becchi:1974md}
C.~Becchi, A.~Rouet, and R.~Stora.
\newblock {Renormalization of the Abelian Higgs-Kibble Model}.
\newblock {\em Commun.Math.Phys.}, 42:127--162, 1975.

\bibitem{Tyutin:1975qk}
I.V. Tyutin.
\newblock {Gauge Invariance in Field Theory and Statistical Physics in Operator
  Formalism}.
\newblock 1975, 0812.0580.

\bibitem{Difrancesco:1999}
Philippe Di~Francesco, Pierre Mathieu, and David Senechal.
\newblock {\em Conformal Field Theory}.
\newblock Springer, corrected edition, January 1999.

\bibitem{Ginsparg:1988ui}
Paul~H. Ginsparg.
\newblock {Applied Conformal Field Theory}.
\newblock 1988, hep-th/9108028.

\bibitem{Blumenhagen:2009zz}
Ralph Blumenhagen and Erik Plauschinn.
\newblock {Introduction to conformal field theory}.
\newblock {\em Lect. Notes Phys.}, 779:1--256, 2009.

\bibitem{Govaerts1989186}
Jan Govaerts.
\newblock String theory constructions: An introduction to modern methods.
\newblock {\em Nuclear Physics B - Proceedings Supplements}, 11(0):186 -- 222,
  1989.

\bibitem{Polchinski:1998rq}
J.~Polchinski.
\newblock {\em {String Theory Vol. 1: An Introduction to the Bosonic String}}.
\newblock Cambridge Univ. Pr., 1998.

\bibitem{Green:1987sp}
Michael~B. Green, J.~H. Schwarz, and Edward Witten.
\newblock {\em {Superstring Theory Vol. 1: Introduction}}.
\newblock Cambridge Monographs On Mathematical Physics. Cambridge Univ. Pr.,
  1987.

\bibitem{Klauder:1968dq}
J.~R. Klauder and G.~Sudarshan.
\newblock {\em {Fundamentals of Quantum Optics}}.
\newblock Benjamin, New York, 1968.

\bibitem{Meyer:2005lr}
R.~Meyer.
\newblock Constraints in two-dimensional dilaton gravity with fermions.
\newblock 2005, 0512267.

\bibitem{Ruegg:2003ps}
Henri Ruegg and Marti Ruiz-Altaba.
\newblock {The Stuckelberg field}.
\newblock {\em Int.J.Mod.Phys.}, A19:3265--3348, 2004, hep-th/0304245.

\bibitem{kucharbrown:1995}
J.~David Brown and Karel~V. Kuchar.
\newblock Dust as a standard of space and time in canonical quantum gravity.
\newblock {\em Phys. Rev. D}, 51:5600--5629, 1995, gr-qc/9409001.

\bibitem{Cianfrani:2008p574}
Francesco Cianfrani, Giovanni Montani, and Simone Zonetti.
\newblock Definition of a time variable with entropy of a perfect fluid in
  canonical quantum gravity.
\newblock {\em Class.Quant.Grav}, 26(125002), Jan 2009, 0807.3281v2.

\bibitem{schutz:1970}
Bernard~F. Schutz.
\newblock Perfect fluids in general relativity: Velocity potentials and a
  variational principle.
\newblock {\em Phys. Rev. D}, 2(12):2762--2773, Dec 1970.

\bibitem{schutz:1971}
Bernard~F. Schutz.
\newblock Hamiltonian theory of a relativistic perfect fluid.
\newblock {\em Phys. Rev. D}, 4(12):3559--3566, Dec 1971.

\bibitem{Murray:2010p3917}
Sean Murray, Christophe Ringeval, and Simone Zonetti.
\newblock Graviton confinement inside hypermonopoles of any dimension.
\newblock {\em J. Cosmol. Astropart. Phys., JCAP}, 1009:015, 2010, 1002.5021v2.

\bibitem{Klauder:1999ba}
John~R. Klauder.
\newblock {Noncanonical quantization of gravity. I. foundations of affine
  quantum gravity}.
\newblock {\em J.Math.Phys.}, 40:5860--5882, 1999, gr-qc/9906013.

\bibitem{Klauder:2011ue}
John~R. Klauder.
\newblock {The Utility of Affine Variables and Affine Coherent States}.
\newblock 2011, 1108.3380.

\bibitem{Fanuel:2012fk}
Michael Fanuel and Simone Zonetti.
\newblock {Affine Quantization and the Initial Cosmological Singularity}.
\newblock 2012, 1203.4936.

\end{thebibliography}
\end{document}